\documentclass[11pt]{article}
\usepackage{graphicx}
\usepackage{amsmath,amssymb}
\usepackage{authblk}
\usepackage{cite}
\usepackage{color}

\usepackage{makeidx}
\usepackage{fixmath}
\usepackage{amsfonts}
\newcommand{\beqa}{\begin{eqnarray}}
\newcommand{\eeqa}[1]{\label{#1}\end{eqnarray}}
\newcommand{\bequ}{\begin{equation}}
\newcommand{\eequ}[1]{\label{#1}\end{equation}}

\newcommand{\beq}{\begin{equation}}
\newcommand{\eeq}{\end{equation}}
\newcommand{\overliner}{\begin{eqnarray}}
\newcommand{\earr}{\end{eqnarray}}
\newcommand{\beqn}{\begin{equation*}}
\newcommand{\eeqn}{\end{equation*}}
\newcommand{\overlinern}{\begin{eqnarray*}}
\newcommand{\earrn}{\end{eqnarray*}}

\usepackage{comment}
\usepackage{float}

\usepackage{authblk}
\usepackage{booktabs}
\usepackage{mathrsfs}

\usepackage{subfig}

\usepackage{tikz}
\usetikzlibrary{decorations.pathmorphing}
\usetikzlibrary{decorations.markings}
\usetikzlibrary{shadows}
\usepackage{pgfplots}
\usetikzlibrary{shapes.misc}
\usetikzlibrary{shapes,snakes}
\usepackage[abs]{overpic}

\begin{document}
\title{Semi-infinite herringbone waveguides in elastic plates}
\label{firstpage}

\date{}
\author{\normalsize{S.G. Haslinger$^1$, I.S. Jones$^2$, N.V. Movchan$^1$ \& A.B. Movchan$^1$} \\
\footnotesize{$^1$ Department of Mathematical Sciences, University of Liverpool, Liverpool L69 7ZL, UK} \\
\footnotesize{$^2$ Mechanical Engineering and Materials Research Centre, Liverpool John Moores University,} \\ 
\footnotesize{Liverpool L3 3AF, UK }\\ }

\maketitle

\begin{abstract}
\noindent
The paper includes novel results for the scattering and localisation of a time-harmonic flexural wave by a semi-infinite herringbone waveguide of rigid pins embedded within an elastic Kirchhoff plate. The analytical model takes into account the orientation and spacing of the constituent parts of the herringbone system, and incorporates dipole approximations for the case of closely spaced pins. Illustrative examples are provided, together with the predictive theoretical analysis of the localised waveforms.

\end{abstract}
\section{Introduction}
Herringbone systems are a source of great interest in the scientific community across a broad spectrum of fields encompassing branches of physics, chemistry and biology. In crystallography and solid-state physics, the preference for herringbone-type close packing of crystals has long been a topic of study for many groups of researchers (see for example the article by Arlt and Sasko~\cite{arlt} concerning the domain configuration in barium titanate ceramics). 

The subject of domain structure to minimise the energy of crystals and grains is still highly relevant today in modern technologies involving ferroelectric and piezoelectric materials. A formal classification of all the rank-2 laminate arrangements for a ferroelectric single crystal was given in~\cite{tsou}, with half featuring herringbone patterns on at least one surface. The patterns are significant in such materials because the geometric arrangement of domains influences both the macroscopic and microscopic properties, including elastic moduli and dielectric permittivity. The characteristic switching behaviour of a ferroelectric is also strongly dependent on the domain pattern~\cite{glasgow}. 
\maketitle

Another related application area is the study of organic semiconductors (OSCs) which are used in organic field effect transistors, organic solar cells and organic LEDs~\cite{vyas}. All of these devices are dependent on the solid state packing of the OSCs, with, for example, the charge transport of the organic field effect transistors being one important property governed by the molecular packing arrangement~\cite{wang}.

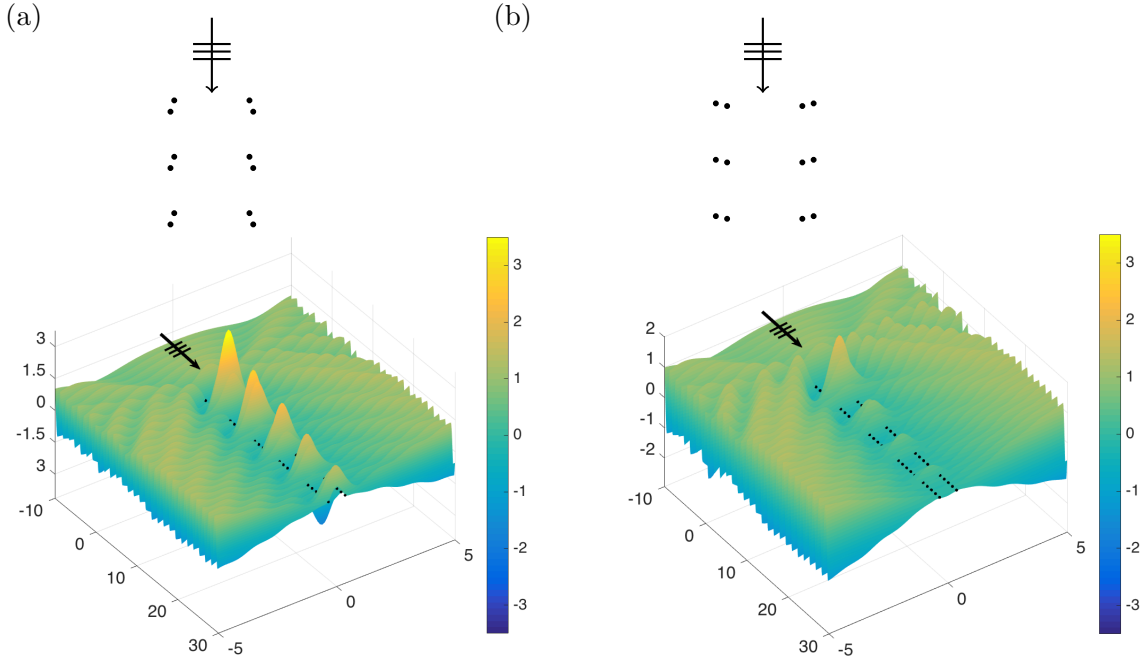
\begin{figure}[h]
\begin{tikzpicture}[thick,scale=0.5] 
\draw [fill] (2.0,1.3) circle [radius=0.05];
\draw [fill] (1.9,1.0) circle [radius=0.05];
\draw [fill] (4.0,1.3) circle [radius=0.05];
\draw [fill] (4.1,1.0) circle [radius=0.05];
\draw [fill] (2.0,2.8) circle [radius=0.05];
\draw [fill] (1.9,2.5) circle [radius=0.05];
\draw [fill] (4.0,2.8) circle [radius=0.05];
\draw [fill] (4.1,2.5) circle [radius=0.05];
\draw [fill] (2.0,-0.2) circle [radius=0.05];
\draw [fill] (1.9,-0.5) circle [radius=0.05];
\draw [fill] (4.0,-0.2) circle [radius=0.05];
\draw [fill] (4.1,-0.5) circle [radius=0.05];
\draw  [->, thick] (3.0,5.0) -- (3.0,3.0);
\draw  [-, thick] (2.5,4.3) -- (3.5,4.3);
\draw  [-, thick] (2.5,4.1) -- (3.5,4.1);
\draw  [-, thick] (2.5,3.9) -- (3.5,3.9);

\node at (-2.0,5){(a)};

\node at (11.0,5){(b)};

\draw [fill] (19.0,1.22) circle [radius=0.05];
\draw [fill] (18.7,1.15) circle [radius=0.05];
\draw [fill] (16.7,1.15) circle [radius=0.05];
\draw [fill] (16.4,1.22) circle [radius=0.05];

\draw [fill] (19.0,2.72) circle [radius=0.05];
\draw [fill] (18.7,2.65) circle [radius=0.05];
\draw [fill] (16.7,2.65) circle [radius=0.05];
\draw [fill] (16.4,2.72) circle [radius=0.05];

\draw [fill] (19.0,-0.28) circle [radius=0.05];
\draw [fill] (18.7,-0.35) circle [radius=0.05];
\draw [fill] (16.7,-0.35) circle [radius=0.05];
\draw [fill] (16.4,-0.28) circle [radius=0.05];

\draw  [->, thick] (17.65,5.0) -- (17.65,3.0);
\draw  [-, thick] (17.15,4.3) -- (18.15,4.3);
\draw  [-, thick] (17.15,4.1) -- (18.15,4.1);
\draw  [-, thick] (17.15,3.9) -- (18.15,3.9);

\end{tikzpicture}

\centerline{
         \includegraphics[width=.45\columnwidth]{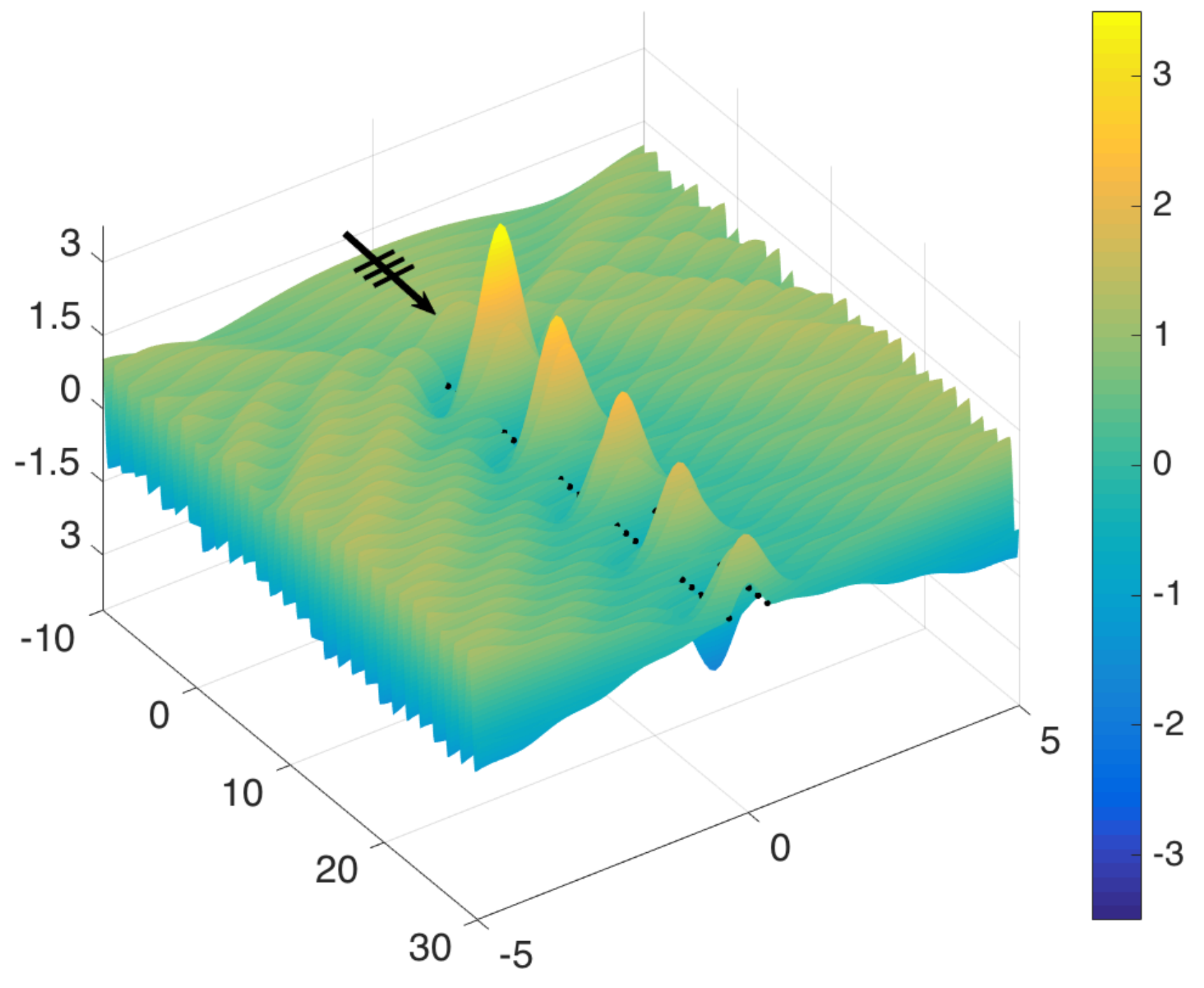}~~~~~~~~
          \includegraphics[width=.45\columnwidth]{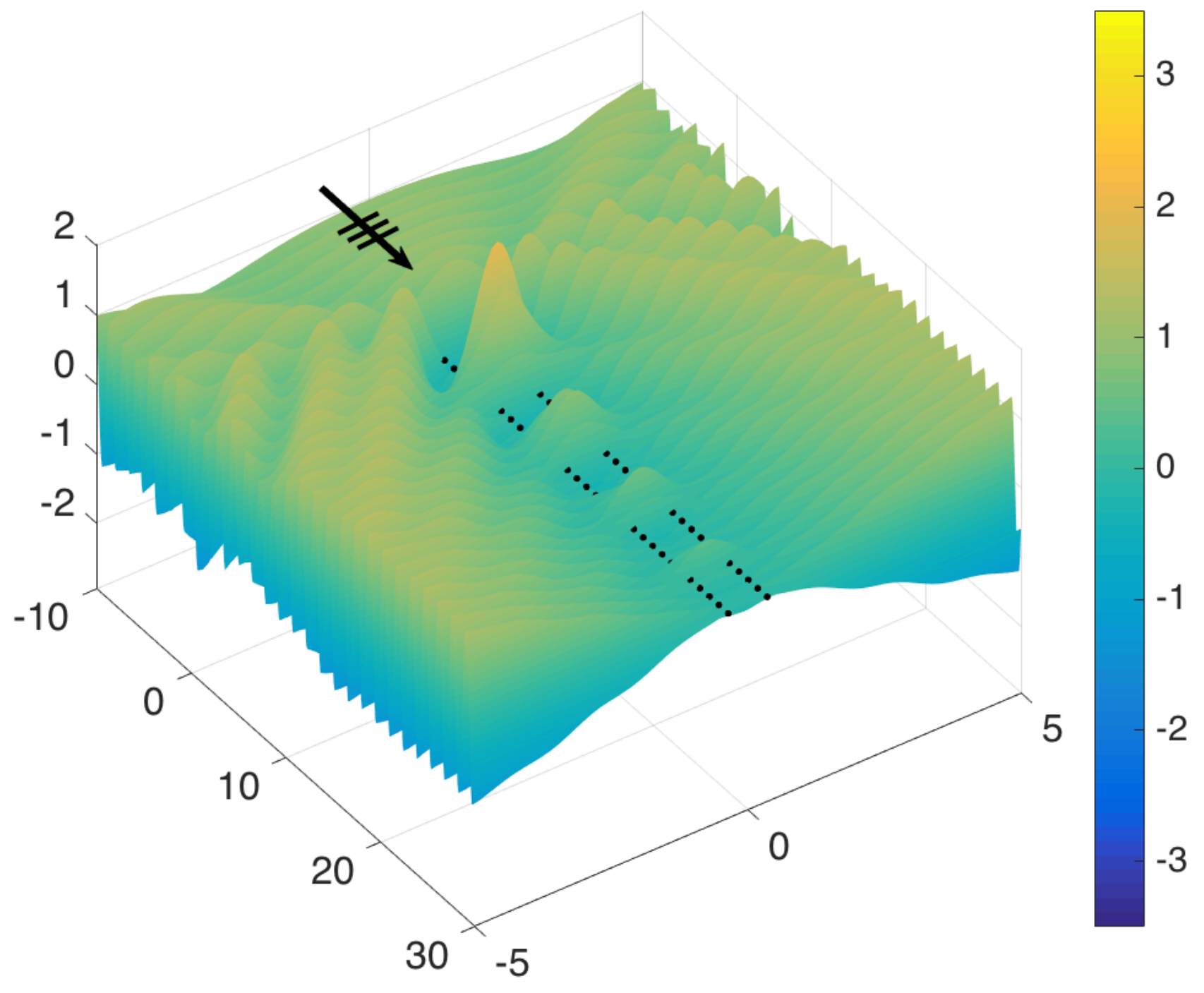}
         }
\caption{ 
Displacement fields for two flexural herringbone waveguides with all parameters prescribed identically except for the orientation of the dipoles, for the case of normal incidence. The pairs are aligned to form (a) convex entrance, (b) concave entrance. For an aluminium plate of thickness 5mm, the frequency is 23.3 Hz in both cases.}   
\label{introfig}
\end{figure}

In this paper, we design and model a novel herringbone system constructed within an elastic Kirchhoff plate. We show how the parameters governing the herringbone geometry may be tuned to optimise waveguiding and localisation of flexural waves within a structured plate, e.g. see Fig.~\ref{introfig}. In a similar way to the herringbone arrangement of crystals being favourable for thin film transport~\cite{zhang}, we show that the concept of cladding a simple two-grating waveguide with an additional pair of appropriately located gratings enhances the waveguiding effect. Adopting a wave scattering method and a dipole approximation assumption, we demonstrate an elegant mathematical formulation and solution to the scattering of flexural plane waves by an elastic plate pinned in a herringbone fashion.

In recent years, there has been substantial interest in wave propagation in structured elastic plates, motivated by the abundance of potential applications in engineering and materials science. The ability to customise systems to control the direction and amplification of flexural waves is important for the design of metamaterials and micro-structured systems that possess special properties unattainable with natural materials.

An attraction of structured Kirchhoff plates is that many of the methods and ideas associated with photonic crystals can be applied to platonic crystals~\cite{RCM_ABM_NVM}. The fourth order biharmonic operator introduces mathematical subtleties to the analysis of the scattering of flexural waves compared with the usual second-order derivatives for the wave equation of optics and acoustics, but with an added advantage. The biharmonic two-dimensional Green's function is bounded, rather than diverging logarithmically at the source point (as in the case of the Helmholtz operator). This feature is of particular importance for the special case of periodically pinned elastic plates. 

Numerous methods have been implemented, including Fourier series expansions by, amongst others, Mead (see \cite{mead} for a review), and the use of Green's functions by Evans and Porter\cite{Evans} and Smith {\it et al.}~\cite{MJAS_RP_TDW}, and more recently by Antonakakis and Craster~\cite{anton1}, \cite{anton2}. 
A complementary approach using multipole methods has been employed by McPhedran, Movchan and Movchan in a series of papers~\cite{RCM_ABM_NVM}, ~\cite{ABM_NVM_RCM} -\cite{Dirac}, where the limiting case of a small hole with clamped edge was considered, recovering the case of rigid pins. Dispersion surfaces and band diagrams were used to predict a variety of interesting wave phenomena, including shielding, focussing and negative refraction.

Bloch-Floquet analysis for an elementary cell is a common approach to the modelling of infinite periodic systems, but for a semi-infinite platonic crystal (an infinite plate containing a semi-infinite periodic array, not to be confused with a pinned semi-infinite plate as in e.g.~\cite{Evans_Porter}) this technique is no longer directly applicable. Recent analysis for flexural wave scattering has been conducted by Haslinger {\it et al.}~\cite{SGH_RVC_ABM_NVM_ISJ}, \cite{SGH_NVM_ABM_ISJ_RVC} and Jones {\it et al.}~\cite{ISJ_NVM_ABM}. The methods of solution adopted by these authors included a discrete Wiener-Hopf method and a wave scattering technique inspired by the classical papers of, respectively, Hills and Karp~\cite{hills1} and Foldy~\cite{foldy}, for related problems of two-dimensional membrane waves in discrete semi-infinite clusters. A review of the problem of acoustic scattering by a two-dimensional semi-infinite periodic array of isotropic point scatterers is provided by Linton and Martin~\cite{linton}.

There has also been extensive interest in the scattering of plane waves by semi-infinite crystals in electromagnetism. Early investigations include the detailed coverage provided by Mahan and Obermair~\cite{mahan} and Mead~\cite{mead_ca}, where nearest-neighbour and dielectic approximations were analysed and discussed. More recently, research has been conducted for applications in the design of metamaterials, for example in~\cite{belov}, where a point dipole approximation for sufficiently small scatterers was implemented. In addition, the authors of~\cite{belov} used knowledge of the eigenmodes for infinite crystals to give insight to the problems of scattering of plane waves by analogously composed semi-infinite crystals. Another recent study~\cite{albani} considered the wave dynamics at the interface of a homogeneous half-space and a half-space of plasmonic nanospheres, using a discrete Wiener-Hopf technique incorporating the assumption that each nanosphere may be described by the single dipole approximation.

The present paper addresses the scattering of flexural plane waves propagating in a structured Kirchhoff plate. A novel design depicts a waveguide consisting of a pair of pinned gratings, which is augmented by an extra pair of gratings, each of which is positioned exterior, but close, to the original set. Defining shift vectors 
for each of the upper and lower pairs, a herringbone system is constructed. A natural configuration to consider is a regular double-pinned structure (symmetric herringbone). One may also classify the special cases for which the leading pair of gratings is either the inner pair (convex entrance) or the outer pair (concave entrance), as illustrated in Fig.~\ref{introfig}. The twin parameters of magnitude and orientation of the shift vectors are used to design herringbone systems to guide and direct waves. 

The proximity of the constituent members of the shifted pairs promotes the use of a dipole approximation for the pairs of closely spaced pins. The first part of the paper analyses the case of a shifted pair of semi-infinite gratings in detail, with an emphasis on using the dipole approximation. This idea is taken further by considering the replacement of each of the dipole pairs by an array of points with two prescribed boundary conditions, zero displacement and zero directional derivative. 

We first consider the problem of a single shifted pair in Sections~\ref{alg_sys} and~\ref{kerher}, followed by the semi-infinite line array of sources and dipoles in Section~\ref{sec:dipole1}. Illustrative examples and comparisons of the two approaches are demonstrated in Section~\ref{sec:illus}. We then formulate the problem for the herringbone system, in both its exact form in Section~\ref{alg_sys_hb}, and with the dipole approximations in Sections~\ref{sec:dipole1HB}, \ref{hb_new}. We illustrate the waveguiding and localisation capabilities, combining the concepts of the dipole approximation and waveguide analysis in Section~\ref{hb_illus}. The latter method recalls the ideas used for both the infinite grating waveguide in~\cite{has2014}, and the semi-infinite waveguide in~\cite{ISJ_NVM_ABM}. The former paper incorporates the eigenvalue problem for finite stacks of shifted infinite gratings, whilst the latter article uses a similar technique to identify blocking and trapping regimes for a pair of parallel semi-infinite gratings. Concluding remarks are drawn together in Section~\ref{conc}.

\section{Kirchhoff plate with a pair of shifted semi-infinite rows of pins}
\label{formulation}

\begin{figure}[h]
\centerline{
         \includegraphics[width=.75\columnwidth]{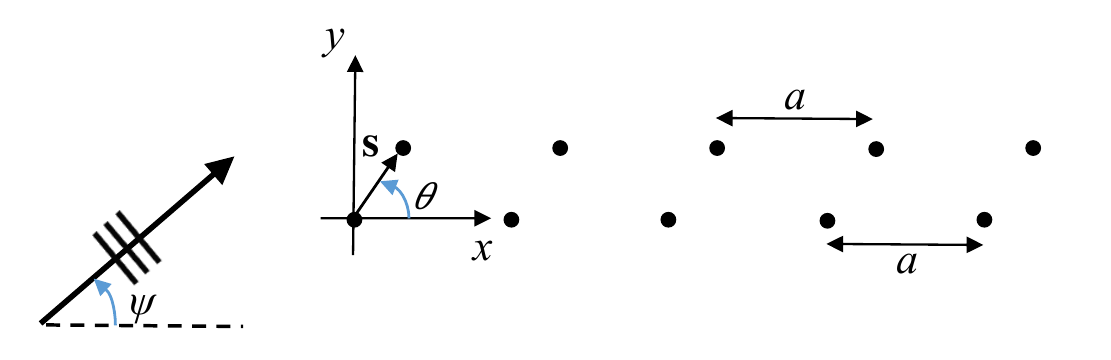}
         \put(8,12) {(I)}
         \put(8,23) {(II)}
         }
\caption{ 
Two semi-infinite horizontal lines of rigid pins with spacing $a$ in an elastic Kirchhoff plate. A plane wave is incident at an angle $\psi$ and the lines are shifted relative to one another according to the shift vector ${\bf s}$ with orientation $\theta$.}
\label{diag}
\end{figure}

We introduce a model problem for a pair of shifted semi-infinite rows of rigid pins embedded in a Kirchhoff elastic plate (see Fig.~\ref{diag}). 
Two-dimensional axes are chosen as shown and one line of pins is shifted relative to the other by the shift vector ${\bf s}
=s_1 {\bf i } + s_2 {\bf j} $. 
The horizontal spacing between the pins in a single grating is represented by $a > |{\bf s}| > 0$. A plane wave is incident at an angle $\psi$ on the lines of pins, which occupy the positive half plane $x \geq 0$ (Fig.~\ref{diag}).

\subsection{Algebraic system}
\label{alg_sys}
In the time-harmonic regime, the amplitude $u$  of the flexural displacement of a homogeneous Kirchhoff plate satisfies the governing equation
\begin{equation}
\Delta^2 u - \beta^4 u = 0, \label{eq1}
\end{equation}
with $\beta^4 = \rho h \omega^2 /D$; here $h$ is the plate thickness, $\rho$ is its density (mass per unit volume), $\omega$ is the radian frequency, and $D = Eh^3/(12 (1-\nu^2))$ is the flexural rigidity of the plate; $E$ and $\nu$ are the Young's modulus and the Poisson ratio, respectively, of the plate.  

Consider an incident field at the point ${\bf r}=(x,y)$ that possesses the amplitude $u_{\scriptsize{\mbox{inc}}}({\bf r})$ given by 
\begin{equation}
u_{\scriptsize{\mbox{inc}}}({\bf r}) = u_{\scriptsize{\mbox{inc}}}(x,y)=e^{i\beta (x\cos \psi + y\sin\psi)}. 
\end{equation} 
In the two-dimensional case, the single-source Green's function satisfying the equation
\begin{equation}
\Delta^2 g(\beta;{\bf r}; {\bf r'}) - \beta^4 g(\beta;{\bf r}; {\bf r'}) = \delta ({\bf r} - {\bf r'}),
\end{equation}
is expressed as
\begin{equation}
g(\beta;{\bf r}; {\bf r'})=\frac{i}{8\beta^2}\Bigg[H_0^{(1)}(\beta  |{\bf r}-{\bf r'}|)+\frac{2i}{\pi}K_0(\beta |{\bf r}-{\bf r'}|)\Bigg],
\label{gf}
\end{equation}
with respect to the source point ${\bf r'} = (x',y')$. The Green's function is finite at this point, rather than diverging logarithmically as is the case for the two-dimensional Helmholtz operator. This non-singular property is useful in the wave scattering approach we employ, whereby the scattered field is expressed as a sum of Green's functions, a method widely reported for acoustics by, amongst others,~\cite{hills1} -\cite{linton}. 

The classical work on the scattering of acoustic waves by semi-infinite gratings is a paper by Hills and Karp\cite{hills1}, which addressed the case of small sound-soft circular cylinders. They followed the technique of~\cite{foldy}, which assumed isotropic point scatterers so that in the neighbourhood of an individual defect, the scattered field may be represented by a term incorporating an unknown coefficient and the free-space Green's function for the Helmholtz operator centred at that scatterer. 
This method is particularly well adapted to the problem presented here, since the assumption that the grating's elements are small, and therefore scatter isotropically, is not formally required for the pinned biharmonic plate. The additional attribute of the non-singular Green's function~(\ref{gf}) is also advantageous for this wave scattering method.

The total field is represented by the superposition of the incident and scattered fields, taking into account all of the scatterers whose unknown intensities are to be determined. Thus, 
the total flexural displacement $u(x,y)$ for the pair of platonic gratings is expressed as
\begin{equation}
u(x,y)=u_{\scriptsize{\mbox{inc}}}(x,y)+\sum_{n=0}^\infty A_n^{\scriptsize \mbox{(I)}} g(\beta;x,y;na,0) + \sum_{m=0}^\infty A_m^{\scriptsize \mbox{(II)}} g(\beta;x,y;s_1+ma,s_2),
\label{totfield}
\end{equation}
where the scattered field coefficients $A_n^{\scriptsize \mbox{(I)}}$ and $A_m^{\scriptsize \mbox{(II)}}$ are to be determined for, respectively, gratings I and II (see Fig.~\ref{diag}).  

Setting the displacement $u(x,y)$ to vanish at the rigid pins located at $(ja, 0)$ and $(s_1+la,s_2)$, we obtain a system of linear algebraic equations for the coefficients $A_n^{\scriptsize \mbox{(I)}}$ and  $A_m^{\scriptsize \mbox{(II)}}$:
\begin{equation}
\begin{split}
\hspace*{-2cm} -e^{i\beta j a \cos\psi}=\sum_{n=0}^\infty A_n^{\scriptsize \mbox{(I)}} g(\beta;ja,0;na,0) + \sum_{m=0}^\infty A_m^{\scriptsize \mbox{(II)}} g(\beta;ja,0;s_1+ma,s_2), \\ j=0,1,2, \ldots  
\end{split}
\label{alg_sys_1}
\end{equation}
\begin{equation}
\begin{split}
 -e^{i\beta[(s_1+la)\cos\psi+s_2\sin\psi]} =
\sum_{n=0}^\infty A_n^{\scriptsize \mbox{(I)}} g(\beta;s_1+la,s_2;na,0) ~~~~~~~~~~~~~~~~~~~~~~~~~~~~~~~~~~\\+ 
\sum_{m=0}^\infty A_m^{\scriptsize \mbox{(II)}} g(\beta;s_1+
la,s_2;s_1+ma,s_2),~~~   l=0,1,2, \ldots 
\end{split}
\label{alg_sys_2}
\end{equation}

\subsection{The kernel  function}
\label{kerher}
For the semi-infinite system of scatterers, we employ a discrete Wiener-Hopf approach, as first implemented by~\cite{hills1}, and recently in a Kirchhoff plate setting in~\cite{SGH_RVC_ABM_NVM_ISJ} -~\cite{ISJ_NVM_ABM}. 
We extend the semi-infinite domain to infinity in the negative direction by introducing the following notations for $N, M \in \mathbb{Z}$:
\begin{equation}
u(Na, 0)=\Bigg \{ 
\begin{array}{c l}  
0,  & \quad\quad N \ge 0 \\
\\
B_N^{\scriptsize \mbox{(I)}}, &  \quad\quad N  <  0
\label{b1n}
\end{array} 
\end{equation}
\begin{equation}
\hspace{-8mm}
u(s_1+Ma, s_2)=\Bigg \{ 
\begin{array}{c l}  
0, & \quad \,\, M \ge 0 \\
\\
B_M^{\scriptsize \mbox{(II)}}, & \quad\,\, M  <  0
\end{array}  
\label{b2m}
\end{equation}
\begin{equation}
u_{\scriptsize{\mbox{inc}}}(Na, 0)=F_N^{\scriptsize \mbox{(I)}},~~~~~~\quad
u_{\scriptsize{\mbox{inc}}}(s_1+Ma, s_2)=F_M^{\scriptsize \mbox{(II)}}.
\label{f12}
\end{equation} 
Here $B_N^{\scriptsize \mbox{(I)}}$ and $B_M^{\scriptsize \mbox{(II)}}$ represent the unknown amplitudes of the total flexural displacement at the points $(Na, 0)$ and $(s_1+Ma, s_2)$ for, respectively, $N,M < 0$ i.e. in the region to the left of the pair of gratings. The field incident at the array points is denoted by $F_N^{\scriptsize \mbox{(I)}}, F_M^{\scriptsize \mbox{(II)}}$ for all $N,M \in \mathbb{Z}$. 

Consider the displacement at two field points  ${\bf r}_N = (Na,0)$ and ${\bf r}_M = (s_1+Ma,s_2)$.  Using definitions~(\ref{b1n})-(\ref{f12}), we extend the evaluation of the coefficients $A_n^{\scriptsize \mbox{(I)}}, A_m^{\scriptsize \mbox{(II)}}$ to $m,n < 0$, setting them to be zero since the pins are not present in this region. Similarly, the notations of $B_N^{\scriptsize \mbox{(I)}}$, $B_M^{\scriptsize \mbox{(II)}}$ are extended to all $N, M \in \mathbb{Z}$, assuming that $B_N^{\scriptsize \mbox{(I)}}$, $B_M^{\scriptsize \mbox{(II)}} = 0$ for $N, M \ge 0$. 

Applying the discrete Fourier Transform to equations~(\ref{alg_sys_1})-(\ref{f12}) and using the Fourier variable $k$, we deduce, respectively,
\begin{equation}
\begin{split}
\sum_{N=-\infty}^\infty B_N^{\scriptsize \mbox{(I)}}e^{ikNa}=\sum_{N=-\infty}^\infty F_N^{\scriptsize \mbox{(I)}}e^{ikNa}+\sum_{N=-\infty}^\infty \sum_{n=-\infty}^\infty A_n^{\scriptsize \mbox{(I)}} g(\beta;Na,0;na,0)e^{ikNa} \\+\sum_{N=-\infty}^\infty \sum_{m=-\infty}^\infty A_m^{\scriptsize \mbox{(II)}} g(\beta;Na,0;s_1+ma,s_2)e^{ikNa},\quad\quad\quad
\end{split}
\label{eq1}
\end{equation}
and
\begin{equation}
\begin{split}
\sum_{M=-\infty}^\infty B_M^{\scriptsize \mbox{(II)}}e^{ikMa}=\sum_{M=-\infty}^\infty F_M^{\scriptsize \mbox{(II)}}e^{ikMa}+\sum_{M=-\infty}^\infty \sum_{n=-\infty}^\infty A_n^{\scriptsize \mbox{(I)}} g(\beta;s_1+Ma, s_2;na,0)e^{ikMa} \\+\sum_{M=-\infty}^\infty \sum_{m=-\infty}^\infty A_m^{\scriptsize \mbox{(II)}} g(\beta;s_1+Ma, s_2;s_1+ma,s_2)e^{ikMa}. \quad\quad\quad
\end{split}
\label{eq2}
\end{equation}
By a change of indices of summation, the above equations can be rewritten in the form:

\begin{equation}
\begin{split}
\sum_{N=-\infty}^\infty B_N^{\scriptsize \mbox{(I)}}e^{ikNa}=\sum_{N=-\infty}^\infty F_N^{\scriptsize \mbox{(I)}}e^{ikNa}+\sum_{n=-\infty}^\infty  A_n^{\scriptsize \mbox{(I)}}e^{ikna} \sum_{j=-\infty}^\infty g(\beta;ja,0;0,0)e^{ikja} \\+\sum_{m=-\infty}^\infty  A_m^{\scriptsize \mbox{(II)}} e^{ikma}\sum_{j=-\infty}^\infty g(\beta;ja,0;s_1,s_2)e^{ikja}, \quad\quad\quad\quad\quad
\end{split}
\label{eq12}
\end{equation}

\begin{equation}
\begin{split}
\sum_{M=-\infty}^\infty B_M^{\scriptsize \mbox{(II)}}e^{ikMa}=\sum_{M=-\infty}^\infty F_M^{\scriptsize \mbox{(II)}}e^{ikMa}+\sum_{n=-\infty}^\infty  A_n^{\scriptsize \mbox{(I)}}e^{ikna}\sum_{j=-\infty}^\infty g(\beta;s_1+ja, s_2;0,0)e^{ikja} \\+\sum_{m=-\infty}^\infty \ A_m^{\scriptsize \mbox{(II)}}e^{ikma} \sum_{j=-\infty}^\infty g(\beta;s_1+ja, s_2;s_1,s_2)e^{ikja}. \quad\quad\quad\quad\quad
\end{split}
\label{eq13}
\end{equation}
Here we adopt the notation of~\cite{SGH_RVC_ABM_NVM_ISJ} -~\cite{ISJ_NVM_ABM}:
\begin{equation}
  \hat B_-^{(\alpha)} = \sum_{N=-\infty}^\infty B_N^{(\alpha)}e^{ikNa}, ~ ~~\hat F^{(\alpha)} = \sum_{N=-\infty}^\infty F_N^{(\alpha)}e^{ikNa},
~~~\hat A_+^{(\alpha)} = \sum_{n=-\infty}^\infty A_n^{(\alpha)}e^{ikna}, \,\,\quad \alpha =  \mbox{I},  \mbox{II}.
\label{nots}
\end{equation}
Using the quasi-periodicity of the gratings, we define the grating Green's function by
\begin{equation}
\hat G(\beta,k;{\boldsymbol \xi}^{(1)};{\boldsymbol \xi}^{(2)})=\sum_{j=-\infty}^\infty  g(\beta;ja{\bf e_1}+{\boldsymbol \xi}^{(1)}; {\boldsymbol \xi}^{(2)})e^{ikja} . 
\label{short1}
\end{equation}
We see from~(\ref{eq12}) and~(\ref{eq13}) that the only choices of ${\boldsymbol \xi}^{(1)}$ and ${\boldsymbol \xi}^{(2)}$ that are required are the zero vector and ${\bf s}$, which identify the front pins of the two shifted gratings. The additional summation over the coefficients $A_n^{(\alpha)}$ in~(\ref{eq12}),~(\ref{eq13}) is then applied and we obtain the functional equation

\begin{equation}
\left(
\begin{array}{c}
\hat B_-^{\scriptsize \mbox{(I)}}\\
\hat B_-^{\scriptsize \mbox{(II)}}\\
\end{array}
\right)=
\left(
\begin{array}{cc}
\hat{G}(\beta,k;{\bf 0};{\bf 0})  & \hat G(\beta,k;{\bf 0};{\bf s}) \\
\hat G(\beta,k;{\bf s};{\bf 0}) & \hat G(\beta,k;{\bf s};{\bf s})  
\end{array}
\right)
\left(
\begin{array}{c}
\hat A_+^{\scriptsize \mbox{(I)}}\\
\hat A_+^{\scriptsize \mbox{(II)}}\\
\end{array}
\right)+
\left(
\begin{array}{c}
\hat F^{\scriptsize \mbox{(I)}}\\
\hat F^{\scriptsize \mbox{(II)}}\\
\end{array}
\right),
\label{whopf}
\end{equation}
whose kernel function is the matrix of grating Green's functions. With reference to~(\ref{short1}) and the symmetry relations:
\begin{equation}
\hat G(\beta,k;{\bf 0};{\bf s}) = \hat G(\beta,k;{\bf -s};{\bf 0}) \quad \mbox{and} \quad  \hat G(\beta,k;{\bf s};{\bf s}) = \hat G(\beta,k;{\bf 0};{\bf 0}),
\label{symm_whopf_sys}
\end{equation}
the system~(\ref{whopf}) can be rewritten so that all the elements are referenced to the origin (${\boldsymbol \xi}^{(2)} = {\bf 0}$): 
\begin{equation}
\left(
\begin{array}{c}
\hat B_-^{\scriptsize \mbox{(I)}}\\
\hat B_-^{\scriptsize \mbox{(II)}}\\
\end{array}
\right)=
\left(
\begin{array}{cc}
\hat{G}(\beta,k;{\bf 0};{\bf 0})  & \hat G(\beta,k;{\bf -s};{\bf 0}) \\
\hat G(\beta,k;{\bf s};{\bf 0}) & \hat G(\beta,k;{\bf 0};{\bf 0})  
\end{array}
\right)
\left(
\begin{array}{c}
\hat A_+^{\scriptsize \mbox{(I)}}\\
\hat A_+^{\scriptsize \mbox{(II)}}\\
\end{array}
\right)+
\left(
\begin{array}{c}
\hat F^{\scriptsize \mbox{(I)}}\\
\hat F^{\scriptsize \mbox{(II)}}\\
\end{array}
\right).
\label{whopf0}
\end{equation}

Multiple gratings extend the case of the scalar Wiener-Hopf equation, analysed in~\cite{SGH_RVC_ABM_NVM_ISJ} and\cite{ISJ_NVM_ABM}, to the matrix form. A thorough study of the interaction of a time-harmonic plane wave with a semi-infinite lattice of identical circular cylinders was provided in~\cite{tymis}, \cite{tymis2}, whereby the assumption that finite-sized cylinders do not scatter isotropically led to a matrix Wiener-Hopf equation. Tymis and Thompson~\cite{tymis2} adopted a method using the truncation of multipole expansions to derive an approximate system, solved by matching poles and residues on opposing sides. In this way, the necessity to factorise the matrix kernel was avoided. 

Factorisation of the matrix kernel is also not required in our approach. As discussed in~\cite{SGH_RVC_ABM_NVM_ISJ} -\cite{ISJ_NVM_ABM}, an important feature of the discrete Wiener-Hopf method for semi-infinite platonic crystals is the direct connection between the kernel function and quasi-periodic Green's functions for analogous infinite systems. Zeros of these functions correspond to Bloch modes meaning that analysis of the kernel matrix provides information to identify special frequency regimes that support transmission and reflection effects in semi-infinite grating pairs. 

For the $2 \times 2$ system, we employ a natural approximation approach, exploiting the dipole-like structure of both the underlying geometry and the matrix kernel in equations~(\ref{whopf}),~(\ref{whopf0}). The off-diagonal entries $\hat G(\beta,k;{\boldsymbol \xi}^{(1)};{\boldsymbol \xi}^{(2)})$ are approximated by expanding about the origin. The derivation of these representations, and illustrative examples demonstrating the efficiency of the approximation, are presented in Appendix~\ref{sec:offdiag}. In the next section, we present the related concept whereby the pair of closely spaced pinned gratings are replaced by a single line of point scatterers. Two boundary conditions are prescribed at each point, from which we determine expressions for source and dipole coefficients.

\subsection{Single semi-infinite line of sources and dipoles}
\label{sec:dipole1}
The problem for a pair of gratings may be converted to a problem for a single array consisting of both sources and dipoles. The solutions of two auxiliary problems are summed; the intensities of the sources are given by $\hat A_+^{\scriptsize \mbox{(I)}} + \hat A_+^{\scriptsize \mbox{(II)}}$, and those of the dipoles may be represented by the difference $\hat A_+^{\scriptsize \mbox{(II)}} - \hat A_+^{\scriptsize \mbox{(I)}}$. Analysis of the coefficients for the sources and dipoles is used to identify the regimes where one of the sets is dominant. 
Referring to Fig.~\ref{diag}, consider the case when $|{\bf s}| \ll 1$, whereupon the system tends towards an array of directional dipoles located on the $x$-axis. 

To illustrate the concept of the dipole approximation, consider the simple case of a pair of pins positioned at, respectively, ${\boldsymbol \xi}^{(1)}$ and ${\boldsymbol \xi}^{(2)}$. The scattered displacement field $u_{\mbox{\scriptsize sc}}$ may be expressed in the form:
\begin{equation}
u_{\mbox{\scriptsize sc}}({\boldsymbol \xi}) = A_1 g(\beta; {\boldsymbol \xi}; {\boldsymbol \xi}^{(1)}) + A_2 g(\beta; {\boldsymbol \xi}; {\boldsymbol \xi}^{(2)}),
\label{eqsd_prelim}
\end{equation}
where $A_1, A_2$ are scattering coefficients, $g$ is the single-source Green's function~(\ref{gf}) and ${\boldsymbol \xi} = (x, y)$ is a general field point. This representation can be rewritten as
\begin{equation}
u_{\mbox{\scriptsize sc}}({\boldsymbol \xi}) = \frac{1}{2}(A_1 + A_2) \left[ g(\beta; {\boldsymbol \xi}; {\boldsymbol \xi}^{(1)}) + g(\beta; {\boldsymbol \xi}; {\boldsymbol \xi}^{(2)}) \right] + \frac{1}{2} (A_2 - A_1) \left[g(\beta; {\boldsymbol \xi}; {\boldsymbol \xi}^{(2)}) - g(\beta; {\boldsymbol \xi}; {\boldsymbol \xi}^{(1)}) \right]. 
\label{eqsd}
\end{equation}
Consider the case when $| {\boldsymbol \xi}^{(2)} - {\boldsymbol \xi}^{(1)}| \ll 1$, and ${\boldsymbol \xi}$ is fixed. Taking ${\boldsymbol \xi}^{(2)} = {\boldsymbol \xi}^{(1)} + {\bf s}, |{\bf s}| \ll 1$, 
\begin{equation}
g(\beta; {\boldsymbol \xi}; {\boldsymbol \xi}^{(1)} + {\bf s}) - g(\beta; {\boldsymbol \xi}; {\boldsymbol \xi}^{(1)}) = {\bf s}   \cdot \nabla_{{\boldsymbol \xi}^{(1)}}  g(\beta; {\boldsymbol \xi}; {\boldsymbol \xi}^{(1)}) + O(|{\bf s}|^2). 
\end{equation}
Then the dipole approximation is
\begin{eqnarray}
u_{\mbox{\scriptsize sc}}({\boldsymbol \xi}) & \simeq & \frac{1}{2}(A_1 + A_2) \left[ 2 \, g(\beta;{\boldsymbol \xi}; {\boldsymbol \xi}^{(1)}) +  {\bf s}   \cdot \nabla_{{\boldsymbol \xi}^{(1)}}  g(\beta; {\boldsymbol \xi}; {\boldsymbol \xi}^{(1)}) \right] + \frac{1}{2} (A_2 - A_1) {\bf s}   \cdot \nabla_{{\boldsymbol \xi}^{(1)}}  g(\beta; {\boldsymbol \xi}; {\boldsymbol \xi}^{(1)})  \nonumber \\
& \simeq & (A_1 + A_2) \, g(\beta; {\boldsymbol \xi}; {\boldsymbol \xi}^{(1)}) + A_2  \, {\bf s}   \cdot \nabla_{{\boldsymbol \xi}^{(1)}}  g(\beta; {\boldsymbol \xi}; {\boldsymbol \xi}^{(1)}) .
\label{dip_app_1_s}
\end{eqnarray}
\noindent {\bf Remark}: Alternatively, one may expand about the vector ${\boldsymbol \xi}^d$, halfway between the two sources:
\begin{equation}
{\boldsymbol \xi}^d = {\boldsymbol \xi}^{(1)} + \frac{1}{2} {\bf s}; \quad {\boldsymbol \xi}^d = {\boldsymbol \xi}^{(2)} - \frac{1}{2} {\bf s}.
\label{xid}
\end{equation}
Substituting these expressions into~(\ref{eqsd}), we obtain
\begin{equation}
u_{\mbox{\scriptsize sc}}({\boldsymbol \xi}) \simeq  (A_1 + A_2) \,  g (\beta;{\boldsymbol \xi}; {\boldsymbol \xi}^d)  + \frac{A_2 - A_1}{2} \, {\bf s}   \cdot \nabla_{{\boldsymbol \xi}^d}  g(\beta; {\boldsymbol \xi}; {\boldsymbol \xi}^d). 
\end{equation}
We note that this representation is characteristic of a load comprising the sum $A_1+A_2$ for the source coefficients, 
and the difference term $A_2 - A_1$ for the dipole contribution. However, for the subsequent derivations and illustrative examples, we use the representation~(\ref{dip_app_1_s}), where the approximation is determined by expanding about the point ${\boldsymbol \xi}^{(1)}$. These expressions contain the familiar $A_1+A_2$ form for the  source terms, but the dipole coefficient is replaced by $A_2$.

\subsubsection{Dipole approximation for two semi-infinite rows of pins}
\label{sec:dipa}
We consider the alternative formulation for the problem described in Section~\ref{kerher}, whereby the two arrays of pins are replaced by one semi-infinite line, and both a source term $S_n$ and a dipole coefficient $D_n$ are associated with each member of the array. We impose two conditions on the total displacement $u(x,y)$ defined at each point ${\boldsymbol \xi} = (ja, 0)$:
\begin{equation}
u(ja, 0) = 0; \quad \quad   \quad  {\bf s}  \cdot \nabla u\bigg|_{{\boldsymbol \xi} = (ja,0)} = 0 \quad \mbox{or} \quad \frac{\partial u}{\partial {\bf s}} (ja, 0) = 0. 
\label{constraints}
\end{equation}
Recalling equations~(\ref{totfield}) and~(\ref{dip_app_1_s}), we write the total displacement field at a point ${\boldsymbol \xi}$ in the form:
\begin{equation}
u({\boldsymbol \xi}) = u_{\scriptsize{\mbox{inc}}}({\boldsymbol \xi}) + \sum_{n=0}^{\infty} \bigg[S_n g(\beta; {\boldsymbol \xi}; {\boldsymbol \xi^{(1)}}) + D_n {\bf s}  \cdot \nabla_{{\boldsymbol \xi^{(1)}}}  g (\beta; {\boldsymbol \xi}; {\boldsymbol \xi}^{(1)}) \bigg] \bigg|_{{\boldsymbol \xi^{(1)}} = (na,0)}.
\label{dip_eq1_disp}
\end{equation}
Consider ${\boldsymbol \xi} = (ja, 0)$, $ j = 0, 1, 2, \ldots$. 
Referring to equations~(\ref{alg_sys_1}), (\ref{alg_sys_2}), we have a system of linear algebraic equations for the coefficients $S_n$ and $D_n$:
\begin{equation}
\label{dip_als_1}
-e^{i\beta j a\cos\psi} = \sum_{n=0}^{\infty} \bigg[S_n g(\beta;ja, 0; na, 0) + D_n {\bf s}  \cdot \nabla_{{\boldsymbol \xi^{(1)}}} g (\beta;ja, 0; na, 0)  \bigg] , \quad j = 0, 1, 2, \ldots
\end{equation} 
and
\begin{equation}
\begin{split}
-{\bf s}  \cdot \nabla_{\boldsymbol \xi} e^{i\beta {\boldsymbol \xi} \cdot (\cos\psi,\sin\psi)} \bigg|_{{\boldsymbol \xi} = (ja,0)} = \sum_{n=0}^{\infty} \bigg[S_n \, {\bf s}  \cdot \nabla_{\boldsymbol \xi} g(\beta;ja, 0; na, 0) \\+ 
D_n \, {\bf s}  \cdot \nabla_{\boldsymbol \xi} \left( {\bf s}  \cdot \nabla_{{\boldsymbol \xi^{(1)}}} g (\beta;ja, 0; na, 0) \right) \bigg] ,
\quad j = 0, 1, 2, \ldots
\end{split}
\label{dip_als_2}
\quad\quad.
\end{equation}
The notation for the directional derivatives distinguishes between differentiating with respect to the first and second arguments, ${\boldsymbol \xi} = (ja,0)$ and ${\boldsymbol \xi}^{(1)}=(na,0)$.

Following the discrete Wiener-Hopf derivation of Section~\ref{kerher}, we 
recall the notation of~(\ref{b1n})-(\ref{f12}) for the unknown displacement amplitudes in the ``reflection" region for $n < 0$,  and for the incident plane wave coefficients. Consider a specific point ${\boldsymbol \xi} = (Na, 0)$ and extend the definitions of $S_n$ and $D_n$ for $n < 0$. Applying the discrete Fourier Transform to~(\ref{dip_als_1}), (\ref{dip_als_2}),  as in Section~\ref{kerher}, we derive the algebraic system of equations for the Wiener-Hopf formulation for the dipole approximation:

\begin{eqnarray}
\sum_{N=-\infty}^\infty B_Ne^{ikNa} & = &\sum_{N=-\infty}^\infty F_Ne^{ikNa} \label{dip_eq1}  \\
& + & 
\sum_{n=-\infty}^\infty   e^{ikna} \left(S_n +  D_n \,{\bf s}  \cdot \nabla_{{\boldsymbol \xi^{(1)}}} \right)   \bigg( \sum_{j=-\infty}^\infty g(\beta;ja,0;0,0)e^{ikja} \bigg),  \nonumber \\
\sum_{N=-\infty}^\infty B_N^{'}e^{ikNa} & = & \sum_{N=-\infty}^\infty F_N^{'}e^{ikNa} \label{dip_eq2}\\ 
& + & \sum_{n=-\infty}^\infty    e^{ikna} \, {\bf s}  \cdot \nabla_{\boldsymbol \xi} \left(S_n +  D_n \,{\bf s}  \cdot \nabla_{{\boldsymbol \xi^{(1)}}} \right) \bigg( \sum_{j=-\infty}^\infty g(\beta;ja,0;0,0)e^{ikja} \bigg),
\nonumber 
\end{eqnarray}
where we note that now ${\boldsymbol \xi} = (ja, 0)$ and ${\boldsymbol \xi^{(1)}} =  (0, 0)$. 
Here we have the following definitions for $B_N, B_N^{'}, F_N, F_N^{'}$ for $N \in \mathbb{Z}$:
\begin{equation}
u(Na, 0)=\Bigg \{ 
\begin{array}{c l}  
0, &  \quad\quad N \ge 0 \\
\\
B_N, & \quad\quad  N  <  0
\label{b1n2}
\end{array} 
\end{equation}
\begin{equation}
{\bf s}  \cdot \nabla_{\boldsymbol \xi} u\bigg|_{(Na,0)} =\Bigg \{ 
\begin{array}{c l}  
0, &  \quad\quad N \ge 0 \\
\\
B_N^{'}, &  \quad\quad N  <  0
\end{array}  
\label{b2m2}
\end{equation}
\begin{equation}
u_{\scriptsize{\mbox{inc}}}(Na, 0)=F_N,~~~~~~
-{\bf s}  \cdot \nabla_{\boldsymbol \xi} u_{\scriptsize{\mbox{inc}}} \bigg|_{(Na,0)} = -i \beta (s_1 \cos{\psi} + s_2 \sin{\psi}) e^{i\beta \cos{\psi} Na }  = F_N^{'}.
\label{f122}
\end{equation} 
We note that $B_N = 0, B_N^{'} = 0$ for $N \ge 0$. 
We observe that in this Wiener-Hopf formulation for the dipole approximation, we employ only the quasi-periodic Green's function~(\ref{short1}) defining a single array of points $\hat G(\beta,k; {\bf 0};{\bf 0})$ 
in contrast to the full system~(\ref{whopf0}), where the kernel matrix includes three grating Green's functions, shifted relative to one another. 

The grating Green's function $\hat G(\beta,k; {\bf 0};{\bf 0})$ is an important and well-studied object in the analysis of platonic grating systems, and has been used numerous times in the literature, see for example~\cite{Evans} -\cite{Dirac}. Refined accelerated convergence formulae have been derived~\cite{ABM_NVM_RCM}, and are implemented here in finding the zeros of the determinants of the kernel matrices for the corresponding semi-infinite problems. The discrete Wiener-Hopf derivation, and its kernel in particular, provides a highly efficient procedure for determining ranges of frequency for scattering effects and trapped waveforms. 
We note that alternative methods for evaluating the lattice and grating Green's functions in pinned plates have been discussed in~\cite{Evans}, \cite{MJAS_RP_TDW}, \cite{ISJ_NVM_ABM}, \cite{MJAS_MHM_RCM}, amongst others.


\subsection{Illustrative examples}
\label{sec:illus}
In this section, we demonstrate the efficacy of the dipole approximation by comparing both scattering coefficients and the resulting displacement field plots for the approaches described in Sections~\ref{kerher} and~\ref{sec:dipole1}. 
Referring to equation~(\ref{dip_eq1_disp}), the dipole approximation yields coefficients $S_n$ (sources) and $D_n$ (dipoles), where $n$ denotes the pin number. The direct approach for a pair of semi-infinite gratings in Section~\ref{kerher} determines coefficients $A_n^{\scriptsize \mbox{(I)}}$ (for the lower grating) and $A_n^{\scriptsize \mbox{(II)}}$ (upper grating). 

\begin{figure}[h]
\begin{center}
\includegraphics[width=.44\columnwidth]{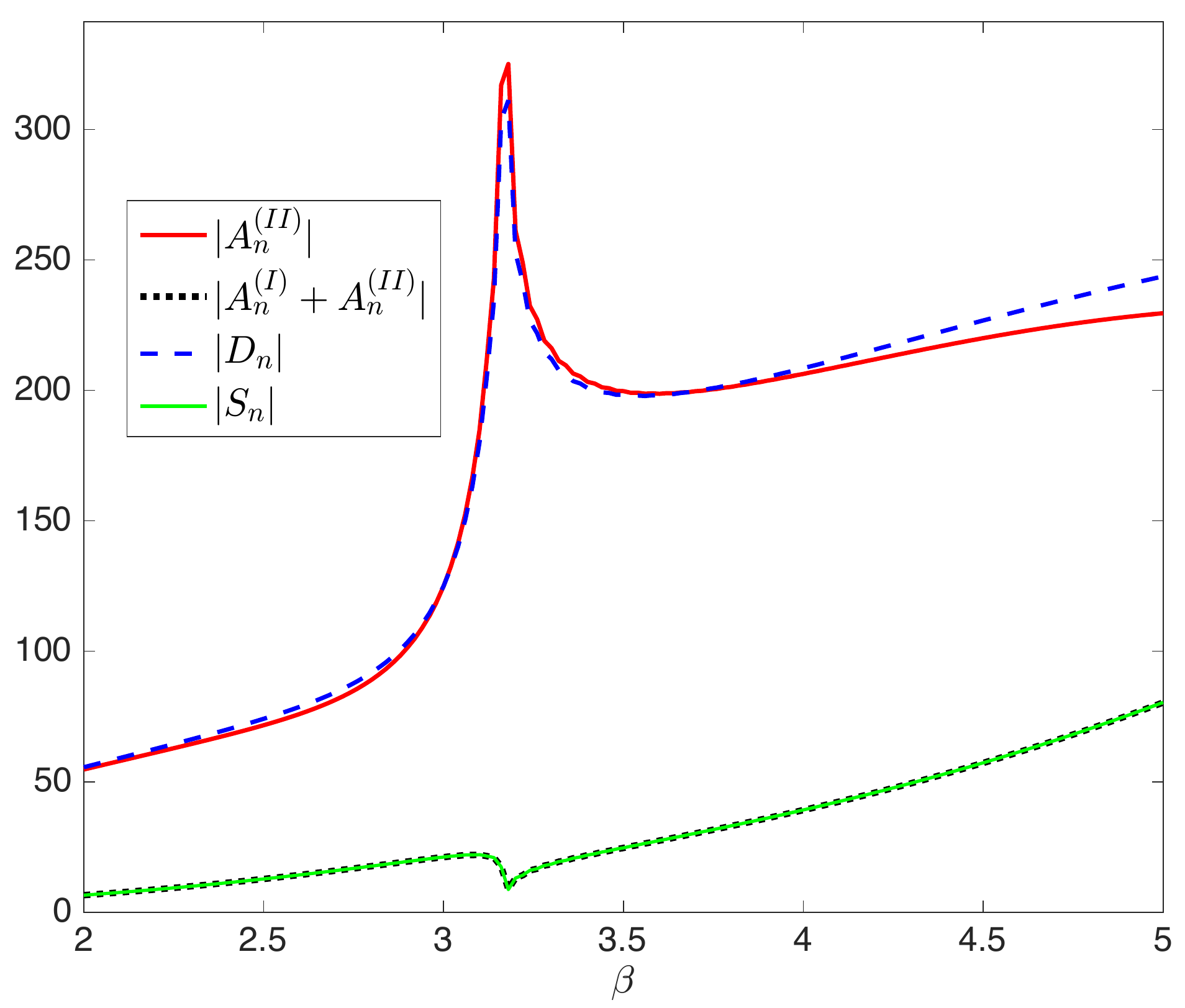}~~
 \put(-70,60) {(a)}
\includegraphics[width=.44\columnwidth]{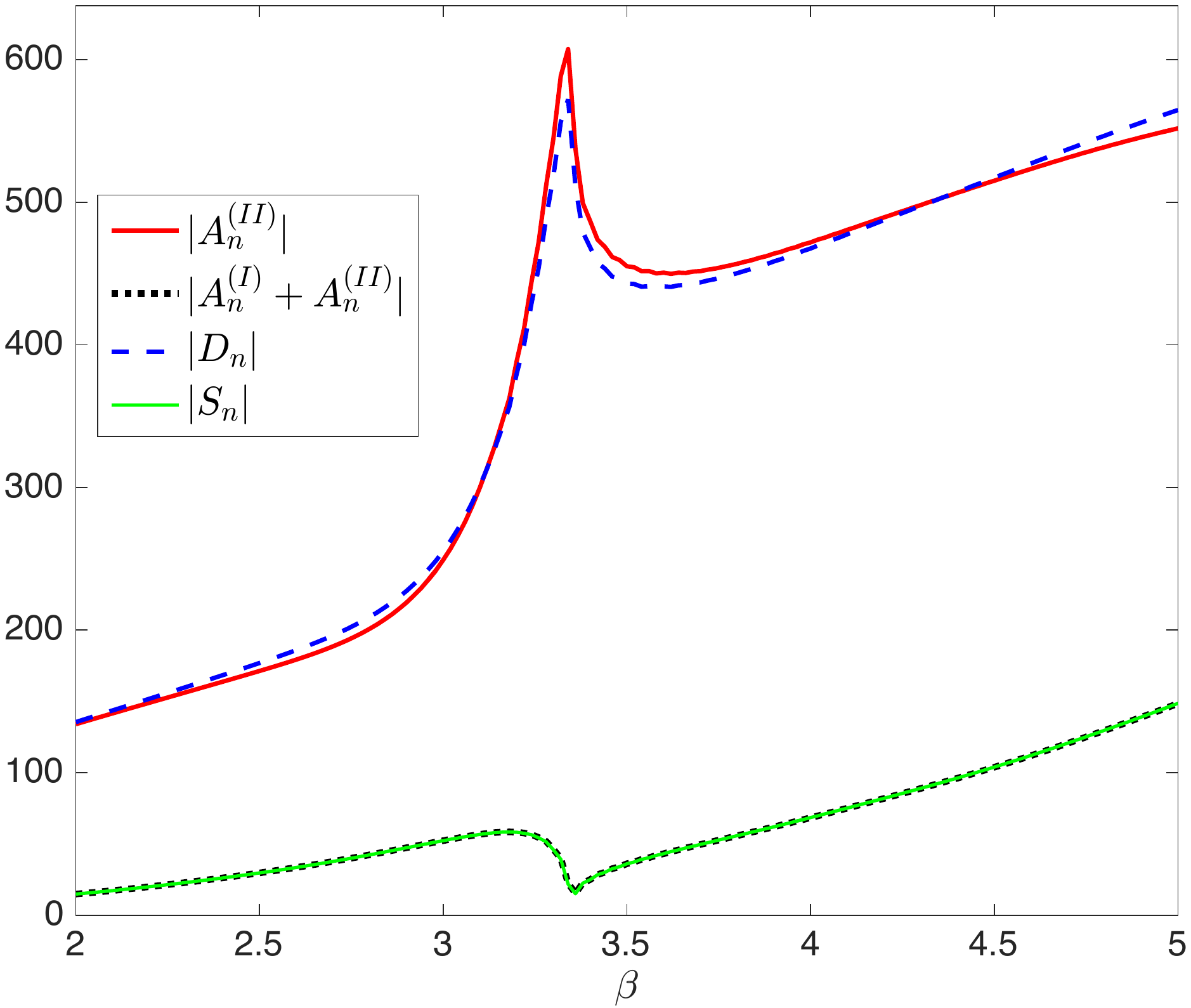}
\put(-70,60) {(b)}
 
\includegraphics[width=.44\columnwidth]{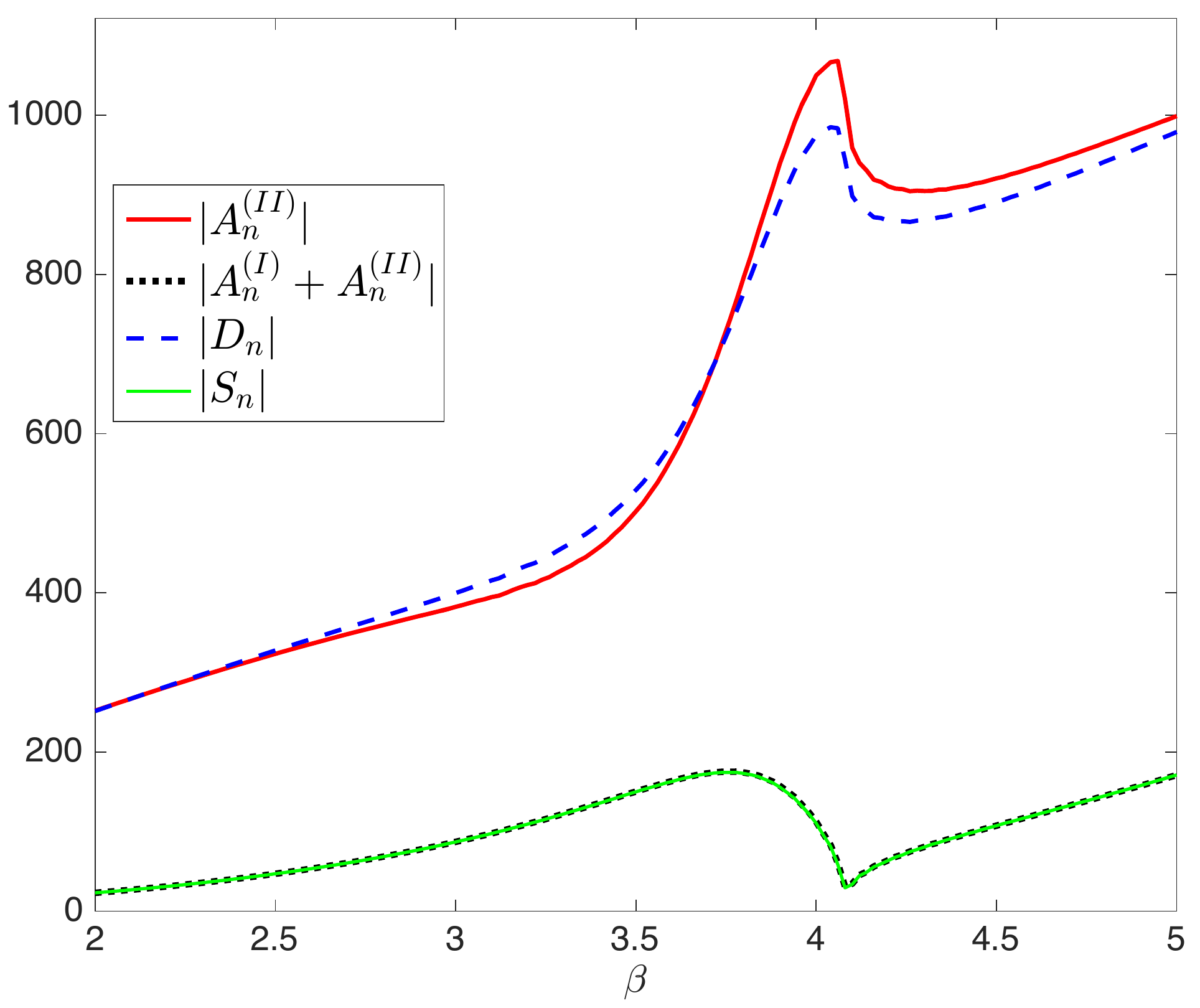}~~
\put(-70,58) {(c)}
\includegraphics[width=.44\columnwidth]{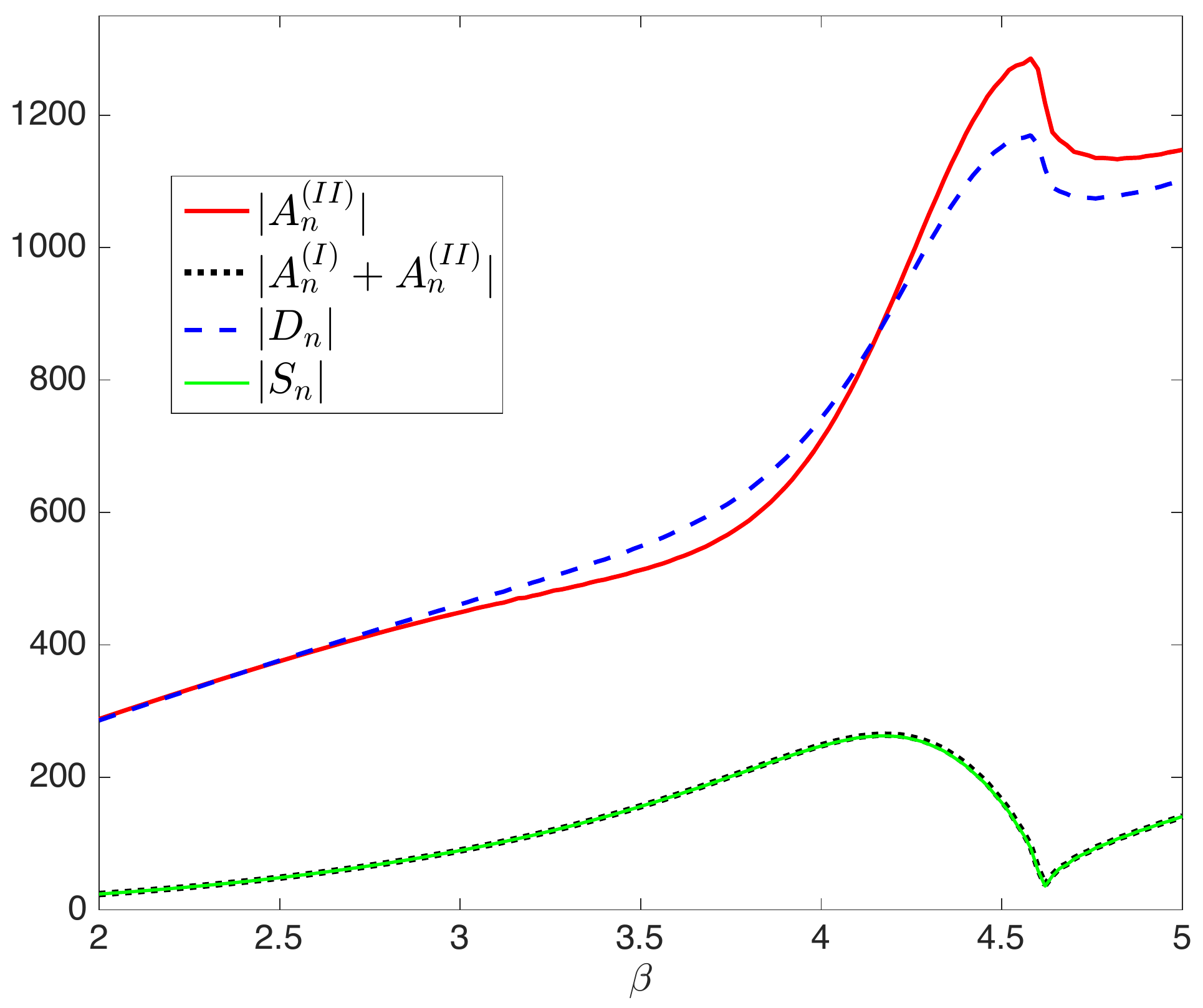}
\put(-70,58) {(d)}
\caption{ 
Comparison of the moduli of scattering coefficients evaluated using the wave scattering method for a pair of shifted gratings (source terms indicated by dotted black curves and dipole terms by solid red curves) and a single semi-infinite line of sources ($|S_n|$, solid green curves) and dipoles ($|D_n|$, dashed blue curves). For shift vector ${\bf s} = (0.005,0.015)$ with $|{\bf s}| = 0.0158$, $\theta = 1.25$, $L=160$, we consider four angles of incidence:
(a) $\psi = 0.2$, (b) $\psi = 0.5$, (c) $\psi = 1.0$, (d) $\psi = 1.2$. 
\label{log_sing}}
\end{center}
\end{figure}

By analogy with equation~(\ref{dip_app_1_s}), the coefficients $S_n$, $D_n$ may be associated with the coefficient terms $A_n^{\scriptsize \mbox{(I)}} + A_n^{\scriptsize \mbox{(II)}}$ and $A_n^{\scriptsize \mbox{(II)}}$, respectively. Here we include several numerical examples, with a selection of graphical plots of coefficients for a range of parameter settings, noting that periodicity/spacing $a$ is set to unity unless stated.  The method for solving the system~(\ref{dip_als_1}),~(\ref{dip_als_2}) for a truncated semi-infinite system is outlined in detail in Appendix~\ref{sec:eval_dipole}, where the truncation parameter $L$ indicates the number of points in the array. The analysis involves the use of an asymptotic term for the logarithmic singularity arising for the second derivative terms in~(\ref{dip_als_2}) when $j = n$.

\subsubsection{Scattering and transmission resonance}
In Fig.~\ref{log_sing}, we plot curves of the moduli for all four sets of the coefficients, two from each of the formulations, for truncation parameter $L=160$. The shifted pair coefficients $|A_n^{\scriptsize \mbox{(II)}}|$ and $|A_n^{\scriptsize \mbox{(I)}} +A_n^{\scriptsize \mbox{(II)}}|$ are plotted with solid (red) curves and dotted (black) curves, respectively. 
The dipole coefficients $|D_n|$ and source terms $|S_n|$ are illustrated by dashed (blue) curves and solid (green) curves. For a fixed shift vector ${\bf s} = (0.005,0.015)$, with $|{\bf s}| = 0.0158$ and dipole angle $\theta = 1.25$, we consider four angles of incidence in radians: $\psi = 0.2, 0.5, 1.0$ and $1.2$. The coefficients are plotted for $\beta$ in the range $2 \le \beta \le 5$. 

\begin{figure}[h]
\begin{center}
\includegraphics[width=.43\columnwidth]{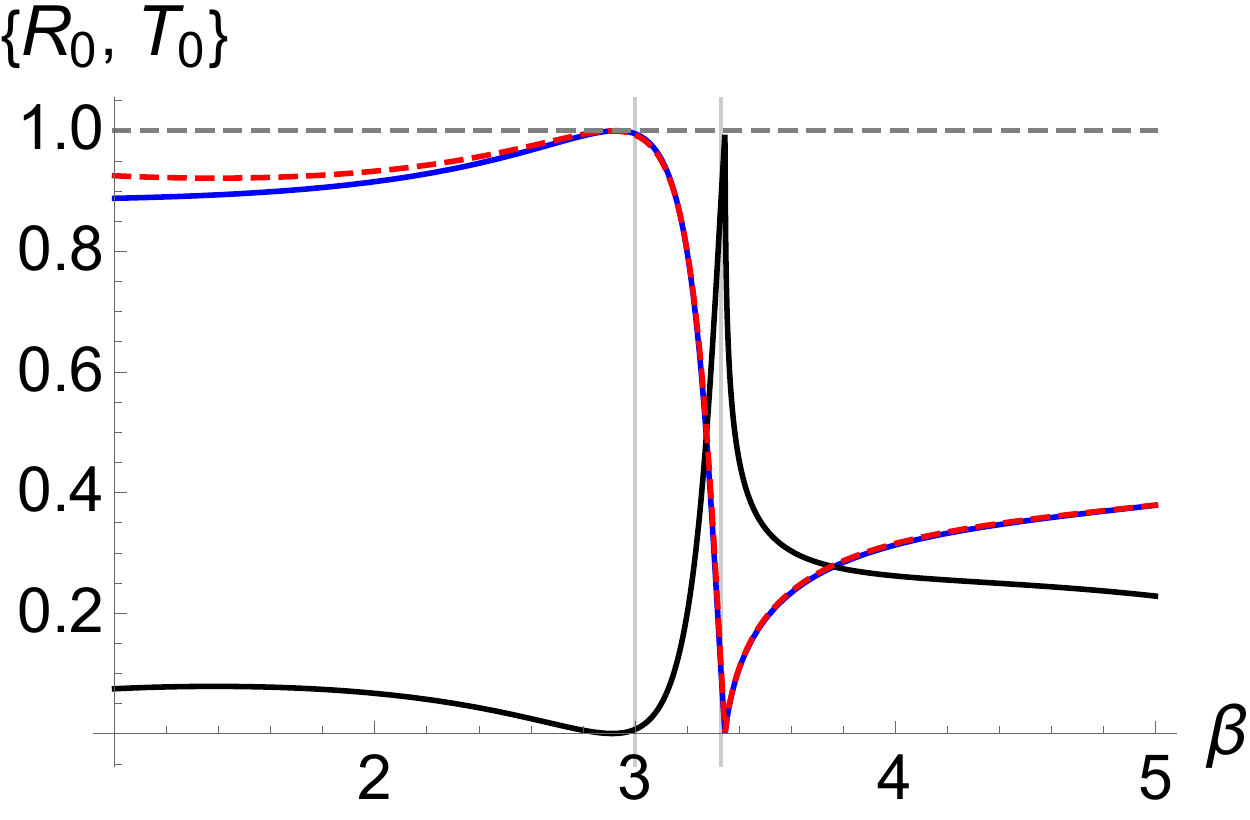}
 \put(-65,50) {(a)}
\includegraphics[width=.4\columnwidth]{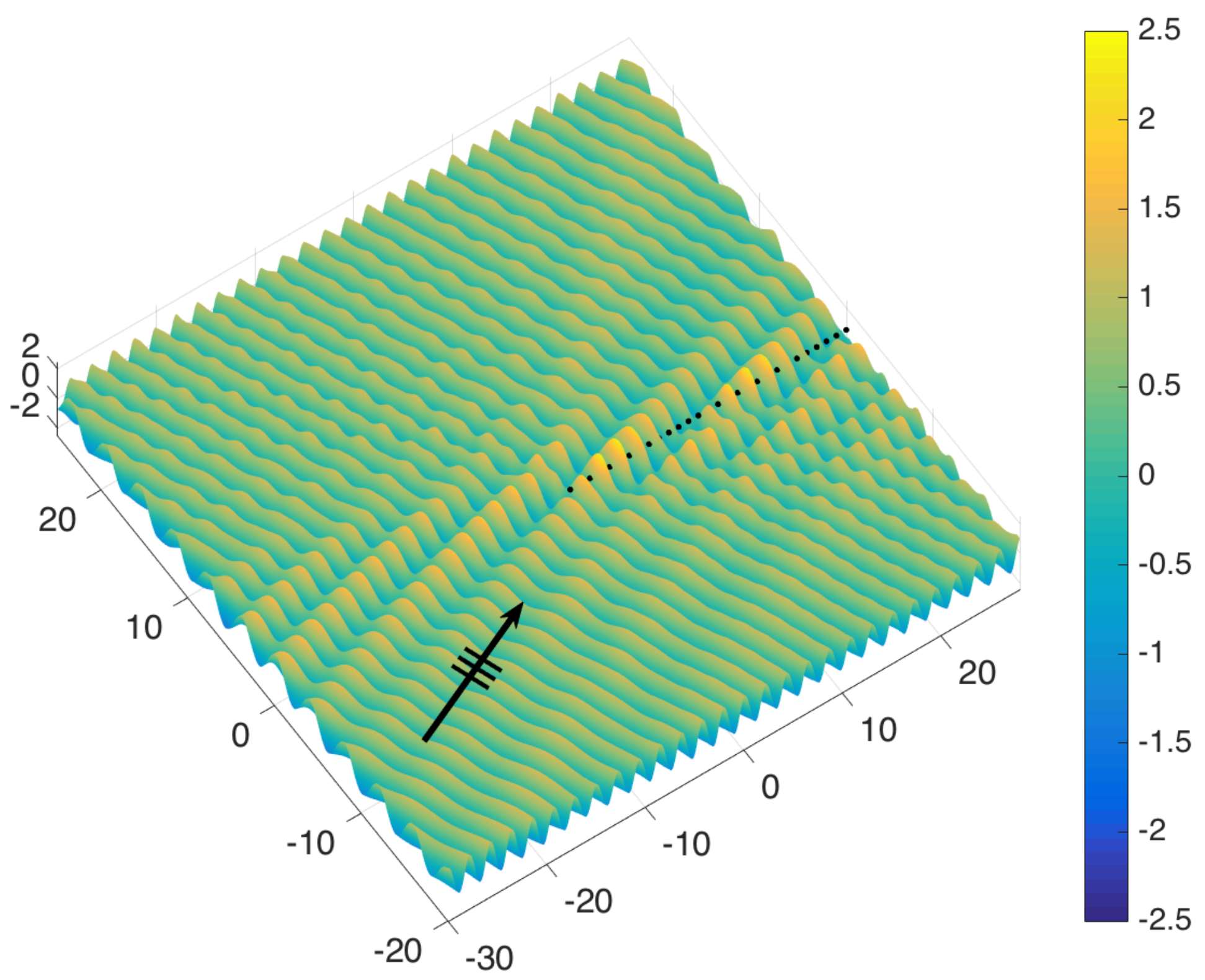}
 \put(-65,50) {(b)}

\includegraphics[width=.4\columnwidth]{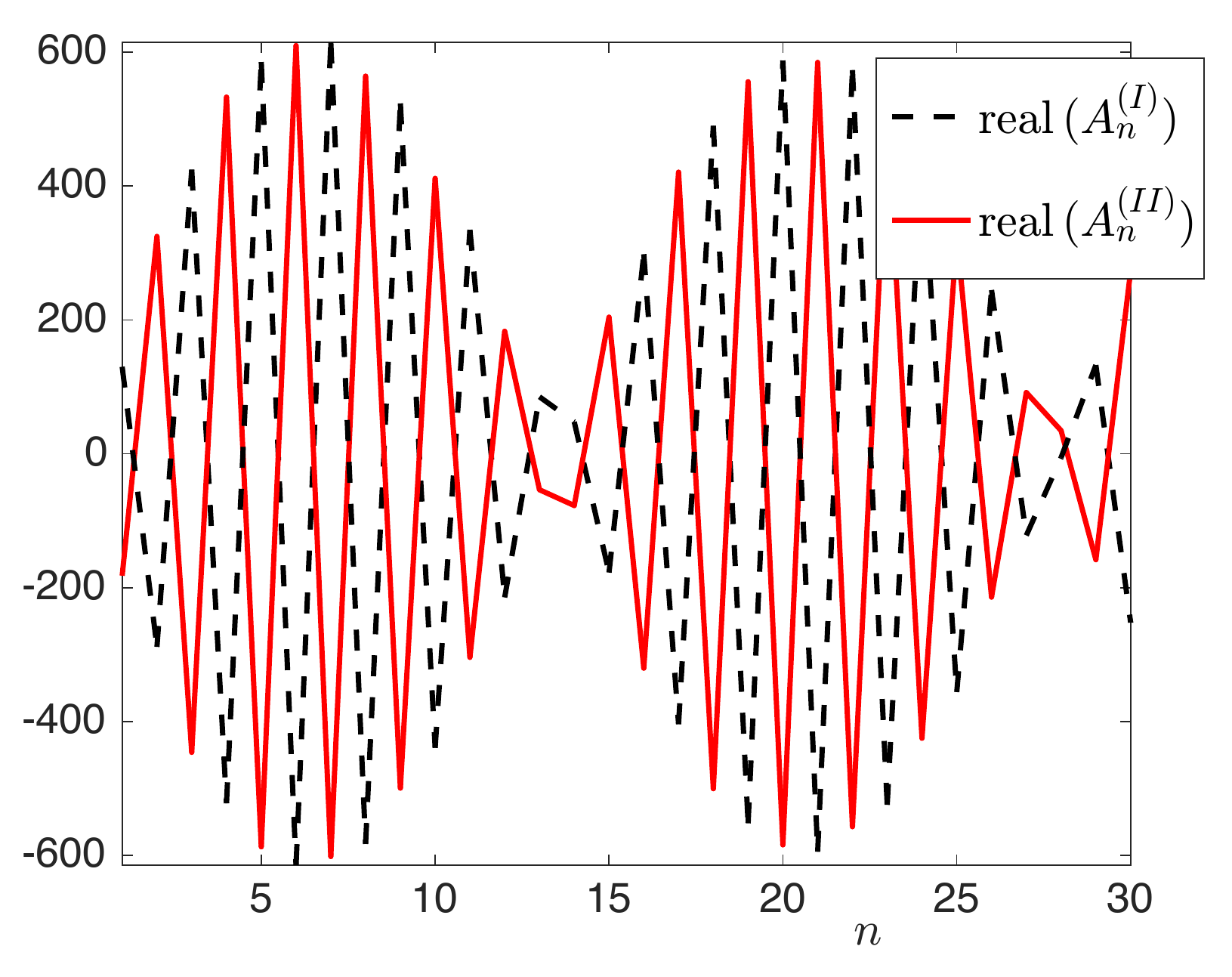}~~~
 \put(-64,50) {(c)}
\includegraphics[width=.4\columnwidth]{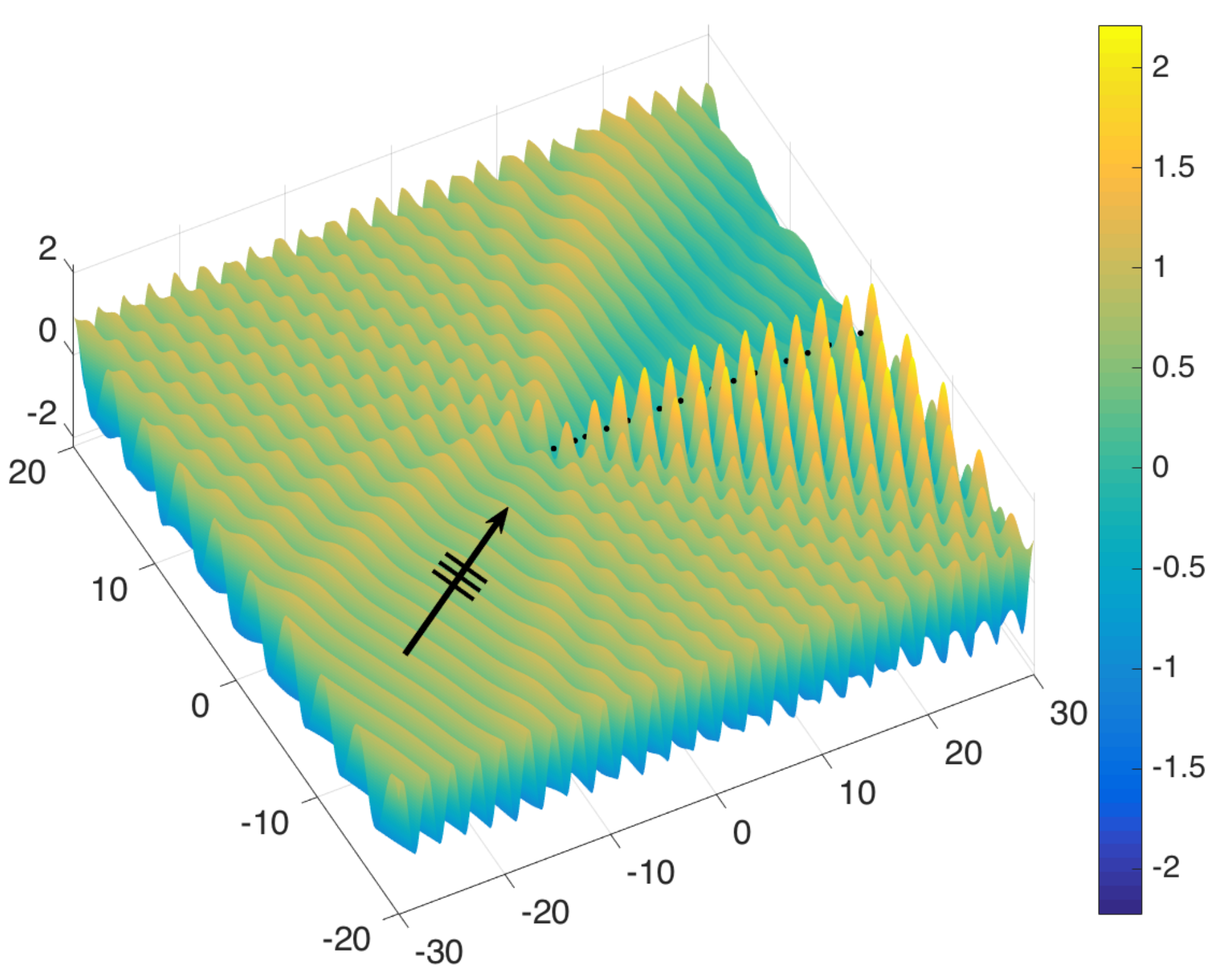}
 \put(-64,50) {(d)}
\caption{Plane wave with $\psi = 0.5$ incident on a pair of gratings with $a = 1.0$, ${\bf s} = (0.005,0.015)$. 
\newline (a) Normalised reflected energy for the zeroth propagating order versus $\beta$ for a single line of pins (solid blue) and a pair of shifted infinite gratings defined by ${\bf s}$ (dashed red), and normalised transmitted energy for the pair (solid black). 
\newline Semi-infinite system (b) Real part of the total displacement field for the first 30 pairs of pins for $\beta = 3.33$, $L = 100$. (c) Real parts of the scattering coefficients for the first 30 pins. (d) Real part of the total displacement field for $\beta = 3.0$, all other parameters are the same as in (b).
\label{TR}
}
\end{center}
\end{figure}

The agreement for the coefficients representing sources (that is $S_n$ and $A_n^{\scriptsize \mbox{(I)}} + A_n^{\scriptsize \mbox{(II)}}$) is excellent for all cases as shown in Fig.~\ref{log_sing}, where the two curves are visually indistinguishable in each of parts (a) to (d). There is also very good agreement between $|D_n|$ and $|A_n^{\scriptsize \mbox{(II)}}|$ 
for all choices of $\psi$. We note that as $\psi$ increases from Fig.~\ref{log_sing}(a) to (d), the moduli of the dipole coefficients increase, and that the shape of the curve is qualitatively the same for all values of $\psi$ considered. 
Note that the typical sharp peak in the dipole coefficient $|D_n|$ (or $|A_n^{\scriptsize \mbox{(II)}}|$ for the pair) occurs for a higher value of $\beta$ as $\psi$ is increased, and its frequency is in the neighbourhood of that of the dip for the source coefficient curves.

In this platonic setting, the sharp peaks are associated with additional spectral orders becoming propagating (rather than evanescent), and are linked to transmission resonances. For this example of $|{\bf s}| = 0.0158$, the dipole approximation appears to be robust since the singularity is well approximated for both formulations. We illustrate an example in Fig.~\ref{TR} for one of the incident angles $\psi = 0.5$ for the spectral parameter $\beta = 3.33$, which is just below the peak frequency arising for $\beta = 3.35$ in Fig.~\ref{log_sing}(b).

In Fig.~\ref{TR}(a), we illustrate the transmission resonance using energy plots for the zeroth order for both an infinite single grating and a pair of shifted gratings (defined by ${\bf s} = (0.005,0.015)$). Normalised reflected ($R_{\scriptsize 0}$) and transmitted ($T_{\scriptsize 0}$) energies for the zeroth order are plotted versus the spectral parameter $\beta$. The reflected energy for the single grating is shown with the solid (blue) curve, and the Wood anomaly at $\beta \simeq 3.35$ signifies the additional order $-1$ passing from evanescence to propagation. This frequency coincides with both a resonance in transmission (solid black) and a zero in reflection (dashed red) for the {\it pair} for the zeroth order (by the conservation of energy). Note that for $\beta > 3.35$, the energy for the zeroth order no longer sums to unity owing to the additional contributions (not shown here in Fig.~\ref{TR}(a)) to the total energy from the new propagating order.

In Figs.~\ref{TR}(b) and (c), we show how this resonance for the infinite system is manifested in the semi-infinite system. For ${\bf s} = (0.005,0.015)$, $\psi = 0.5$, $\beta = 3.33$, we plot the real part of the total displacement field in Fig.~\ref{TR}(b) for the first thirty pinned pairs ($0 \le n \le 29$), with the direction of the incident plane wave indicated by the arrow. The real parts of the corresponding coefficients are shown in Fig.~\ref{TR}(c). We observe the transmission resonance associated with the transition of an evanescent to a propagating order. Although scattering effects are present in the vicinity of the leading vertex, there is clear evidence of transmission along, and behind, the grating pair.

In particular, the large amplitudes and envelope function for the scattering coefficients for the dipole terms $A_n^{\scriptsize \mbox{(II)}}$ of Fig.~\ref{TR}(c) are matched by the displacement field along the line of the first thirty pairs of pins in Fig.~\ref{TR}(b). This example coincides with the sharp peaks in both Figs.~\ref{log_sing}(b) and~\ref{TR}(a). The zero in transmission (a reflection mode) for the pair at $\beta \simeq 3.0$ in Fig.~\ref{TR}(a) is illustrated for the semi-infinite system in Fig.~\ref{TR}(d), which not only exhibits strong reflection, but on comparison with the field in Fig.~\ref{TR}(b), further emphasises the transmission regime for $\beta = 3.33$. 

\subsubsection{Magnitude and orientation of dipole}
\label{sec242}
\begin{figure}[h]
\begin{center}
\includegraphics[width=.48\columnwidth]{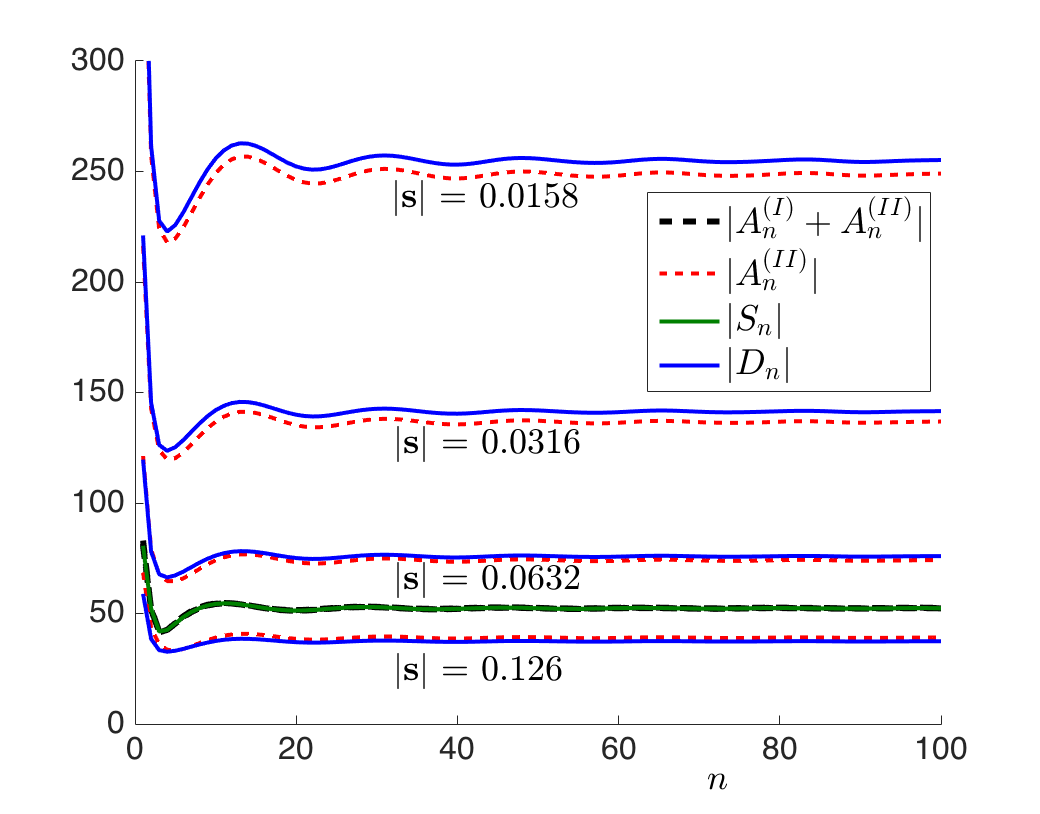}~~
\put(-70,60) {(a)}
\includegraphics[width=.48\columnwidth]{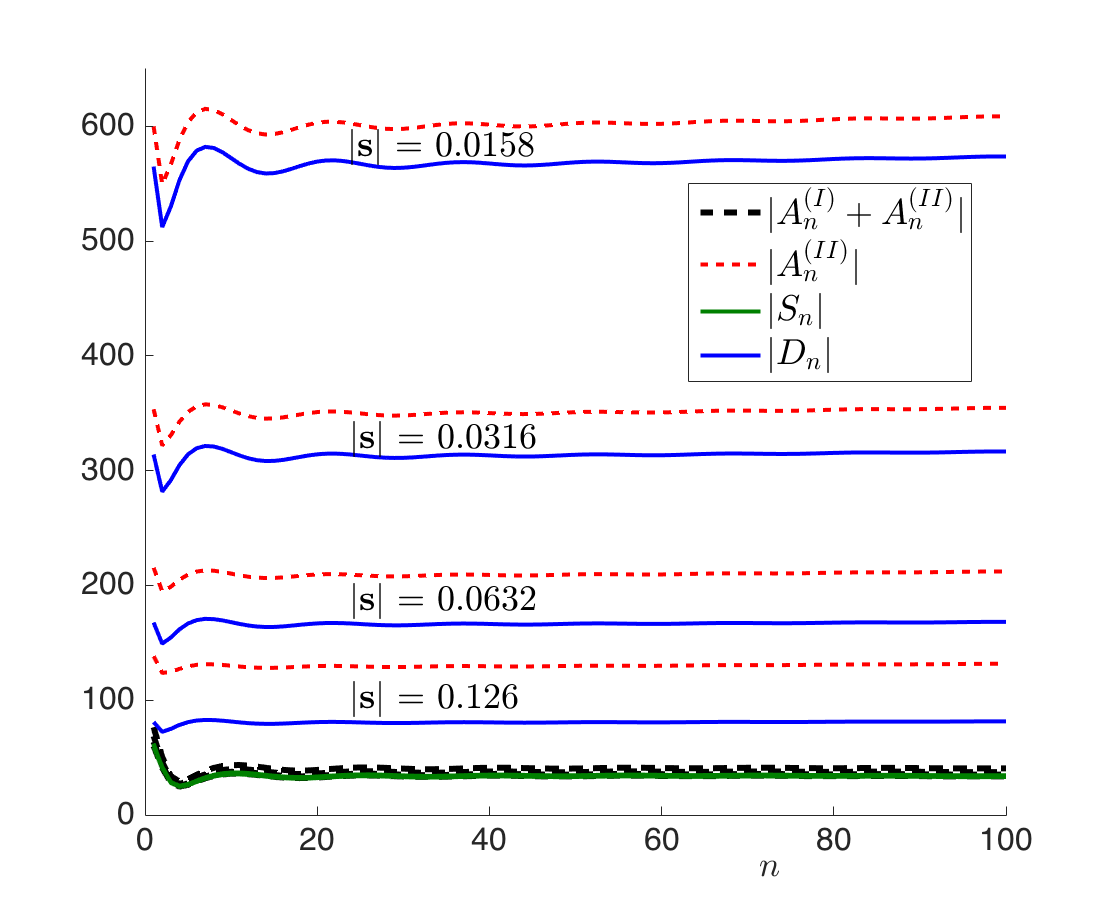}
\put(-70,60) {(b)}
\caption{
Comparison of scattering coefficients for variation of $|{\bf s}|$ for a pair of gratings ($A_n^{\tiny \mbox{(I)}}, A_n^{\tiny \mbox{(II)}}$) and for the line of sources and dipoles ($S_n$ and $D_n$). Fixed parameter settings match those of Fig.~\ref{TR}: $\psi = 0.5, a = 1.0, \theta = \arctan(3), L=100$. (a) Reflection frequency of Fig.~\ref{TR}(d), $\beta = 3.0$. (b) Transmission resonance frequency of Fig.~\ref{TR}(b), $\beta = 3.33$. 
\label{TR2}
}
\end{center}
\end{figure}

The dominance of the dipole terms over the source terms for the reflection regime of Fig.~\ref{TR}(d) is shown in Fig.~\ref{TR2}(a) where four values of $|{\bf s}|$ (the first of which is $|{\bf s}| = 0.0158$) are shown for the parameter settings of $\psi = 0.5$, $\beta = 3.0$. However, as $|{\bf s}|$ is increased the dipole coefficients tend towards those of the sources. We also include a study of magnitude $|{\bf s}|$ for the transmission case $\beta = 3.33$ in Fig.~\ref{TR2}(b). Note that for the resonant frequency, the agreement of $D_n$ and $A_n^{\scriptsize \mbox{(II)}}$ is reduced.

\begin{figure}[h]
\begin{center}
\includegraphics[width=.41\columnwidth]{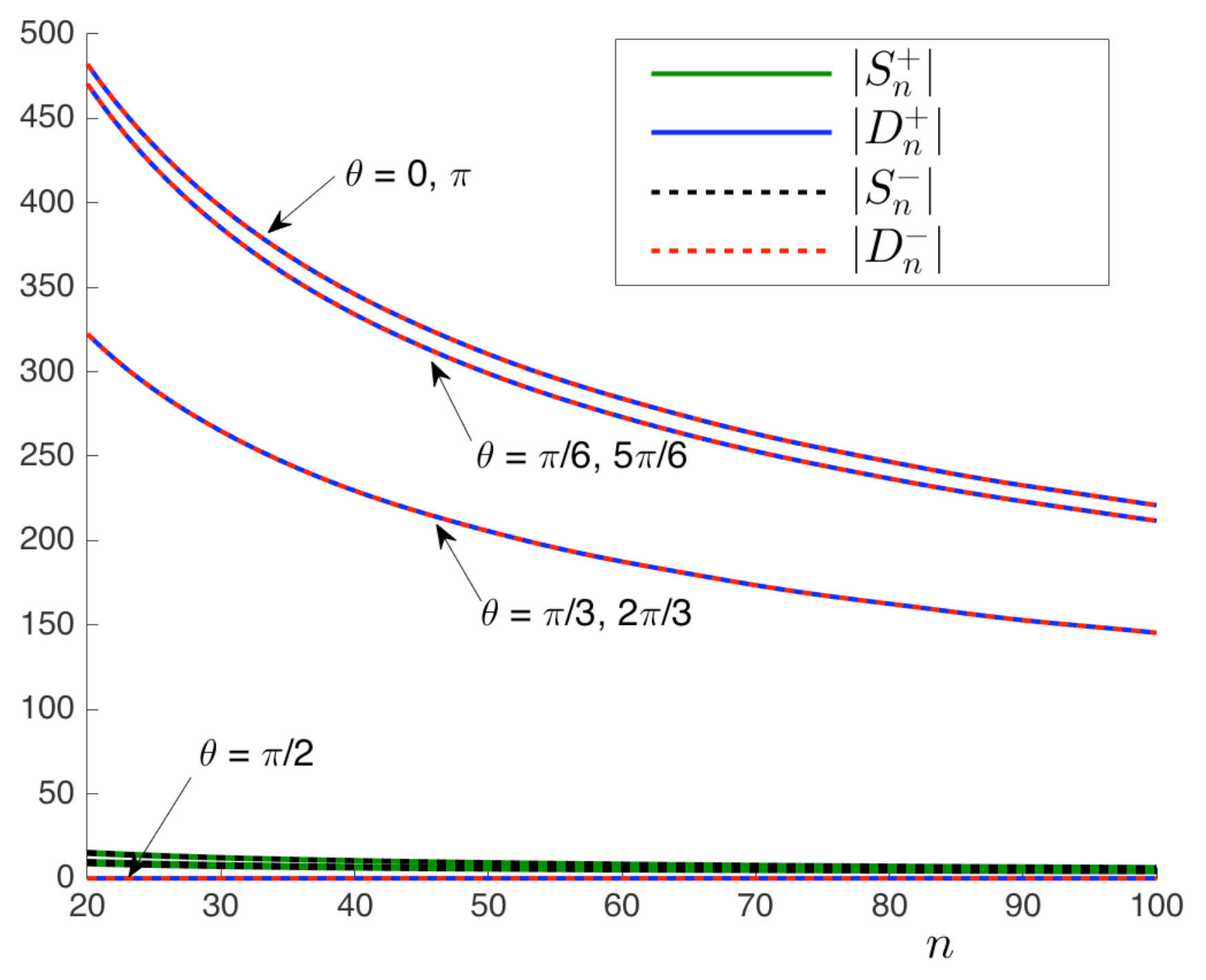}~~~
\put(-70,55) {(a)}
\includegraphics[width=.45\columnwidth]{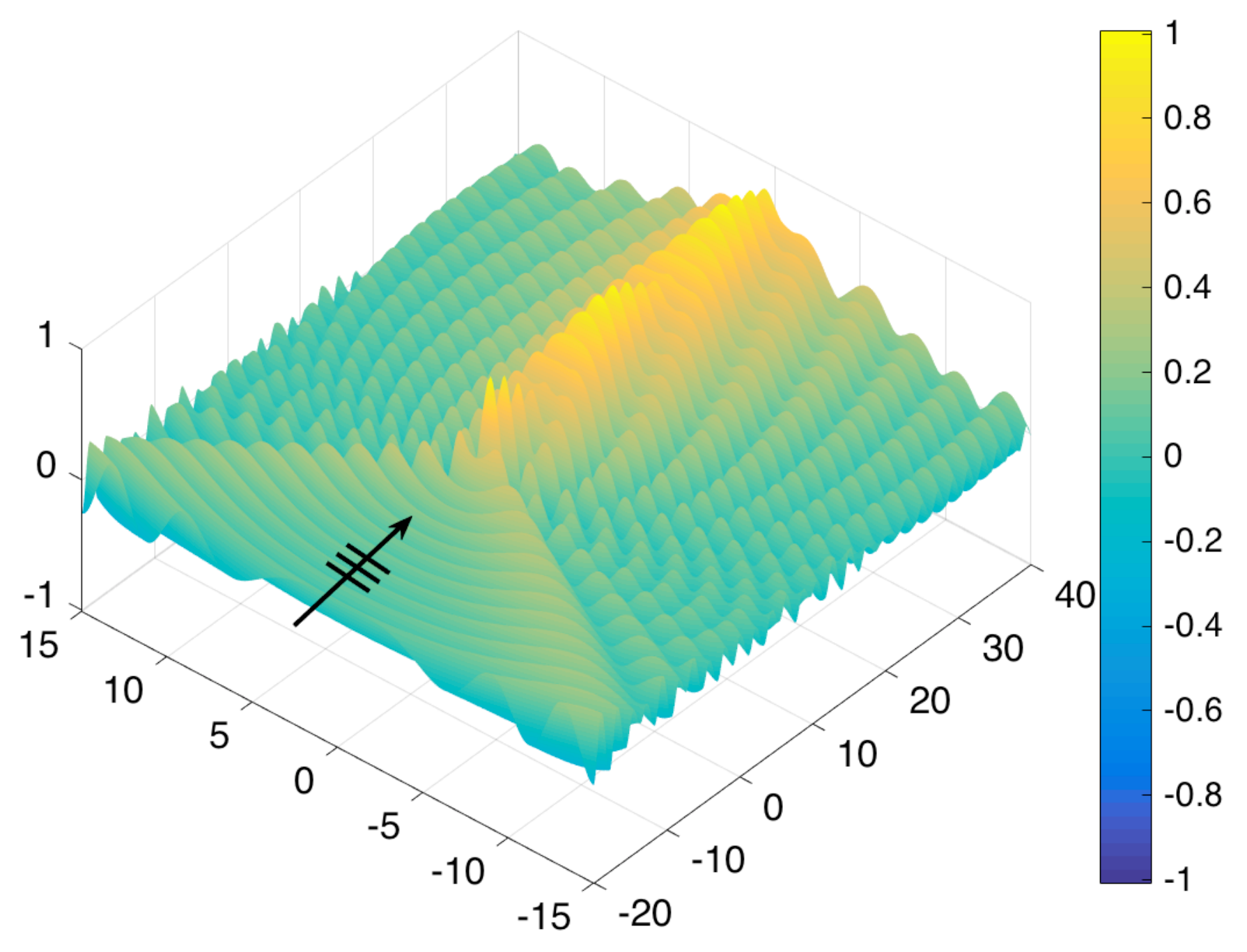}
\put(-70,55) {(b)}
\caption{
(a) Moduli of coefficients for the source and dipole terms for various dipole angles $\theta$ for normal incidence ($\psi = 0$) for upwardly ($S_n^+$, $D_n^+$), and downwardly ($S_n^-$, $D_n^-$), oriented shifted pairs with $|{\bf s}| = 0.005$, $\beta = 3.33$. The angle $\theta$ is varied by increments of $\pi/6$, as indicated by the labels adjacent to the curves of $|D_n^+|/|D_n^-|$. (b) Scattered field for $\theta = 0$, $|{\bf s}| = 0.005$, $\psi = 0, \beta = 3.33$. 
\label{comp_dip_angles}
}
\end{center}
\end{figure}

Clearly, the magnitude of $|{\bf s}|$ is linked to both the efficiency of the dipole approximation and the relationship between the dipole and source coefficients. A natural question to ask is how does the orientation of the dipole affect the scattering properties of the system? We consider the case of fixing $|{\bf s}| = 0.005$ and varying the dipole angle $\theta = \arctan(s_2/s_1)$ in the range $0 \le \theta \le \pi$ (see Fig.~\ref{diag}). Using the same value of $\beta = 3.33$, we investigate the case of normal incidence, $\psi = 0$, in Fig.~\ref{comp_dip_angles}.

We consider two pairs of gratings for each dipole angle $\theta$, one oriented upwards with which we associate the superscript $+$, and one oriented downwards (effectively defined by $-\theta$) with which we associate the superscript $-$.
The moduli of the coefficients for seven choices of $\theta$, multiples of $\pi/6$, are plotted in Fig.~\ref{comp_dip_angles}(a).  We observe that in isolation, there is virtually no difference in the results for the upwardly and downwardly oriented shifted pairs. The formulation of a herringbone system by combining these pairs is considered in the next section, including an analysis of the effect of the direction of the dipoles, where naturally the roles of $\pm \theta$ are much more significant.

In Fig.~\ref{comp_dip_angles}(a), we note that as $\theta$ is increased, the dipole coefficients ($D_n^+$ and $D_n^-$) are reduced, 
and that the dipole terms dominate the source coefficients with the notable exception of $\theta = \pi/2$, which leads to the extreme reduction of all coefficients. This suggests that this orientation of the dipoles for $|{\bf s}| \ll 1$, for which the source and dipole coefficients are comparable, replicates the line of equally spaced pins that does not support Rayleigh-Bloch modes~\cite{Evans}. For the parallel direction with $\theta = 0$, however, some localisation is observed, as indicated by the associated scattered field in Fig.~\ref{comp_dip_angles}(b) for $\psi = 0$, $|{\bf s}| = 0.005$.

 \section{
Herringbone system of rigid pins
}
\label{sec:hbone}
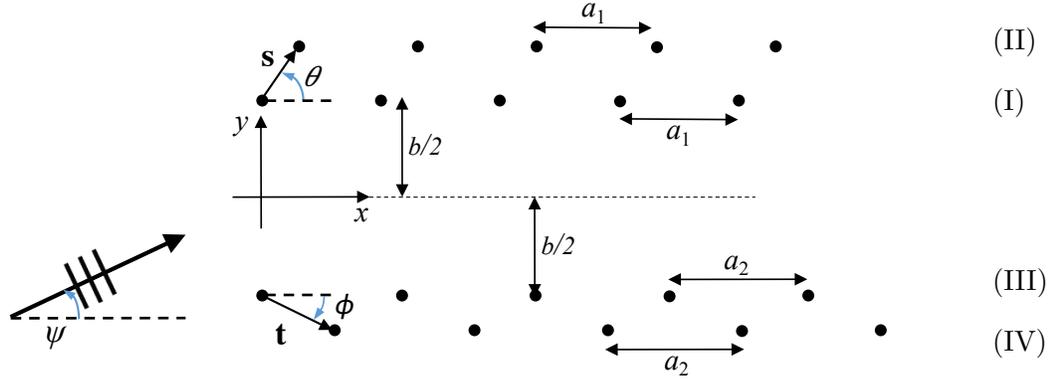
\begin{figure}[h]
\centerline{
         \includegraphics[width=.82\columnwidth]{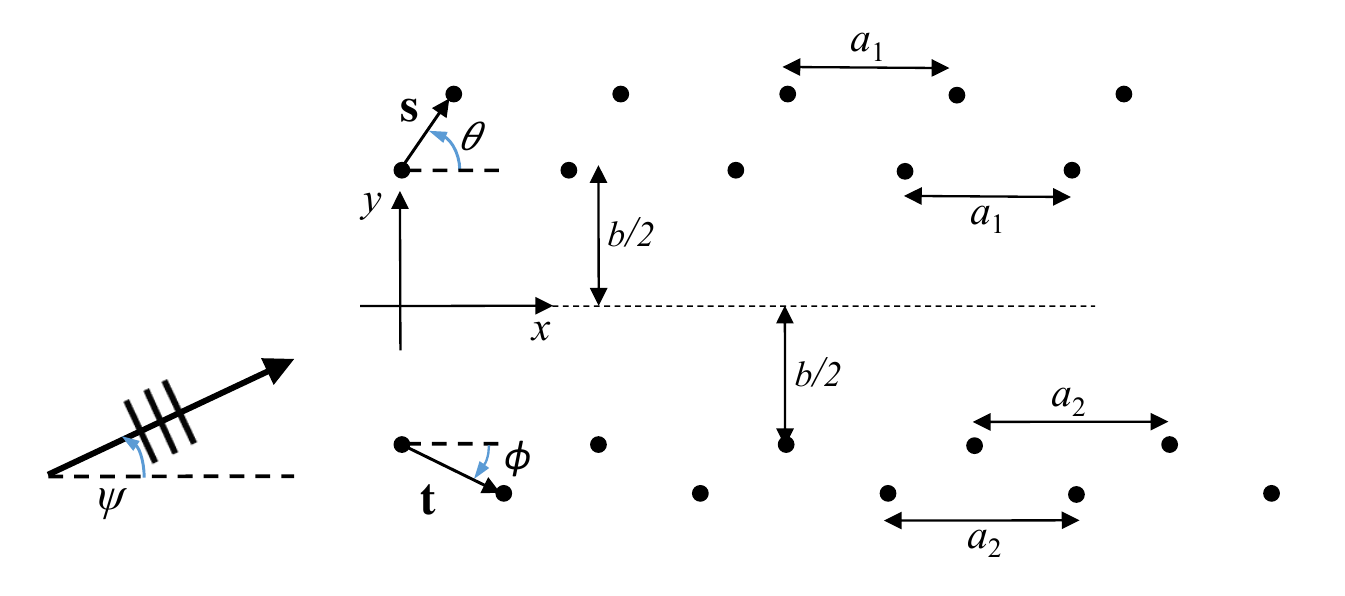}
\put(8,7) {(IV)}
\put(8,15) {(III)}
\put(8,39) {(I)}
\put(8,47) {(II)}
         }
\caption{ 
A semi-infinite ``herringbone" pattern of of rigid pins in an elastic Kirchhoff plate. The upper pair is characterised by the common spacing $a_1$ and shift vector ${\bf s}$, and the lower pair by $a_2$ and ${\bf t}$. The separation of the pairs of gratings is denoted by $b$.
}
\label{herr}
\end{figure}

Consider a herringbone pattern in an elastic Kirchhoff plate,  
as shown in Fig.~\ref{herr}, where the upper and lower pairs are characterised by, respectively, the shift vectors ${\bf s} =(s_1,s_2)$ and ${\bf t}=(t_1,t_2)$, and the spacings $a_1$ and $a_2$. The separation of the pairs of gratings is denoted by $b$.

\subsection{Algebraic system}
\label{alg_sys_hb}
A natural configuration arises for $a_1 = a_2$ and $s_1 = t_1$, $t_2 = -s_2$ for $s_1, s_2 > 0$, $|{\bf s}| \ll 1$, which may be thought of as a symmetric herringbone with a convex entrance, and a model for a regular double-pinned structure. One may also consider the special cases of $s_1 < 0$ (concave entrance) and ${\bf t} = -{\bf s}$, where the origin of the coordinate axes is shifted such that ${\bf s}$ and ${\bf t}$ both lie on the line $y = \theta x$ (wedge).

The formulation of the problem is similar to that outlined in Section~\ref{formulation}. We first consider the general case illustrated by Fig.~\ref{herr}. The total flexural displacement $u(x,y)$ is given by
\begin{multline}
u(x,y)=u_{\scriptsize{\mbox{inc}}}(x,y)+\sum_{n=0}^\infty A_n^{\scriptsize \mbox{(I)}} g(\beta; x,y;na_1,b/2) + \sum_{m=0}^\infty A_m^{\scriptsize \mbox{(II)}} g(\beta;x,y;s_1+ma_1,s_2+b/2)\\+\sum_{c=0}^\infty A_c^{\scriptsize \mbox{(III)}} g(\beta;x,y;ca_2,-b/2) + \sum_{d=0}^\infty A_d^{\scriptsize \mbox{(IV)}} g(\beta;x,y;t_1+da_2,t_2-b/2),
\label{totfieldher}
\end{multline}
where the scattering coefficients $A_n^{\scriptsize \mbox{(I)}}$, $A_m^{\scriptsize \mbox{(II)}}$, $A_c^{\scriptsize \mbox{(III)}}$, $A_d^{\scriptsize \mbox{(IV)}}$  are to be determined.  
In a similar way to Section~\ref{formulation}, boundary conditions are applied so that the total displacement $u(x,y)$ vanishes at the rigid pins. We include the details of the method in Appendix~\ref{app2}. Here we present only the final results for the case $a_1 = a_2$.

Adopting the notations of~(\ref{nots}), but for $\alpha = \mbox{I}-\mbox{IV}$, and the shorthand notation from~(\ref{short1}) together with the additional vector ${\boldsymbol \tau}=(0,b/2)$,
we obtain the 
functional equation

\begin{equation}
\left(
\begin{array}{c}
\hat B_-^{\scriptsize \mbox{(I)}}\\
\hat B_-^{\scriptsize \mbox{(II)}}\\
\hat B_-^{\scriptsize \mbox{(III)}}\\
\hat B_-^{\scriptsize \mbox{(IV)}}\\
\end{array}
\right)={\boldsymbol {\cal K}}
\left(
\begin{array}{c}
\hat A_+^{\scriptsize \mbox{(I)}}\\
\hat A_+^{\scriptsize \mbox{(II)}}\\
\hat A_+^{\scriptsize \mbox{(III)}}\\
\hat A_+^{\scriptsize \mbox{(IV)}}
\end{array}
\right)+
\left(
\begin{array}{c}
\hat F^{\scriptsize \mbox{(I)}}\\
\hat F^{\scriptsize \mbox{(II)}}\\
\hat F^{\scriptsize \mbox{(III)}}\\
\hat F^{\scriptsize \mbox{(IV)}}
\end{array}
\right),
\label{whopf1}
\end{equation}
with matrix kernel 
\begin{equation}
{\footnotesize
{\boldsymbol {\cal K}}= 
 \left(
\begin{array}{cccc}
\hat G(\beta,k;{\boldsymbol \tau};{\boldsymbol \tau})  & \hat G(\beta,k;{\boldsymbol \tau};{\boldsymbol \tau}+{\bf s})& \hat G(\beta,k;{\boldsymbol \tau};-{\boldsymbol \tau}) & \hat G(\beta,k;{\boldsymbol \tau};{\bf t}-{\boldsymbol \tau})\\
\hat G(\beta,k;{\boldsymbol \tau}+{\bf s};{\boldsymbol \tau})  & \hat G(\beta,k;{\boldsymbol \tau}+{\bf s};{\boldsymbol \tau}+{\bf s})& \hat G(\beta,k;{\boldsymbol \tau}+{\bf s};-{\boldsymbol \tau}) & \hat G(\beta,k;{\boldsymbol \tau}+{\bf s};{\bf t}-{\boldsymbol \tau})\\
\hat G(\beta,k;-{\boldsymbol \tau};{\boldsymbol \tau})  & \hat G(\beta,k;-{\boldsymbol \tau};{\boldsymbol \tau}+{\bf s})& \hat G(\beta,k;-{\boldsymbol \tau};-{\boldsymbol \tau}) & \hat G(\beta,k;-{\boldsymbol \tau};{\bf t}-{\boldsymbol \tau})\\
\hat G(\beta,k;{\bf t}-{\boldsymbol \tau};{\boldsymbol \tau})  & \hat G(\beta,k;{\bf t}-{\boldsymbol \tau};{\boldsymbol \tau}+{\bf s})& \hat G(\beta,k;{\bf t}-{\boldsymbol \tau};-{\boldsymbol \tau}) & \hat G(\beta,k;{\bf t}-{\boldsymbol \tau};{\bf t}-{\boldsymbol \tau})\\
\end{array}
\right)}.
\label{herringkernel1a}
\end{equation}
Note that all the elements are referenced to the origin illustrated in Fig.~\ref{herr}. Using ~(\ref{symm_whopf_sys}), the kernel can be rewritten in the simpler form: 
 \begin{equation}
{\boldsymbol {\cal K}}= 
 \left(
\begin{array}{cccc}
\hat G(\beta,k;{\bf 0};{\bf 0})  & \hat G(\beta,k;-{\bf s};{\bf 0})& \hat G(\beta,k;2{\boldsymbol \tau};{\bf 0}) & \hat G(\beta,k;2{\boldsymbol \tau}-{\bf t};{\bf 0})\\
\hat G(\beta,k;{\bf s};{\bf 0})  & \hat G(\beta,k;{\bf 0};{\bf 0})& \hat G(\beta,k;2{\boldsymbol \tau}+{\bf s};{\bf 0}) & \hat G(\beta,k;2{\boldsymbol \tau}+{\bf s}-{\bf t};{\bf 0})\\
\hat G(\beta,k;-2{\boldsymbol \tau};{\bf 0})  & \hat G(\beta,k;-2{\boldsymbol \tau}-{\bf s};{\bf 0})& \hat G(\beta,k;{\bf 0};{\bf 0}) & \hat G(\beta,k;-{\bf t};{\bf 0})\\
\hat G(\beta,k;{\bf t}-2{\boldsymbol \tau};{\bf 0})  & \hat G(\beta,k;{\bf t}-{\bf s}-2{\boldsymbol \tau};{\bf 0})& \hat G(\beta,k;{\bf t};{\bf 0}) & \hat G(\beta,k;{\bf 0};{\bf 0})\\
\end{array}
\right). 
\label{herringkernel2a}
\end{equation}
As previously discussed for the constituent shifted pairs in Sections~\ref{kerher} and~\ref{sec:dipole1}, the zeros of the kernel matrix determine frequency regimes in which the semi-infinite systems may exhibit interesting scattering patterns. The same holds true for the herringbone systems described and illustrated here.

\subsection{Sources and dipoles}
\label{sec:dipole1HB}
\begin{figure}[h]
\centerline{
\includegraphics[width=.95\columnwidth]{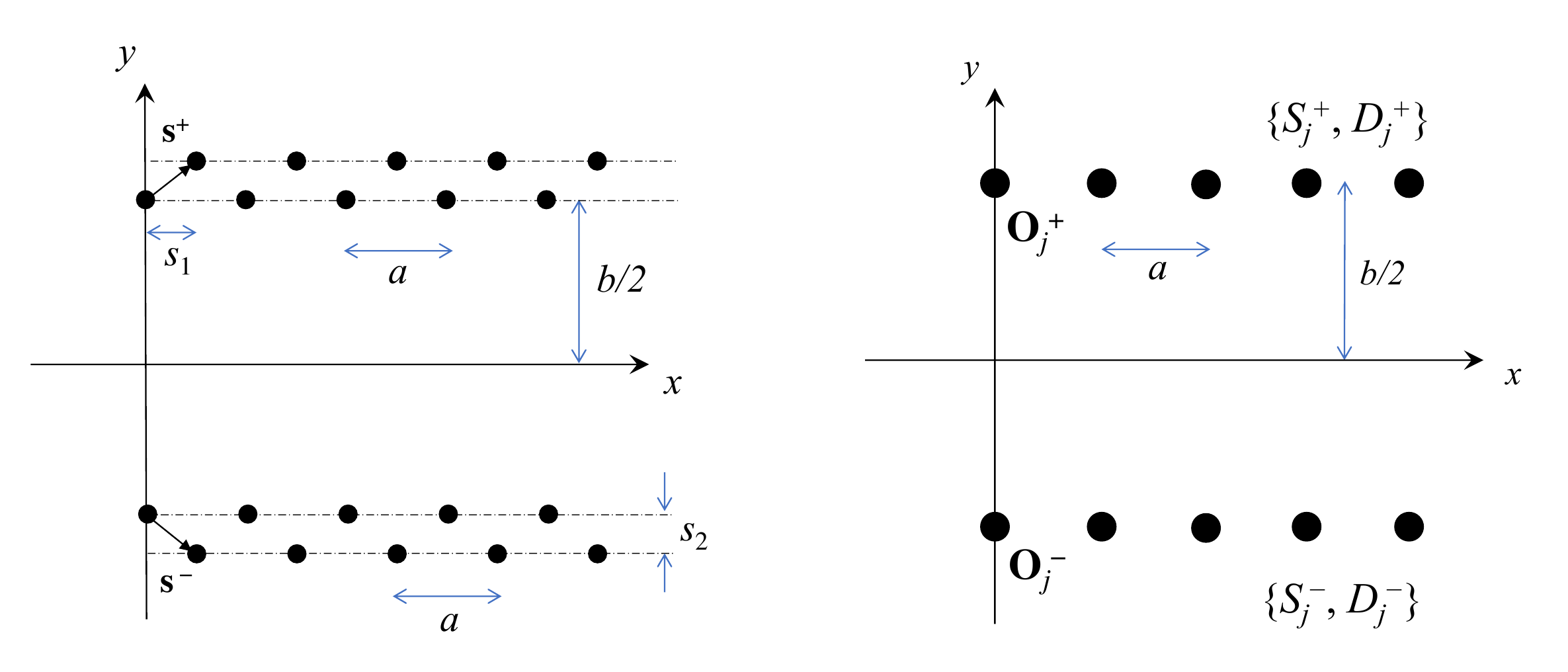}
\put(-145,60) {(a)}
\put(-65,60) {(b)}
}
\caption{ 
(a) Symmetric herringbone system of gratings defined by periodicity $a$ and shift vectors ${\bf s^\pm }$. Part (b) shows the source, dipole approximations.
\label{sources_dip}}
\end{figure}

In a similar way to Section~\ref{sec:dipole1}, we may consider the herringbone system as a pair of line arrays consisting of sources and dipoles. Each pair of gratings is approximated as a single array of point scatterers located at ${\bf O}_j^{\scriptsize \pm} = (ja,\pm b/2)$, as illustrated for the symmetric herringbone with ${\bf t} = {\bf s}^- = (s_1,-s_2)$ in Fig.~\ref{sources_dip}.
Following the approach defined in equations~(\ref{constraints})-(\ref{dip_eq1_disp}), we sketch the formulation for the herringbone. 
We begin by imposing the boundary conditions \begin{equation}
u \bigg|_{{\bf r}={\bf O}^{\pm}_j} = 0, \quad \frac{\partial u}{\partial {\bf s}^\pm}\bigg|_{{\bf r}={\bf O}^\pm_j} = 0,
\label{constr}
\end{equation}
where we employ the expression ${\bf s^+} = {\bf s}$ for ease of notation. The total field is expressed as 
\begin{equation}
u({\bf r}) = u_{\scriptsize{\mbox{inc}}}({\bf r}) + \sum_{\pm} \bigg[ \sum_{j = 0}^\infty S_j^{\pm} g(\beta; {\bf r};{\bf O}^\pm_j) + \sum_{j = 0}^\infty D_j^{\pm} \frac{\partial g}{\partial {\bf s}^\pm} (\beta; {\bf r}; {\bf O}^\pm_j) \bigg].
\label{sd}
\end{equation}
Here we associate the coefficients $S_j^{\pm}$ with the intensities of the sources, and $D_j^{\pm}$ with the intensities of the  dipoles as indicated in Fig.~\ref{sources_dip}, and $g( \beta; {\bf r};{\bf O}^\pm_j)$ is the single source Green's function as defined by equation~(\ref{gf}).

Substituting the constraints~(\ref{constr}) into~(\ref{sd}), we obtain two systems of equations:
\begin{equation}
\sum_{\pm} \bigg[ \sum_{j = 0}^\infty S_j^{\pm} g(\beta; {\bf r};{\bf O}^\pm_j) + \sum_{j = 0}^\infty D_j^{\pm} \frac{\partial g}{\partial {\bf s}^\pm} (\beta; {\bf r}; {\bf O}^\pm_j) \bigg] \bigg|_{{\bf r}={\bf O}^\pm_k} = -u_{\scriptsize{\mbox{inc}}} ({\bf O}^\pm_k), \quad k = 0,1,2,\ldots
\label{LAS1}
\end{equation}
and
\begin{equation}
\sum_{\pm} \bigg[ \sum_{j = 0}^\infty S_j^{\pm} \frac{\partial g}{\partial {\bf s_r}} (\beta; {\bf r}; {\bf O}^\pm_j) + \sum_{j = 0}^\infty D_j^{\pm} \frac{\partial}{\partial {\bf s_r}} \frac{\partial g}{\partial {\bf s}^\pm} (\beta; {\bf r};  {\bf O}^\pm_j) \bigg] \bigg|_{{\bf r}={\bf O}^\pm_k} = - \frac{\partial}{\partial {\bf s_r}} u_{\scriptsize{\mbox{inc}}} ({\bf O}^\pm_k), 
\label{LAS2}
\end{equation}
where 
\begin{equation}
\frac{\partial}{\partial {\bf s_r}}
=\Bigg \{ 
\begin{array}{c l}  
 \frac{\partial}{\partial {\bf s^+}} \quad \quad \mbox{for} \,\, {\bf r} = {\bf O}^+_k, \\
\\
\frac{\partial}{\partial {\bf s^-}} \quad \quad \mbox{for} \,\, {\bf r} = {\bf O}^-_k.
\end{array} 
\label{LAS3}
\end{equation}
For the sake of numerical illustrations, we introduce the truncation parameter $L$ to represent the number of points in each constituent grating of a restricted system. In that case, we have a $4L \times 4L$ linear algebraic system~(\ref{LAS1})-(\ref{LAS3}) for $\{S_j^\pm, D_j^\pm\}_{j=0}^{L-1}$.

\subsection{Algebraic systems in matrix form for wave scattering method}
\label{hb_new}
The full herringbone system comprising four pinned gratings whose point scatterers enforce zero displacement is defined by~(\ref{her1})-(\ref{her4}). Its matrix form is given by
\begin{equation}
\left(
\begin{array}{c}
\textbf{F}^{\scriptsize \mbox{(I)}}\\
\textbf{F}^{\scriptsize \mbox{(II)}}\\
\textbf{F}^{\scriptsize \mbox{(III)}}\\
\textbf{F}^{\scriptsize \mbox{(IV)}}
\end{array}
\right)= 
\left(
\begin{array}{cccc}
\textbf{M}^{(11)}  & \textbf{M}^{(12)} & \textbf{M}^{(13)} & \textbf{M}^{(14)} \\
\textbf{M}^{(21)} & \textbf{M}^{(11)} & \textbf{M}^{(23)} & \textbf{M}^{(24)} \\
\textbf{M}^{(13)} & \textbf{M}^{(32)} & \textbf{M}^{(11)} & \textbf{M}^{(34)} \\
\textbf{M}^{(41)} & \textbf{M}^{(24)} & \textbf{M}^{(43)} & \textbf{M}^{(11)} 
\end{array}
\right)
\left(
\begin{array}{c}
\textbf{A}^{\scriptsize \mbox{(I)}}\\
\textbf{A}^{\scriptsize \mbox{(II)}}\\
\textbf{A}^{\scriptsize \mbox{(III)}}\\
\textbf{A}^{\scriptsize \mbox{(IV)}}
\end{array}
\right),
\label{hbone_me}
\end{equation}
where the various terms $\textbf{M}^{(\mbox{\scriptsize ij})}$ are block matrices. For the truncated semi-infinite system with truncation parameter $L$, these blocks are of size $L \times L$ and each of $\textbf{F}^{\scriptsize \mbox{(I)}}$ to $\textbf{F}^{\scriptsize \mbox{(IV)}}$ and $\textbf{A}^{\scriptsize \mbox{(I)}}$ to $\textbf{A}^{\scriptsize \mbox{(IV)}}$ is an $L \times 1$ column vector. The $2L \times 2L$ matrix equation for the line array of sources and dipoles approximating a shifted pair is presented in equations~(\ref{dipole_me}),~(\ref{dipole_me_entries}) in Appendix~\ref{sec:eval_dipole}, where the $\textbf{F}$ terms incorporate two boundary conditions~(\ref{constraints}) rather than only zero displacement as in~(\ref{hbone_me}). 

For the system~(\ref{LAS1})-(\ref{LAS3}), we derive a similar truncated system where the $4L \times 4L$ matrix may be considered as an array of four $2L \times 2L$ block matrices. The two blocks on the main diagonal are determined using the system~(\ref{dipole_me}),~(\ref{dipole_me_entries}), provided that we replace the arguments $(ja,0)$ with $(ja, \pm b/2)$. Here we shall denote them ${\boldsymbol {\cal M}}^{\scriptsize (++)}({\bf s}^+)$ and ${\boldsymbol {\cal M}}^{\scriptsize (- -)}({\bf s}^-)$,  where $+$ denotes the upper line, and $-$, the lower (see~(\ref{hbone_me_SD}) below).

The off-diagonal block matrices take into account the interaction of the upper and lower line arrays, and therefore require expressions that differ from the isolated shifted pair given by~(\ref{dip_als_1_exp}), (\ref{dip_als_2_exp}). For the first derivative terms in equations~(\ref{LAS1}), (\ref{LAS2}), we deduce~\cite{Abram} 
\begin{equation}
\frac{\partial g}{\partial {\bf s}^\pm} (\beta; {\bf r}; {\bf O}^\pm_j) = \frac{i}{8 \beta} \left[H^{(1)}_1(\beta \rho_{\xi}) + \frac{2i}{\pi}K_1 (\beta \rho_{\xi}) \right] \bigg( \frac{s_1(x-ja)}{\rho_{\xi}} \pm \frac{s_2(y \mp b/2)}{\rho_{\xi}} \bigg), 
\label{array_1stderiv}
\end{equation}
where ${\bf r} = (ka, \mp b/2)$, $k = 0,1,2,\ldots$ and $\rho_{\xi}$ is the distance between ${\bf r}$ and ${\bf O}^\pm_j$, as defined  in~(\ref{details}). The important difference is that the second group of terms involving $s_2$ no longer vanish, since $y = \mp b/2$ is always of opposite sign to the $y$-component of ${\bf O}^\pm_j$. Similarly,
\begin{equation}
\frac{\partial g}{\partial {\bf s_r}^\pm} (\beta; {\bf r}; {\bf O}^\pm_j) = - \frac{i}{8 \beta} \left[H^{(1)}_1(\beta \rho_{\xi}) + \frac{2i}{\pi}K_1 (\beta \rho_{\xi}) \right] \bigg( \frac{s_1(x-ja)}{\rho_{\xi}} \pm \frac{s_2(y \mp b/2)}{\rho_{\xi}} \bigg).
\end{equation}
As one would expect, the second derivatives in~(\ref{LAS2}) also include more terms:
\begin{eqnarray}
\frac{\partial}{\partial {\bf s_r}} \frac{\partial g}{\partial {\bf s}^\pm} (\beta; {\bf r}; {\bf O}^\pm_j) & = & s_1 \bigg\{ \frac{s_1 g_{\xi}^{(1)}}{\rho_{\xi}}  + \frac{\rho_{\xi} \beta \left[ H_0^{(1)}(\beta \rho_{\xi}) - \frac{2i}{\pi} K_0(\beta \rho_{\xi}) \right] -  2 g_{\xi}^{(1)}}{\rho_{\xi}^3} \nonumber \\
& & \times  \left( s_1(x - ja)^2 \pm s_2(x - ja)(y \mp b/2) \right) \bigg \} \nonumber \\
& \pm & s_2 \bigg\{ \pm \frac{s_2 g_{\xi}^{(1)}}{\rho_{\xi}} + \frac{\rho_{\xi} \beta \left[ H_0^{(1)}(\beta \rho_{\xi}) - \frac{2i}{\pi} K_0(\beta \rho_{\xi}) \right] -  2 g_{\xi}^{(1)}}{\rho_{\xi}^3} \nonumber \\
& & \times  \left( s_1(x - ja)(y \mp b/2) \pm s_2(y \mp b/2)^2 \right) \bigg \},
\end{eqnarray}
where once again $\rho_{\xi}$ is the distance between ${\bf r}$ and ${\bf O}^\pm_j$ and $g_{\xi}^{(1)}$ is defined as (see also~(\ref{greens_1_xi})):
\begin{equation}
g_{\xi}^{(1)} (\beta; {\bf r}; {\bf O}^\pm_j)  = H_1^{(1)} (\beta \rho_{\xi}) + \frac{2i}{\pi} K_1 (\beta \rho_{\xi}).
\label{greens_1_mt}
\end{equation} 
The matrix equation for the system may be expressed in the following way
\begin{equation}
\normalsize
\left(
\begin{array}{c}
{\boldsymbol {\cal F}}^+ \\
{\boldsymbol {\cal F}}^-
\end{array}
\right)
=  
\left(
\begin{array}{cc}
{\boldsymbol {\cal M}}^{\scriptsize (++)}({\bf s}^+) & {\boldsymbol {\cal M}}^{\scriptsize (+ -)}({\bf s}^-) \\
{\boldsymbol {\cal M}}^{\scriptsize (- +)}({\bf s}^+)  &  {\boldsymbol {\cal M}}^{\scriptsize (- -)}({\bf s}^-) \\
\end{array}
\right)
\left(
\begin{array}{c}
{\boldsymbol {\cal T}}^+ \\
{\boldsymbol {\cal T}}^-
\end{array}
\right),
\label{hbone_me_SD}
\end{equation}
where the two $2L \times 2L$ block matrices ${\boldsymbol {\cal M}}^{\scriptsize (- +)}$ and ${\boldsymbol {\cal M}}^{\scriptsize (+ -)}$ give the information regarding the interaction of the upper array (governed by the shift vector ${\bf s}^+$) and the lower array (characterised by ${\bf s}^-$), evaluated using equations~(\ref{array_1stderiv})-(\ref{greens_1_mt}). The two column vectors have size $4L \times 1$, consisting of two concatenated $2L \times 1$ vectors defined by
\begin{equation}
{\boldsymbol {\cal F}}^+ = 
\left(
\begin{array}{c}
\textbf{F}_1^+\\
\textbf{F}_2^+
\end{array}
\right)
, \,\,\,\,
{\boldsymbol {\cal F}}^- = 
\left(
\begin{array}{c}
\textbf{F}_1^-\\
\textbf{F}_2^-
\end{array}
\right)
; \,\,\,\,
{\boldsymbol {\cal T}}^+ = 
\left(
\begin{array}{c}
\textbf{S}^+\\
\textbf{D}^+
\end{array}
\right)
, \,\,\,\,
{\boldsymbol {\cal T}}^- = 
\left(
\begin{array}{c}
\textbf{S}^-\\
\textbf{D}^-
\end{array}
\right)
.
\end{equation}
The source and dipole coefficients for the upper and lower arrays in Fig.~\ref{sources_dip}, $S_j^{\pm}, D_j^{\pm}$, are represented by the column vectors on the right-hand side of~(\ref{hbone_me_SD}). The column vectors $\textbf{F}_1^{\pm}$ and $\textbf{F}_2^{\pm}$ represent the two sets of boundary conditions on the right hand sides of~(\ref{LAS1}), (\ref{LAS2}) for each of the corresponding line arrays. In the illustrative examples that follow, both systems~(\ref{hbone_me}) and~(\ref{hbone_me_SD}) are solved and compared.


\subsection{Herringbone systems for waveguiding and localisation}
\label{hb_illus}
We recall the work of Jones {\it et al.}~\cite{ISJ_NVM_ABM} who used the connection between the channelling of trapped modes in a pair of semi-infinite gratings and the Bloch-Floquet analysis for the infinite waveguide. The Bloch modes are obtained by solving the eigenvalue problem for the matrix of governing grating Green's functions, which is equivalent to finding the zeros of the determinant of the kernel matrix of Green's functions presented in~\cite{ISJ_NVM_ABM} and here. Considering the case of normal incidence, the authors presented a contour plot in Fig. 3(b) of~\cite{ISJ_NVM_ABM} that identifies the range of values of the spectral parameter $\beta$ and grating separation $b$ to support waveguide modes. An effective waveguide model for the simply supported boundary condition was derived which can be used in conjunction with the Bloch-Floquet analysis to estimate the wavenumber $k_x$, and hence wavelength $\lambda_w$, for the first order waveguide modes: 
\begin{equation}
b = \frac{\pi}{\sqrt{\beta^2 - k_x^2}}, \,\,\, \lambda_w = \frac{2 \pi}{k_x}, \,\,\, \lambda_{\mbox{\scriptsize ext}} = \frac{2 \pi}{\beta},
\label{wg_model}
\end{equation}
where the wavelength exterior to the gratings is denoted by $\lambda_{\mbox{\scriptsize ext}}$.

A similar approach was implemented by Haslinger {\it et al.}~\cite{has2014} for the connection between the scattering problem and the infinite grating system's waveguide modes. Using the related eigenvalue problem for a governing matrix of grating Green's functions dependent on $\beta$ and $k_x$, and incorporating spacing $b$, solutions are obtained in the form of localised minima of the logarithm of the determinant function (see Fig.~\ref{hb_examp_1}(a) here, for example). In conjunction with an approximate waveguide model for the Helmholtz operator (neglecting the evanescent modes for the biharmonic case), the Table 1 in~\cite{has2014}, and updated here in the electronic supplementary material Appendix~\ref{wg_table}, presents a selection of illustrative parameter settings. For gratings with unit periodicity ($a = 1$), various pairs of $(\beta, k_x)$ values determine resonant trapped modes for spacings $b$, that are well approximated by~(\ref{wg_model}). 

There is good correspondence between the $\beta$, $b$ pairs in that table and those featured in the white strip illustrating the approximate regions of roots in Fig. 3(b) of~\cite{ISJ_NVM_ABM}. 
The additional feature of Table 1 in Appendix~\ref{wg_table}, i.e. providing associated $k_x$ values, may be used to extend this approach from normal to oblique incidence. The projection onto the waveguide's axis of symmetry,
\begin{equation}
k_x = \beta \cos{(\psi)}, \,\,\,\, 0 \le \psi \le \pi, 
\label{kx}
\end{equation}
is used to incorporate oblique angles of incidence $\psi$ that correspond to the wavenumbers $k_x$ for normally incident resonant modes in~\cite{ISJ_NVM_ABM}. Both methods provide invaluable insight for the first order periodic patterns for both the pair and the herringbone systems.

In the examples that follow, the periodicity $a$ of all gratings is taken to be unity, unless otherwise stated. We also refer to various constituent pairs of gratings within the herringbone structures. Recalling Fig.~\ref{herr}, we denote gratings $\mbox{I}$ and $\mbox{III}$, separated by $b$, as the inner pair. The outer pair refers to gratings $\mbox{II}$ and $\mbox{IV}$, and a shifted pair is either $\mbox{I}$, $\mbox{II}$ or $\mbox{III}$, $\mbox{IV}$.

\subsubsection{Waveguide modes}
In this section, we demonstrate two distinct types of waveguide mode. In the first case, we consider a localised mode that is supported by a simple grating pair waveguide, but which is enhanced, in terms of both amplitude and reduced leakage, by adding the extra gratings to produce the herringbone system. The second example features a pair that blocks a specific range of plane waves but, by forming the tuned herringbone system, highly localised waveguide modes are observable. In this way, we illustrate how a simple tuning parameter can be used to convert a reflective mode to a highly localised guided waveform.

\begin{figure}[h]
\begin{center}
\includegraphics[width=.45\columnwidth]{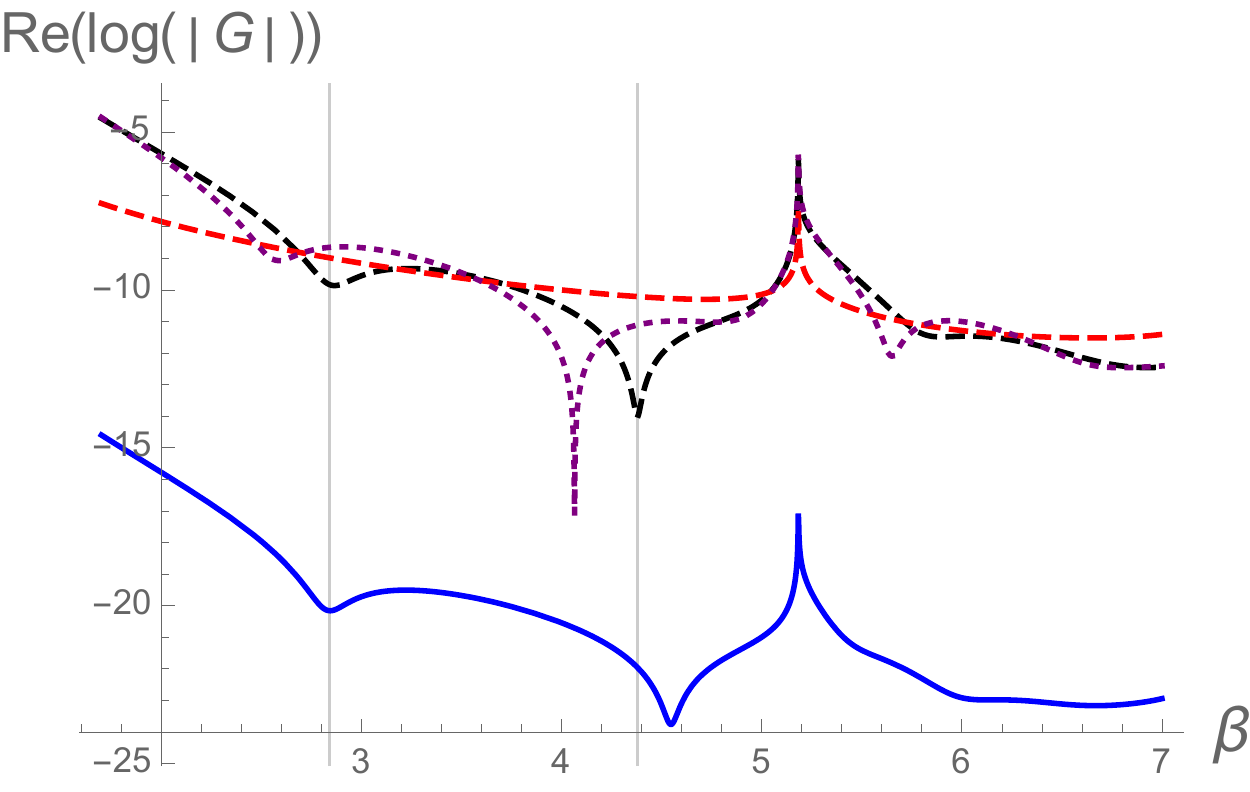}~
\put(-65,50) {(a)}
\includegraphics[width=.45\columnwidth]{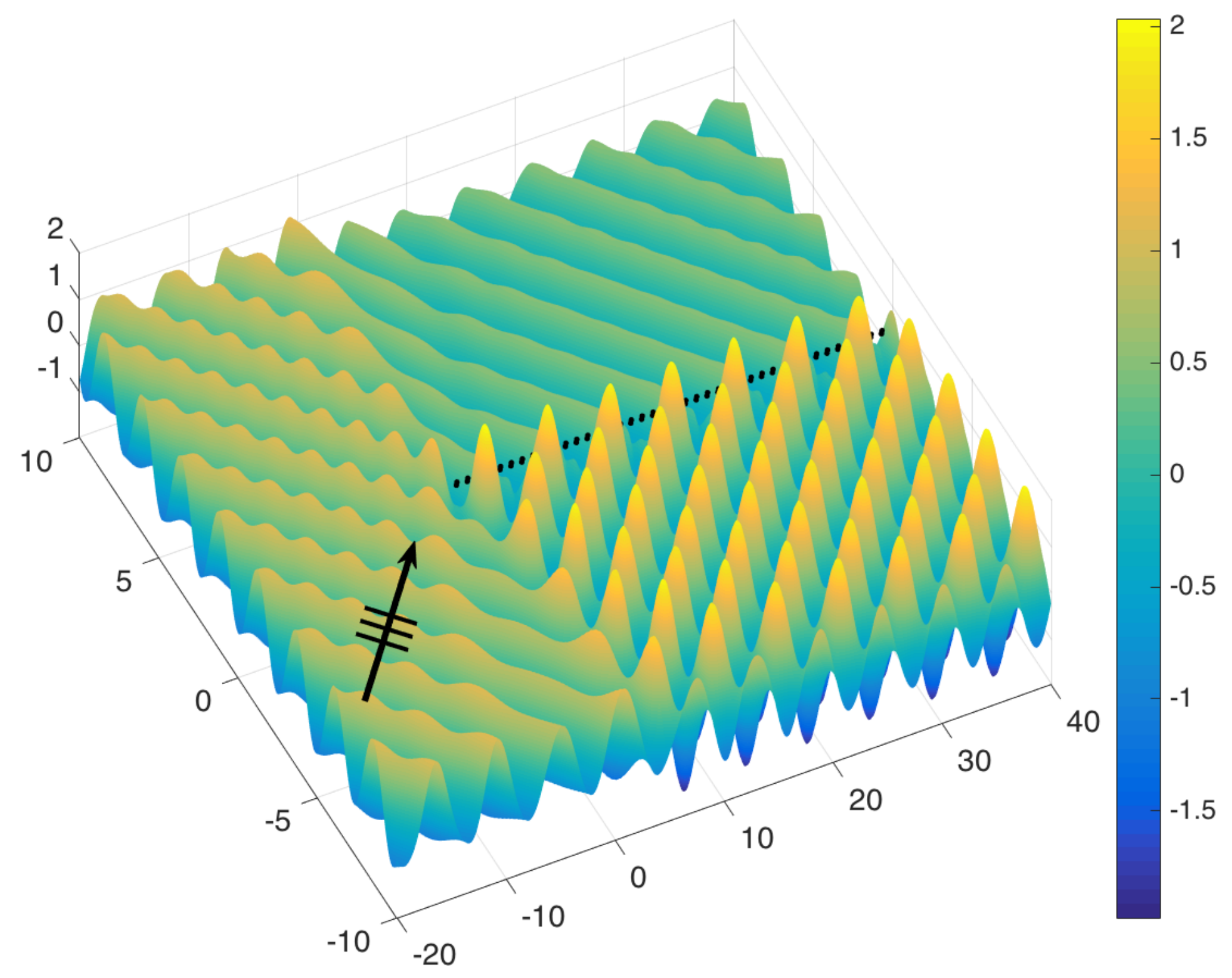}
\put(-65,50) {(b)}
 
\includegraphics[width=.45\columnwidth]{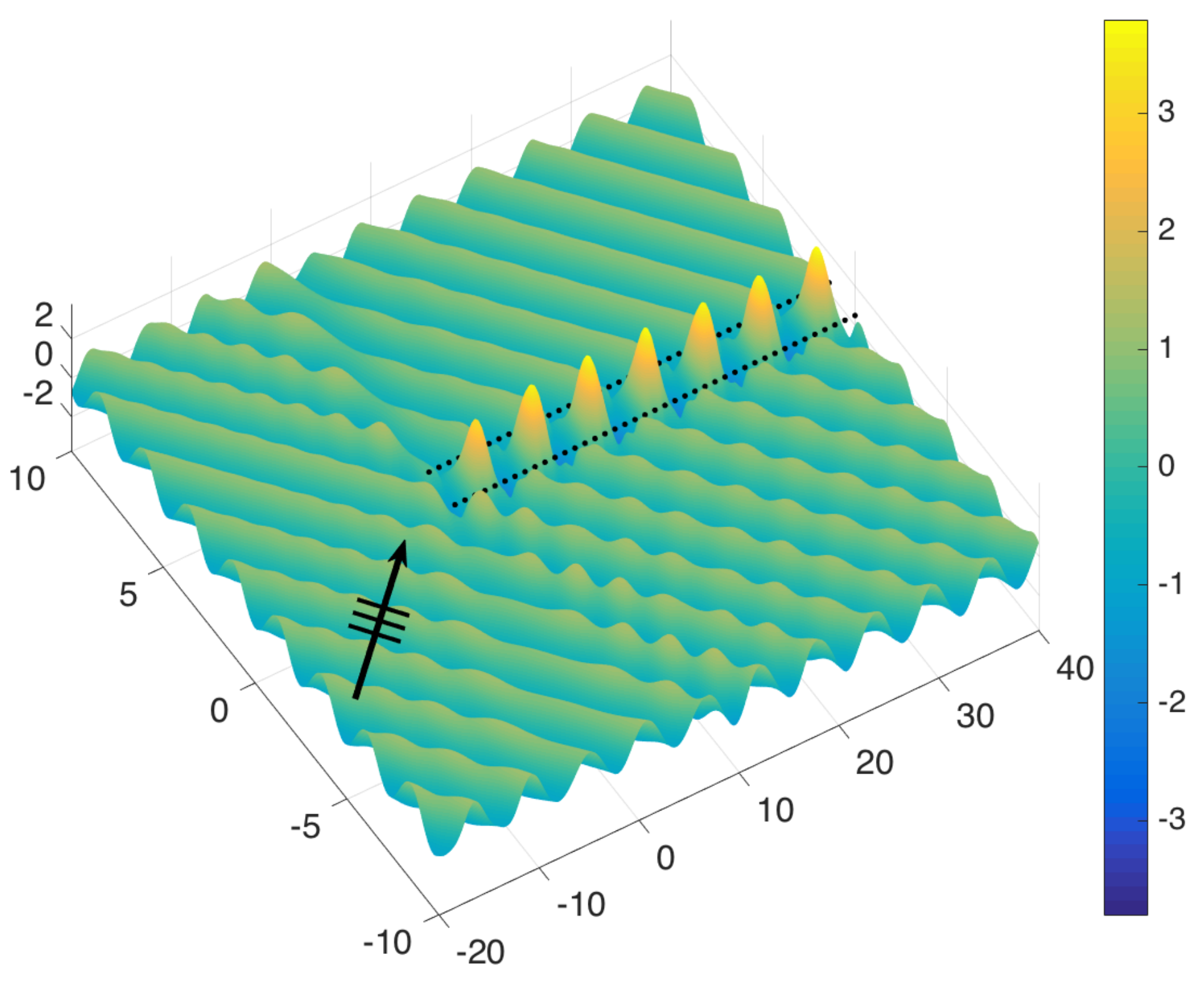}~
\put(-65,50) {(c)}
\includegraphics[width=.45\columnwidth]{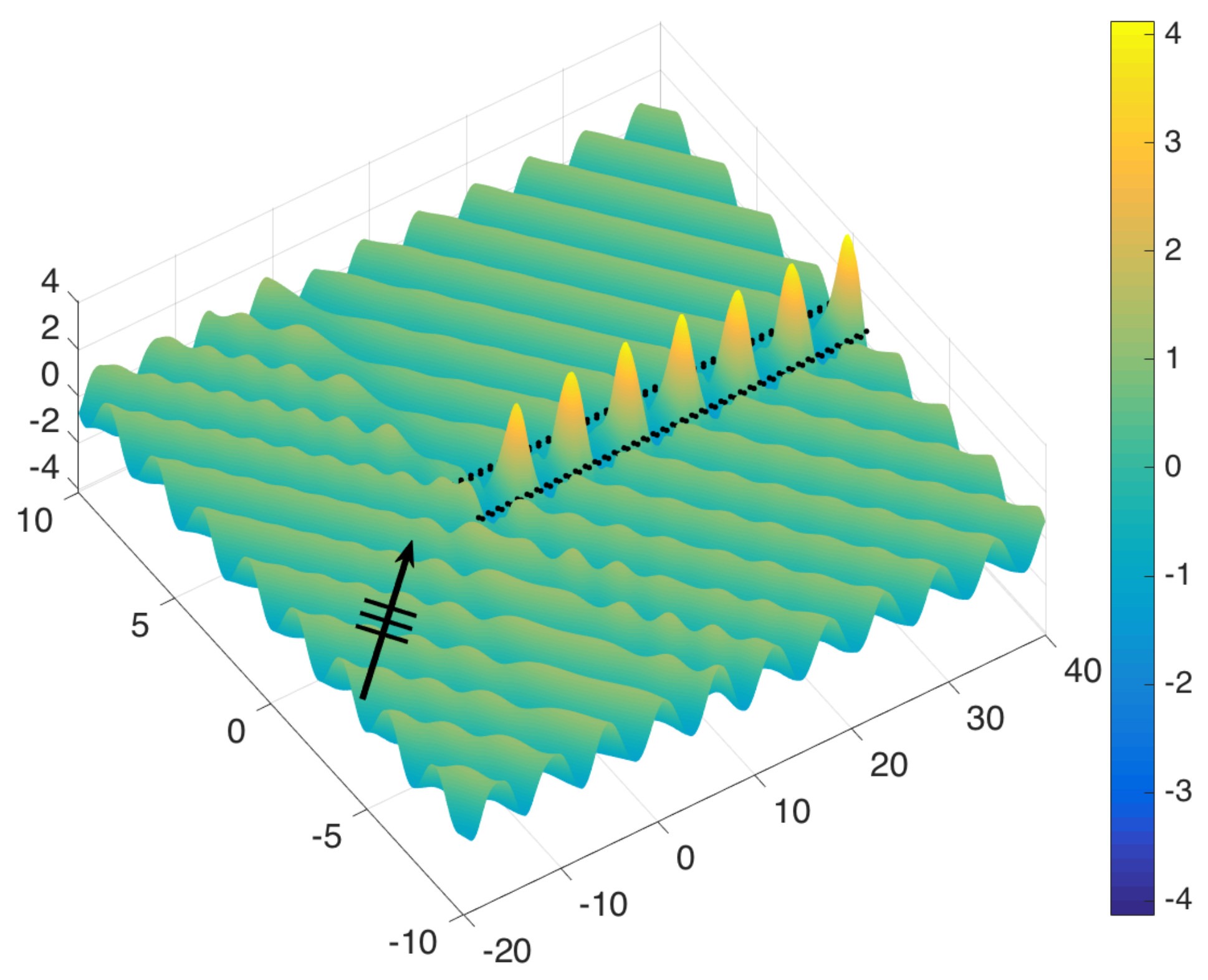}
\put(-65,50) {(d)}
\caption{Symmetric herringbone system with ${\bf s} = (0.2,0.1), b = \sqrt{2}$, $a_1 = a_2 = 1$.
(a) Identification of Bloch modes for the infinite herringbone system for $k_x = 1.1$. Solid curve shows eigenmodes for $\beta \simeq 2.84, 4.55$. Dashed curves show modes for the inner pair with spacing $b = \sqrt{2}$ (black), the shifted pair defined by ${\bf s}$ (red) and the outer pair with spacing $\sqrt{2}+0.2$ (purple). 
Parts (b-d) feature the total displacement field plots for an incident plane wave with $\psi = 1.17$ and $\beta = 2.82$ for $L = 100$, with the first 40 pins shown. (b) Shifted pair defined by ${\bf s} = (0.2,0.1)$. (c) Inner pair separated by $b = \sqrt{2}$. (d) Symmetric herringbone defined by the parameters $b$ and ${\bf s}, {\bf s}^-$. 
\label{hb_examp_1}
}
\end{center}
\end{figure}

We consider a symmetric herringbone defined by the parameter choices ${\bf s} = (0.2,0.1), b = \sqrt{2}$. 
The separation $b = \sqrt{2}$ was inspired by \cite{ISJ_NVM_ABM}, where the motivation was linked to the presence of Dirac-like points on the dispersion surfaces of doubly periodic systems, as explained in \cite{Dirac}.
The Bloch modes of the corresponding infinite herringbone system are obtained by solving the eigenvalue problem for a system of four appropriately positioned pinned gratings. 
The solutions are illustrated in Fig.~\ref{hb_examp_1}(a), where the appearance of localised minima of the logarithm of the determinant of the system's governing matrix indicate the presence of Bloch modes. We plot this function versus the spectral parameter $\beta$ in Fig.~\ref{hb_examp_1}(a) for $k_x = 1.1$.

The solid curve indicates the resonant $\beta$ values for the herringbone structure, the dashed (black) curve for the inner pair  separated by $b = \sqrt{2}$ and the dashed (red) curve for the shifted pair (either $\mbox{I}$, $\mbox{II}$ or $\mbox{III}$, $\mbox{IV}$). For the infinite system, pairs $\mbox{I}$, $\mbox{II}$ and $\mbox{III}$, $\mbox{IV}$ are characterised by the same Bloch modes. The outer pair's modes are shown by the dashed (purple) curve. We observe that both the herringbone and the inner pair possess two clear modes, whereas the shifted pair possesses no modes for this range of frequencies. The first mode occurs for $\beta \simeq 2.84$ and a line has been added to the figure to indicate the coincident frequency for both the inner pair and herringbone. 

The second vertical line at $\beta \simeq 4.38$ highlights the location of the second mode for the inner pair, but the herringbone's second mode arises for a different value, $\beta \simeq 4.55$. 
The implication is that for the higher frequency, the addition of the extra gratings to create the herringbone system induces a resonant mode that would not be apparent for the inner pair at the same frequency. In contrast the first mode is seen regardless of the addition of the extra gratings. 
This analysis is also valid for the semi-infinite herringbone systems, as illustrated in Figs.~\ref{hb_examp_1}(b-d), where we consider the case of $\beta = 2.82$ and hence $\psi = 1.17$ from~(\ref{kx}) for $k_x = 1.1$. Note that the determinant vanishes at $\beta \simeq 2.84$ in Fig.~\ref{hb_examp_1}(a), which indicates the neighbourhood of values of $\beta$ corresponding to waveguide modes in the semi-infinite structure, and for illustrative purposes we use the truncation parameter $L=100$.

We show the strong reflection for this incident wave on a pair of shifted gratings defined by ${\bf s} = (0.2,0.1)$ in Fig.~\ref{hb_examp_1}(b). The shifted pair exhibits no resonance, which is consistent with the dashed red curve in Fig.~\ref{hb_examp_1}(a). However, the inner pair of semi-infinite gratings supports a waveguiding effect consistent with the mode (dashed black curve) displayed in Fig.~\ref{hb_examp_1}(a). The total displacement field is plotted in Fig.~\ref{hb_examp_1}(c). A regular one-dimensional periodic pattern, with seven clear peaks and wavelength $\lambda_w = 2\pi/k_x \simeq 5.71$, is observed, shown here for the first forty pins.

The localised waveguide mode, wherein the incident plane wave (clearly indicated by the arrow in all diagrams) undergoes significant bending to be channelled between the pinned gratings, is observed for the herringbone structure in Fig.~\ref{hb_examp_1}(d). The mode is virtually identical for both cases (c) and (d), with the same seven regular peaks, and same wavelength $\lambda_w = 2\pi/k_x \simeq 5.71$. However, the addition of the outer gratings reduces the scattering and amplifies the waveguiding localisation, as shown in Fig.~\ref{hb_examp_1}(d), thereby enhancing the waveguiding effect. In this specific case, the enhancement factor is 1.09.

\begin{figure}[h]
\begin{center}
\includegraphics[width=.45\columnwidth]{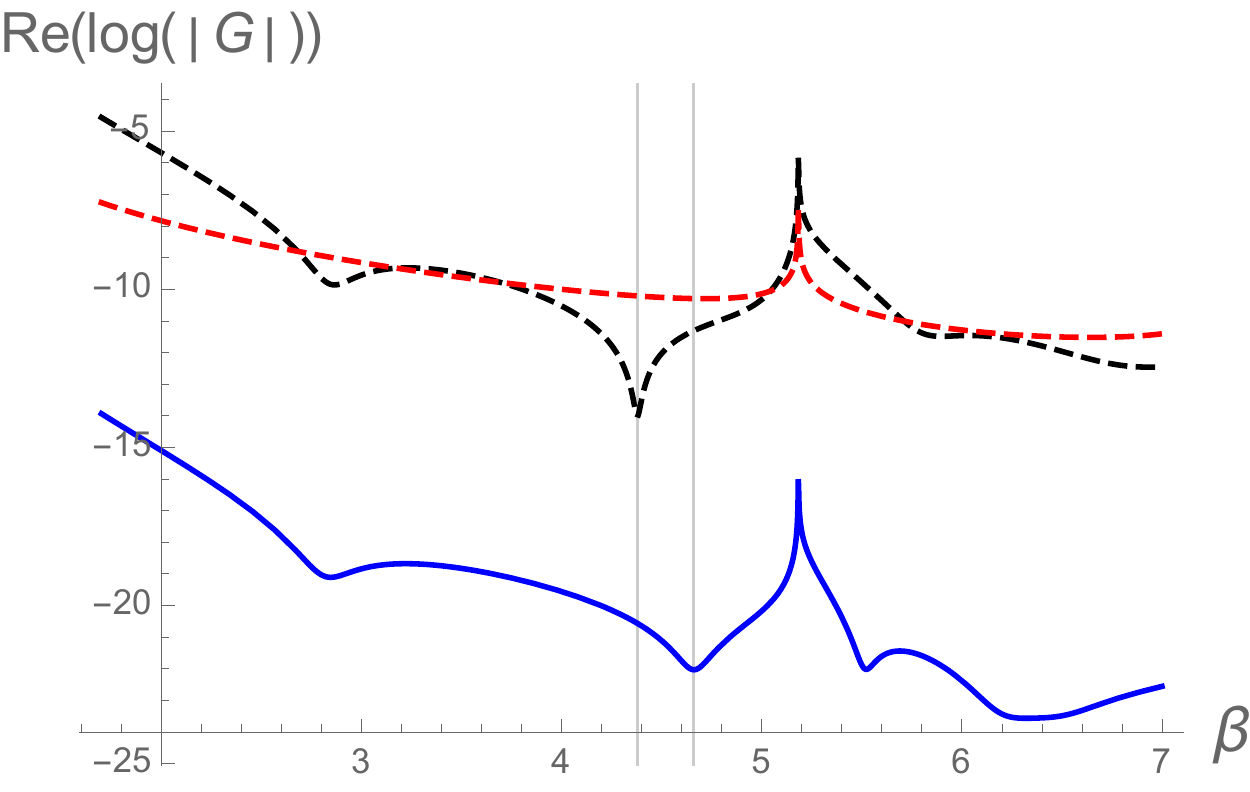}~
 \put(-70,55) {(a)}
\includegraphics[width=.42\columnwidth]{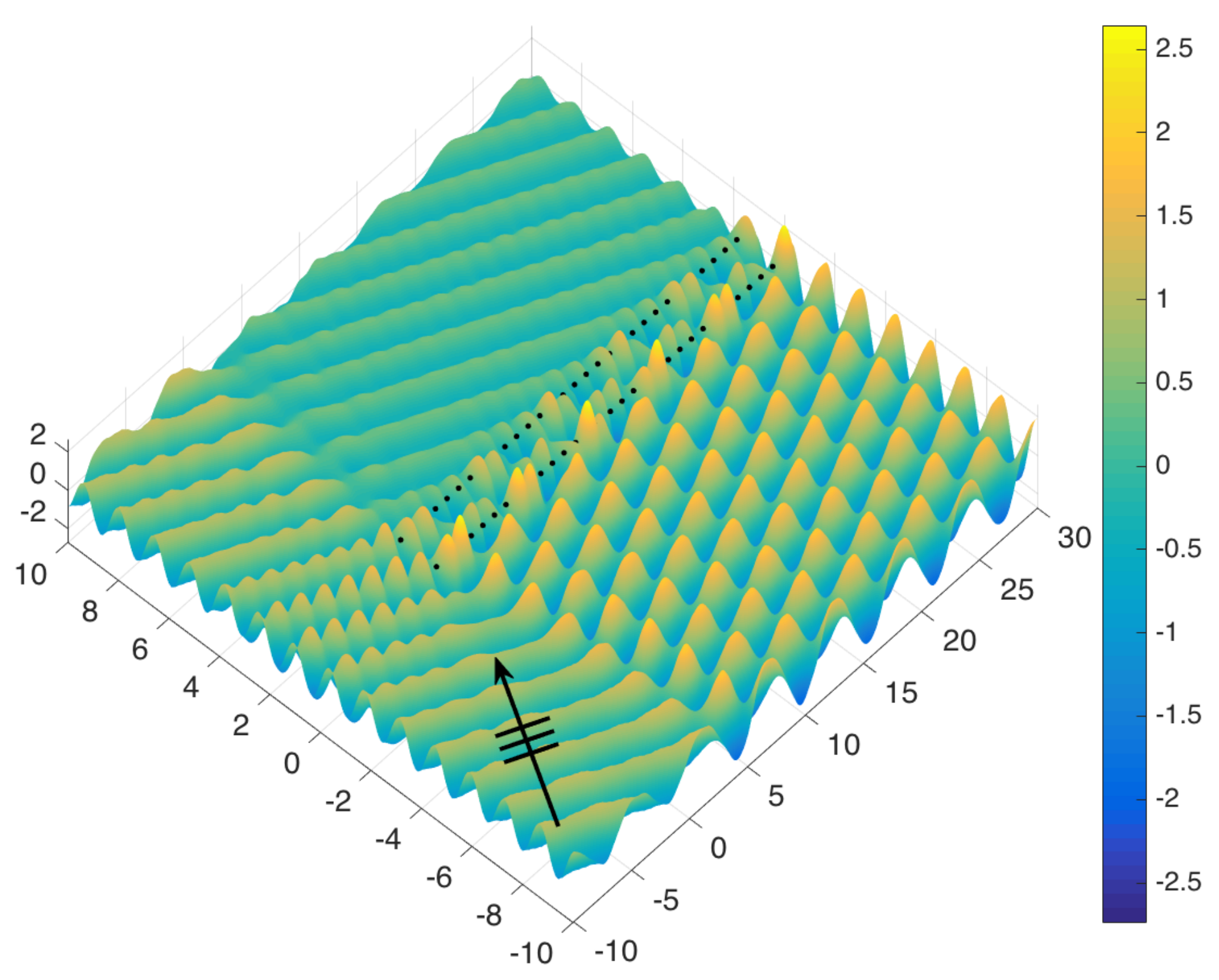}
 \put(-70,55) {(b)}
 
\includegraphics[width=.42\columnwidth]{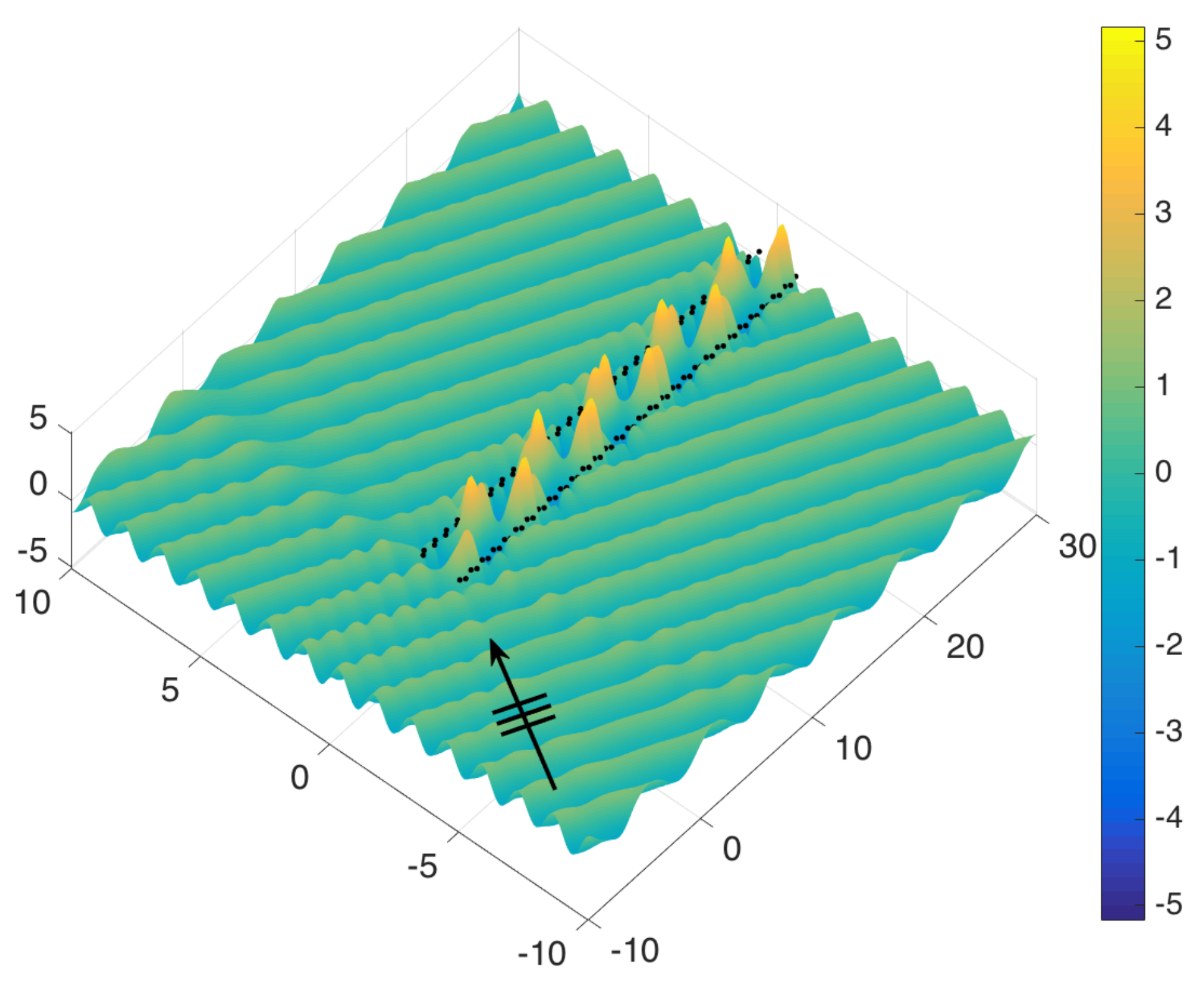}~~~~~
 \put(-70,55) {(c)}
\includegraphics[width=.42\columnwidth]{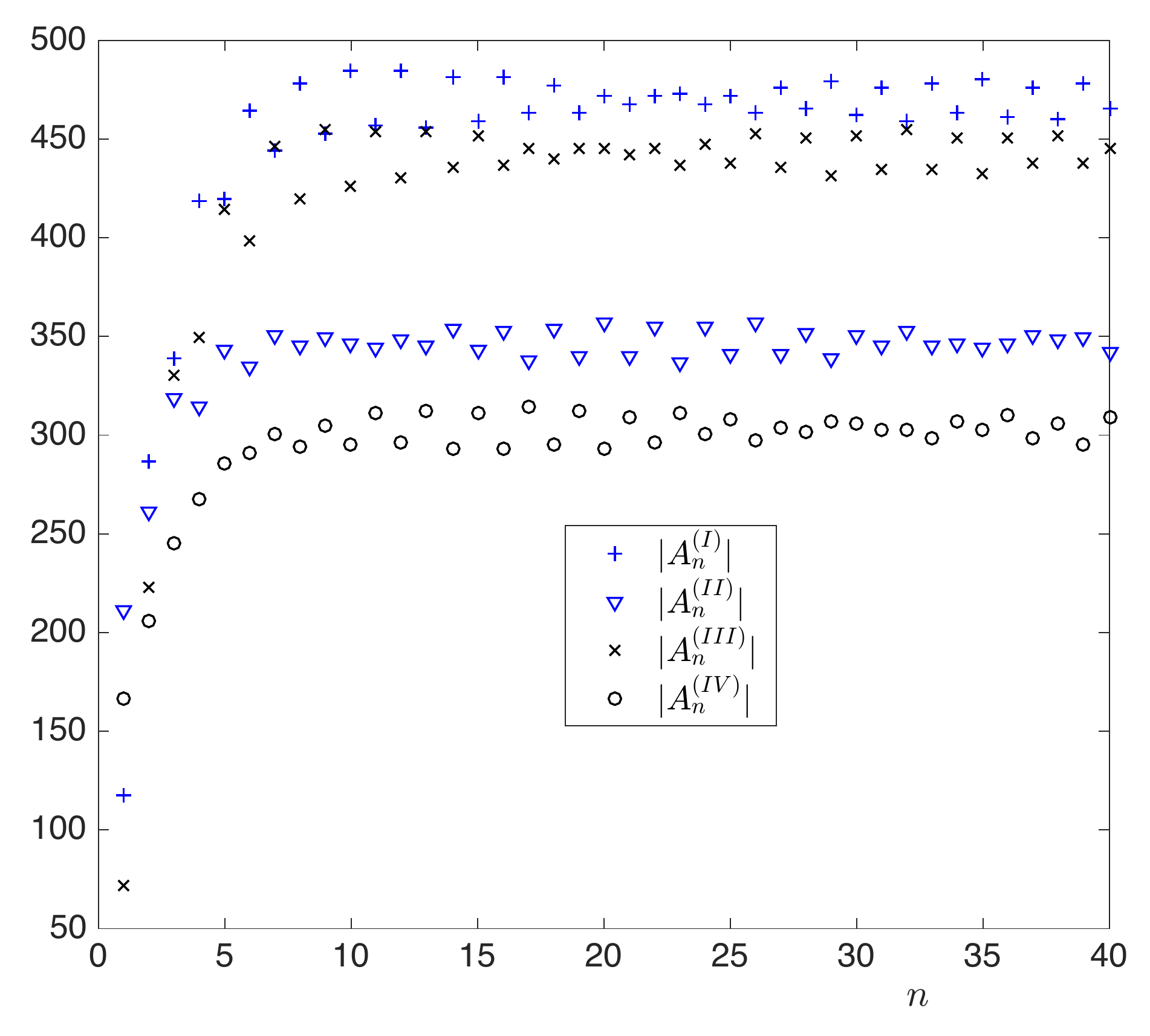}
 \put(-70,55) {(d)}
\caption{Symmetric herringbone system with ${\bf s} = (0.3,0.1), b = \sqrt{2}$.
(a) Identification of Bloch modes for the infinite herringbone system for $k_x = 1.1$. Solid curve shows eigenmodes for $\beta \simeq 2.80, 4.66, 5.53$. Dashed black curve shows inner pair modes, and dashed red curve represents the shifted pair. (b,c) Total displacement field plots for an incident plane wave with $\psi = 1.3325$ and $\beta = 4.66$ for $L = 100$ (first 30 shown). (b) Inner pair with $b = \sqrt{2}$. (c) Symmetric herringbone defined by $b$ and ${\bf s}, {\bf s}^-$. (d) Moduli of the scattering coefficients for the herringbone (first 40 shown).
\label{hb_examp_2}
}
\end{center}
\end{figure}
In contrast, the second mode brings no correspondence for the $\beta$ values of both systems in Fig.~\ref{hb_examp_1}(a). The extra shifted gratings alter the frequency of the second Bloch mode that can be excited by appropriately chosen values of $\beta$ and $\psi$ in the semi-infinite problem. A larger value of $s_1$ increases the difference in $\beta$ values for modes trapped by the inner pair and the herringbone structure. 

To better illustrate the concept of a pair of gratings that reflect waves at a given frequency being enhanced by an additional pair to support a waveguide transmission, we consider the case of ${\bf s} = (0.3,0.1)$ (increasing $s_1$ from the value illustrated in Fig.~\ref{hb_examp_1}). The logarithm of the determinant of the governing matrix for this system is plotted versus $\beta$ in Fig.~\ref{hb_examp_2}(a). This herringbone system's second mode occurs for $\beta = 4.66$, compared with that of $\beta = 4.38$ for the inner pair (both marked by vertical lines in Fig.~\ref{hb_examp_2}(a)). Thus, the addition of the outer gratings to the inner pair converts a reflection mode into a waveguide mode.

We show the total displacement field for the inner pair for a plane wave incident at $\psi = 1.3325$ and $\beta = 4.66$ in Fig.~\ref{hb_examp_2}(b). This choice of $\psi$ is determined using equation~(\ref{kx}) for $\beta = 4.66$ and $k_x = 1.1$. Reflection dominates but there is some localisation within the grating pair, as can be predicted from the dashed black curve in Fig.~\ref{hb_examp_2}(a). However, by adding the extra gratings defined by ${\bf s} = (0.3,0.1)$ above, and ${\bf s}^-$ below, and exciting the system with the same incident wave, we now observe a highly localised waveguide mode within the herringbone structure in Fig.~\ref{hb_examp_2}(c). 

The moduli of the scattering coefficients for all four gratings are plotted in Fig.~\ref{hb_examp_2}(d), with the first forty shown. The amplitudes for the inner pair coefficients $A_n^{\scriptsize \mbox{(I)}}$ and $A_n^{\scriptsize \mbox{(III)}}$ are very similar, which is also true for the first example with $\beta = 2.82$. The important factor here is that they are also of similar order to the outer pair coefficients $A_n^{\scriptsize \mbox{(II)}}$ and $A_n^{\scriptsize \mbox{(IV)}}$, which emphasises that all four gratings are required to support the localised mode, whereas only two gratings were sufficient for the previous case. In this case, the herringbone structure supports a unique waveguide mode, attainable only with the shifted grating structure. 

\subsubsection{Herringbone systems: dipole contribution}
The examples of waveguiding demonstrated in Figs.~\ref{hb_examp_1} and~\ref{hb_examp_2} arise for, respectively, $|{\bf s}| = 0.22$ and $|{\bf s}| = 0.32$, choices of ${\bf s}$ that are not well approximated using the array of sources and dipoles method outlined in Sections~\ref{sec:dipole1} and~\ref{sec:dipole1HB}. In this section, we consider an example for $|{\bf s}| = 0.032$, where the dipole approximations are found to be valid and the dipole terms dominate the source terms.

\begin{figure}[H]
\begin{center}
\includegraphics[width=.4\columnwidth]{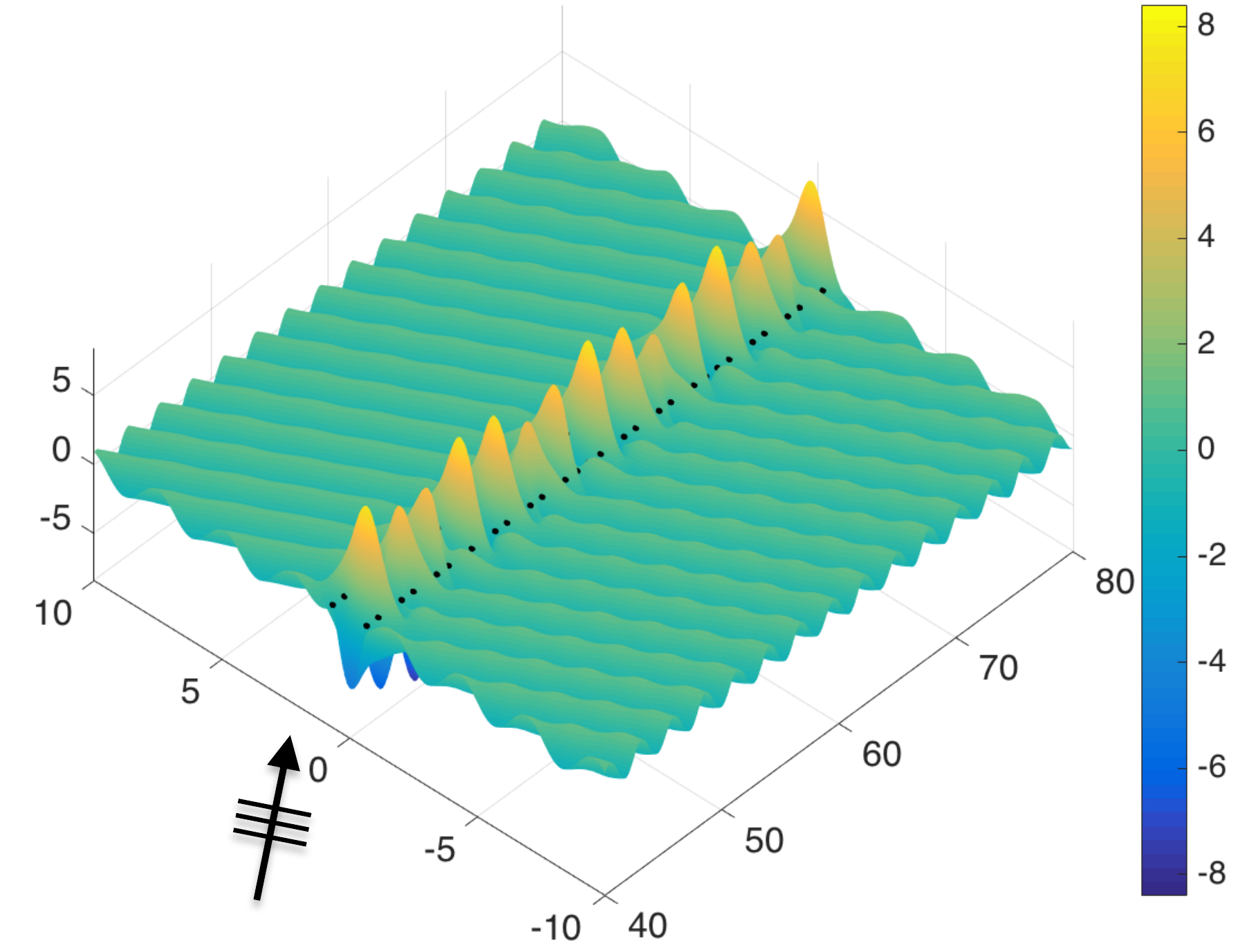}~~~~~~
 \put(-70,50) {(a)}
 \includegraphics[width=.4\columnwidth]{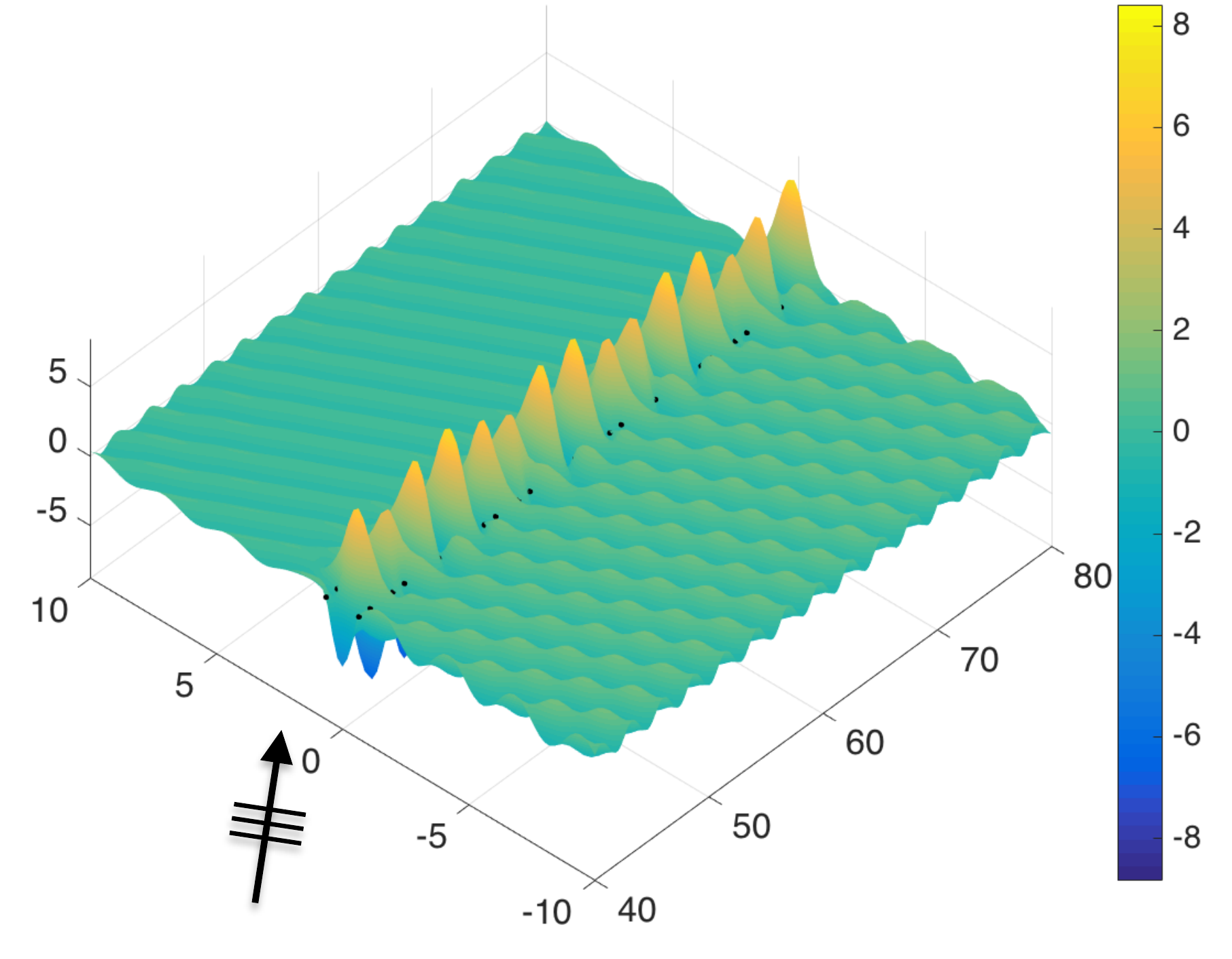}
 \put(-70,50) {(b)}
 
\includegraphics[width=.4\columnwidth]{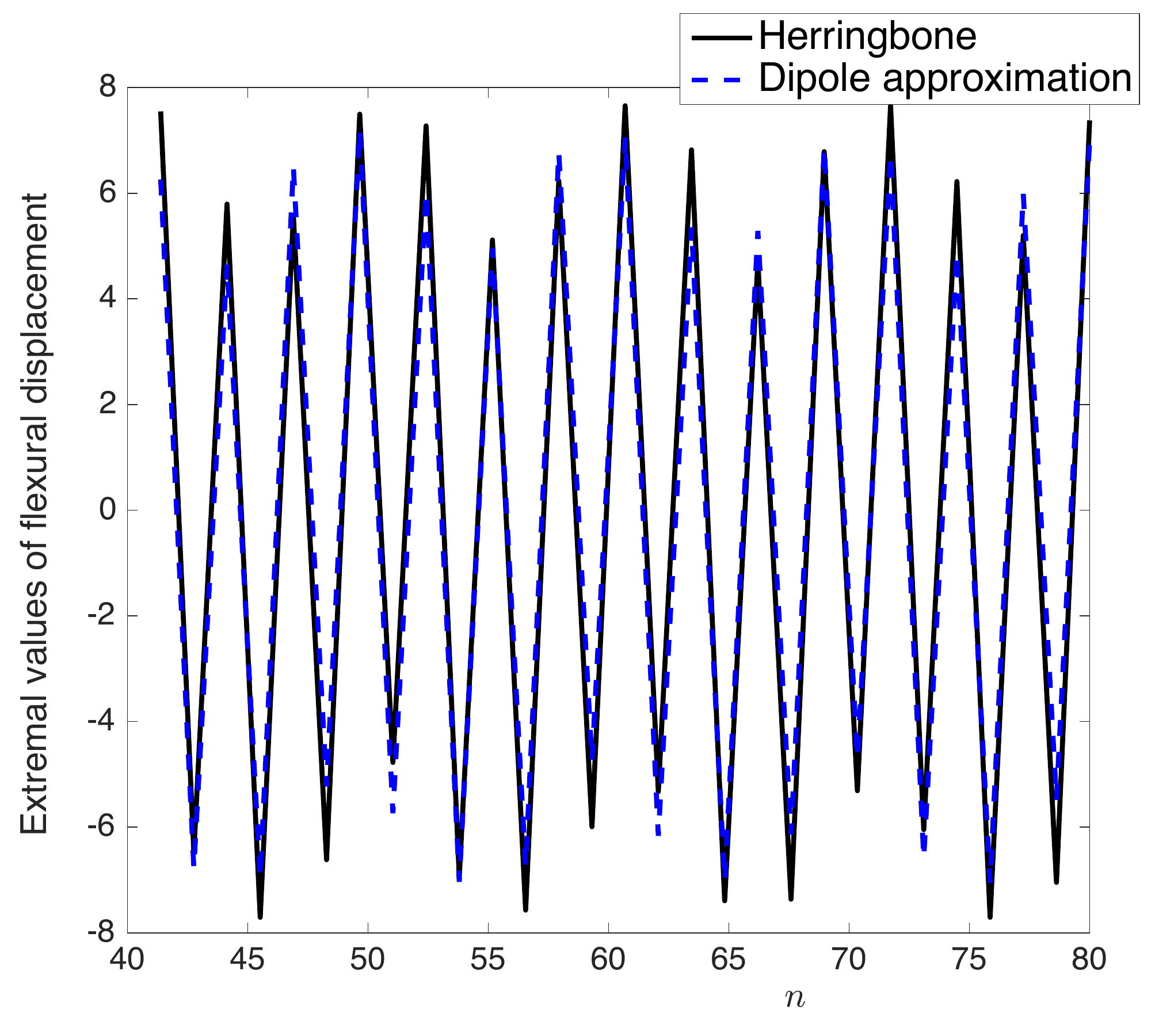}~~~~~~~
 \put(-70,50) {(c)}
\includegraphics[width=.4\columnwidth]{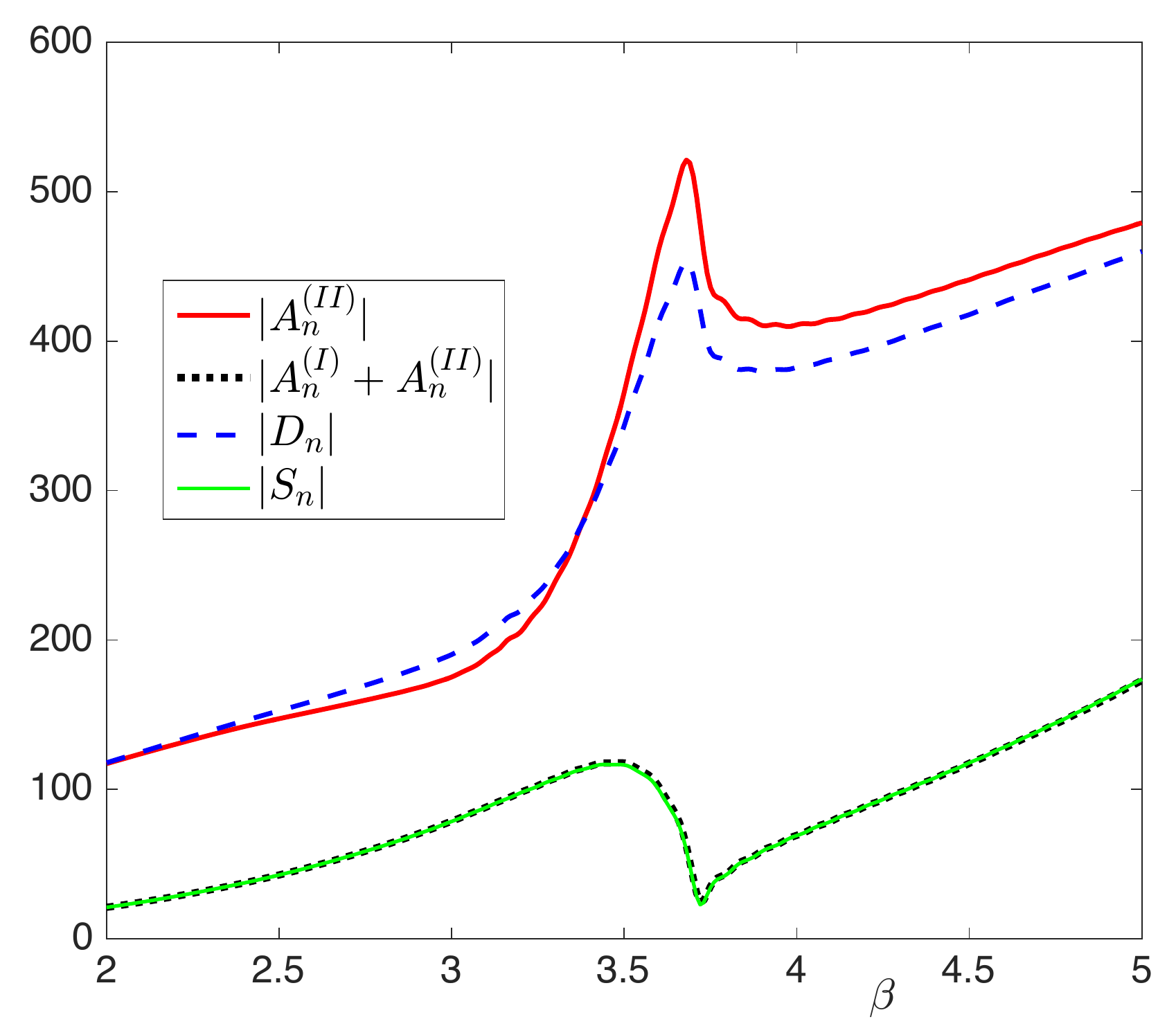}
 \put(-70,50) {(d)}
 
 \includegraphics[width=.4\columnwidth]{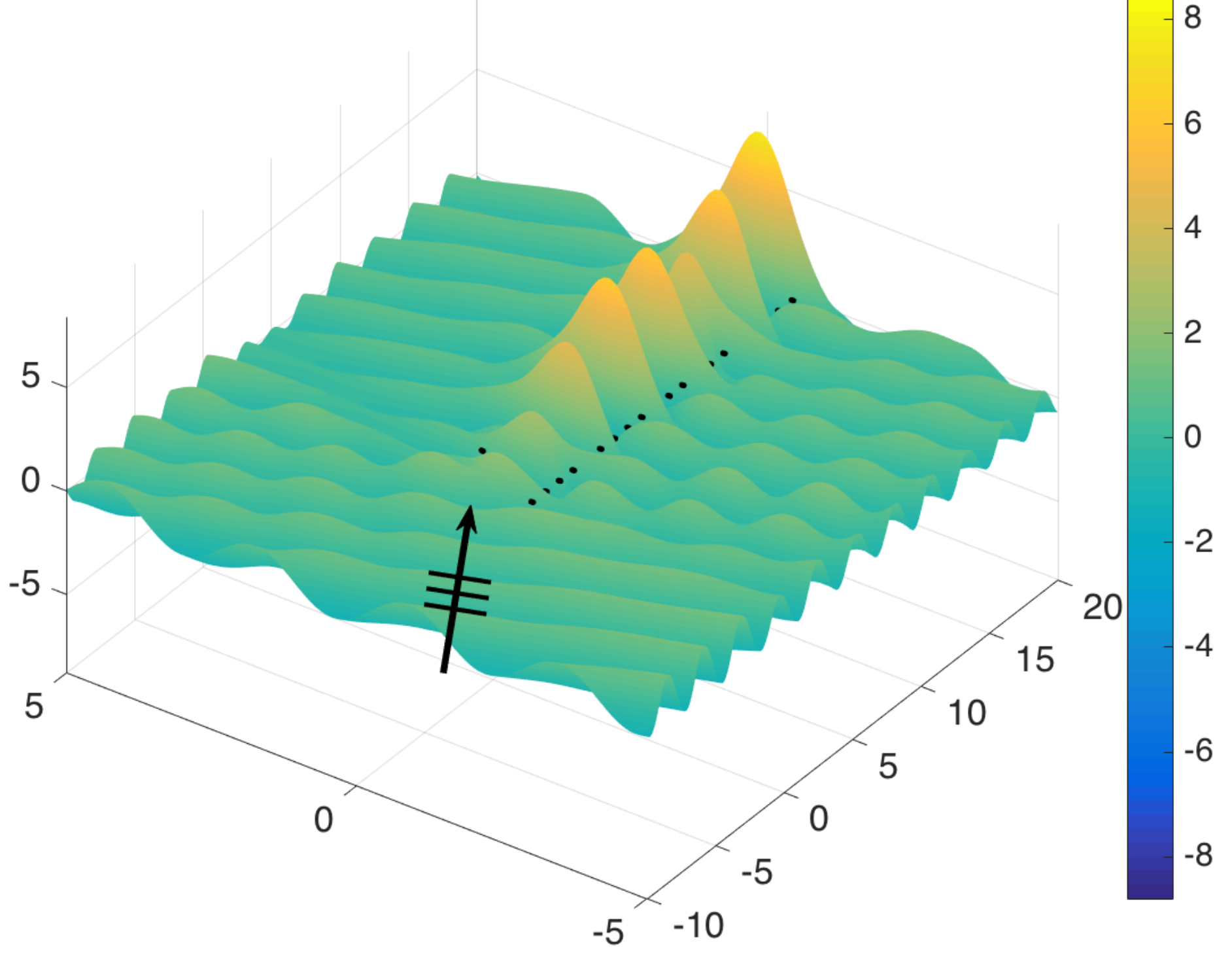}~~~
 \put(-67,50) {(e)}
\includegraphics[width=.4\columnwidth]{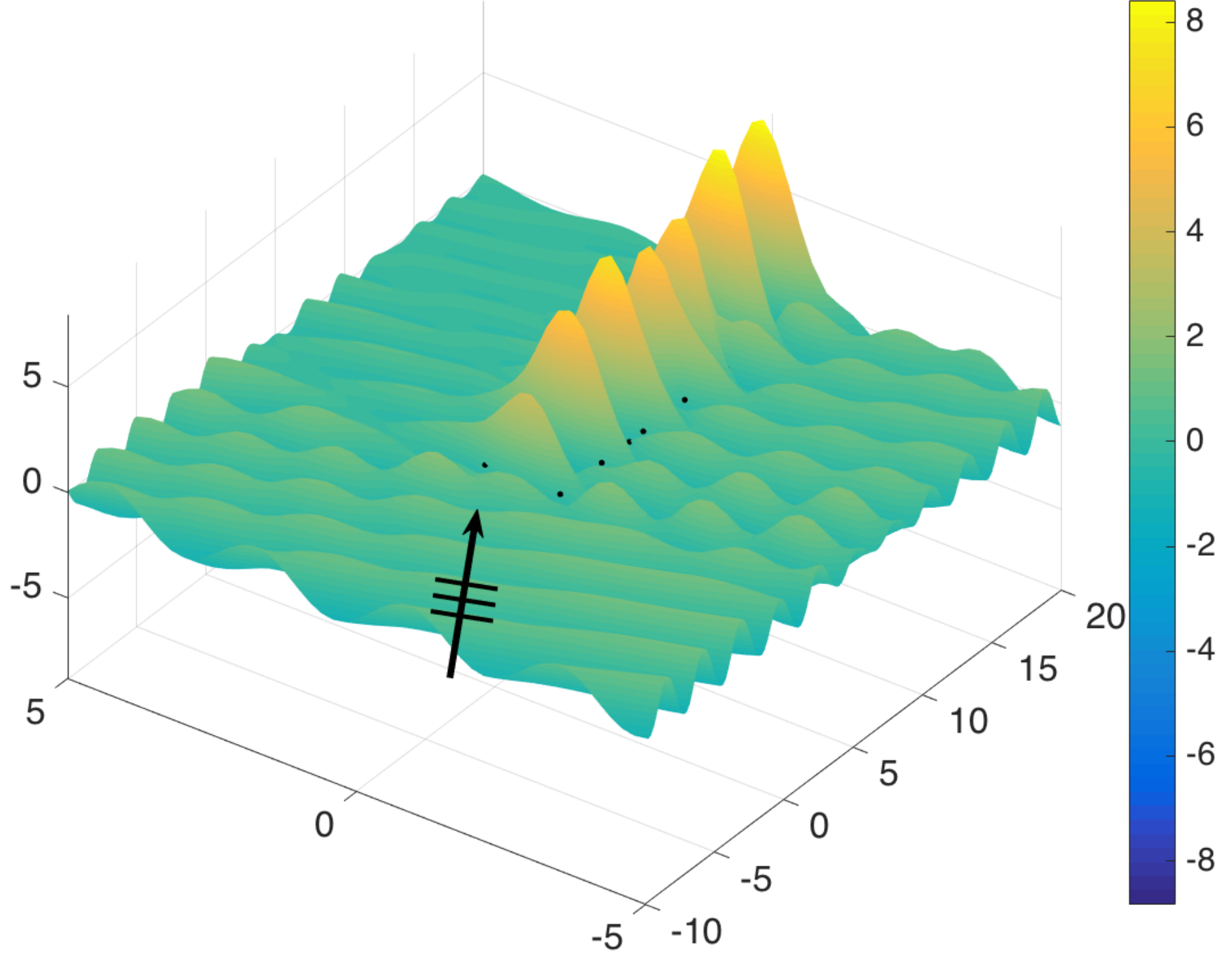}
 \put(-67,50) {(f)}
\caption{Herringbone system with ${\bf s} = (0.01,0.03), b = 1.3$, $a_1 = a_2 = 1$.
Total displacement field plots for an incident plane wave with $\psi = 0.805$ and $\beta = 3.318$ for $L = 160$ (pins 40 to 80 shown) for (a) herringbone system comprising four rows of pins. (b) two line arrays of sources and dipoles with coefficients $S_n$ and $D_n$. (c) Extremal values (above and below the gratings) of the flexural displacements for herringbone (solid black) and two arrays of sources/dipoles (dashed blue) for the fields in parts (a) and (b). (d) Comparison of coefficients for the upper pair of the herringbone system defined by ${\bf s}$. (e, f) Comparison of end effects for (e) herringbone and (f) two line arrays. 
\label{hb_examp_3}
}
\end{center}
\end{figure}

In Fig.~\ref{hb_examp_3}(a) we plot the total displacement field for the herringbone system comprising four rows of pins, with central spacing $b = 1.3$ and shift vectors ${\bf s} = (0.01,0.03$), ${\bf t} = (0.01,-0.03)$, using the truncation parameter $L = 160$ (pins 40 to 80 shown). The incident plane wave is defined by $\psi = 0.805$ for $k_x = 2.3$ and $\beta = 3.318$ from~(\ref{kx}). Here the scattering coefficients $A_n^{\scriptsize \mbox{(I)}}$ to $A_n^{\scriptsize \mbox{(IV)}}$ are determined using the wave scattering method. The bending and waveguiding localisation are quite striking, with amplitudes reaching more than eight times those of the incident plane wave. The field shows transmission above the grating system in Fig.~\ref{hb_examp_3}(a), with only very minimal scattering effects evident around the lower pair of gratings, thereby demonstrating the focussing and waveguiding capabilities of a tuned herringbone system.

The relatively small value of $|{\bf s}| = 0.032$ ensures that the source-dipole approach approximates the system well, as illustrated in Figs.~\ref{hb_examp_3}(b-f). The displacement field plotted using the source and dipole coefficients $S_n^+$, $D_n^+$ for ${\bf s}$, and those for ${\bf s}^-$ of $S_n^-$, $D_n^-$ calculated using the system~(\ref{LAS1})-(\ref{LAS3}) described in Sections~\ref{sec:dipole1HB} and \ref{hb_new}, is shown in Fig.~\ref{hb_examp_3}(b) for truncation parameter $L = 160$. A very similar localisation effect is observed within the system, with the envelope function and location of peaks consistent with Fig.~\ref{hb_examp_3}(a). 

Fig.~\ref{hb_examp_3}(c) provides further evidence for investigating the source-dipole approximation approach. The extremal values (above and below the gratings) of the flexural displacements are plotted for the fields depicted in parts (a) and (b), with those of the full herringbone system shown using the solid (black) curve, and those of the dipole approximation, using the dashed (blue) curve. The excellent correspondence is clearly evident, with both the magnitudes and the distribution of the coefficient terms matching well. 

The upper pair coefficients, shown in Fig.~\ref{hb_examp_3}(d), where the values for the pair $A_n^{\scriptsize \mbox{(I)}}, A_n^{\scriptsize \mbox{(II)}}$ (solid curve) are compared with the coefficients $D_n^+$ (dashed curve) for the dipole approximation's upper array, display excellent correspondence. In particular, there is a very good match in the vicinity of the operating frequency $\beta = 3.318$, where the dashed curve crosses the solid $|A_n^{\scriptsize \mbox{(II)}}|$ curve.

However, there are some small but discernible differences. As well as the increased reflection visible in front of the system in Fig.~\ref{hb_examp_3}(b), some of the details of the end effects in the vicinity of the system's entrance are lost with the dipole approximation, as shown in Figs.~\ref{hb_examp_3}(e,f). The full herringbone system Fig.~\ref{hb_examp_3}(e) has slightly lower amplitudes, and there appears to be a small phase difference when comparing Fig.~\ref{hb_examp_3}(f) with (e). The small discrepancies are likely to have arisen from the size of $|{\bf s}|$ and that this is a resonant example. 

\subsubsection{Angular dependence for dipole contributions}
In Section~\ref{sec242}, and Fig.~\ref{comp_dip_angles} in particular, we looked at how scattering patterns depend on the orientation of the dipole contributions. As Fig.~\ref{comp_dip_angles} illustrates for normal incidence, the increase of $\theta$ from being aligned with the incident plane wave to becoming perpendicular, reduces the scattering coefficients to zero. 
\begin{figure}[h]
\begin{center}
\includegraphics[width=.45\columnwidth]{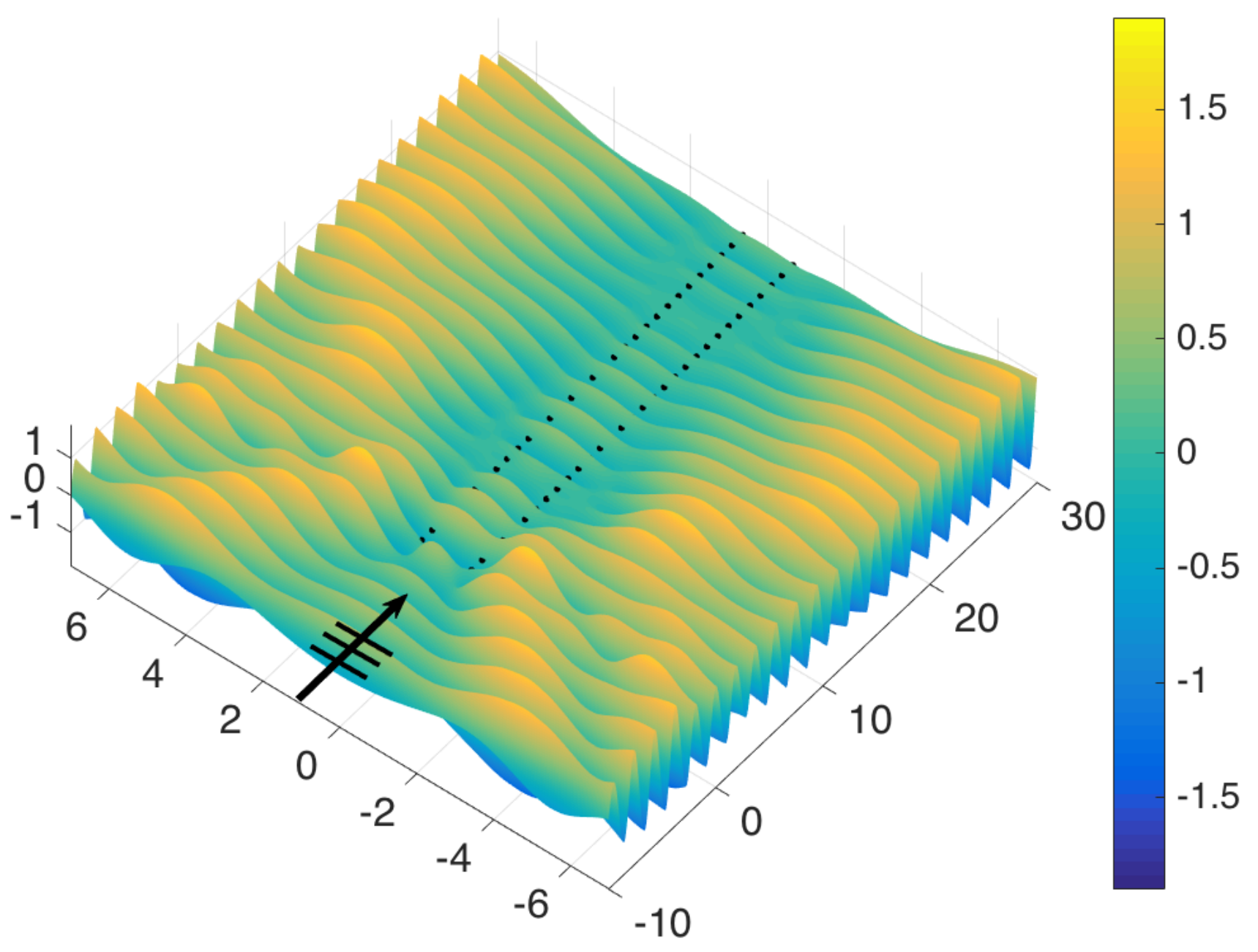}~~~~
 \put(-70,50) {(a)}
\includegraphics[width=.45\columnwidth]{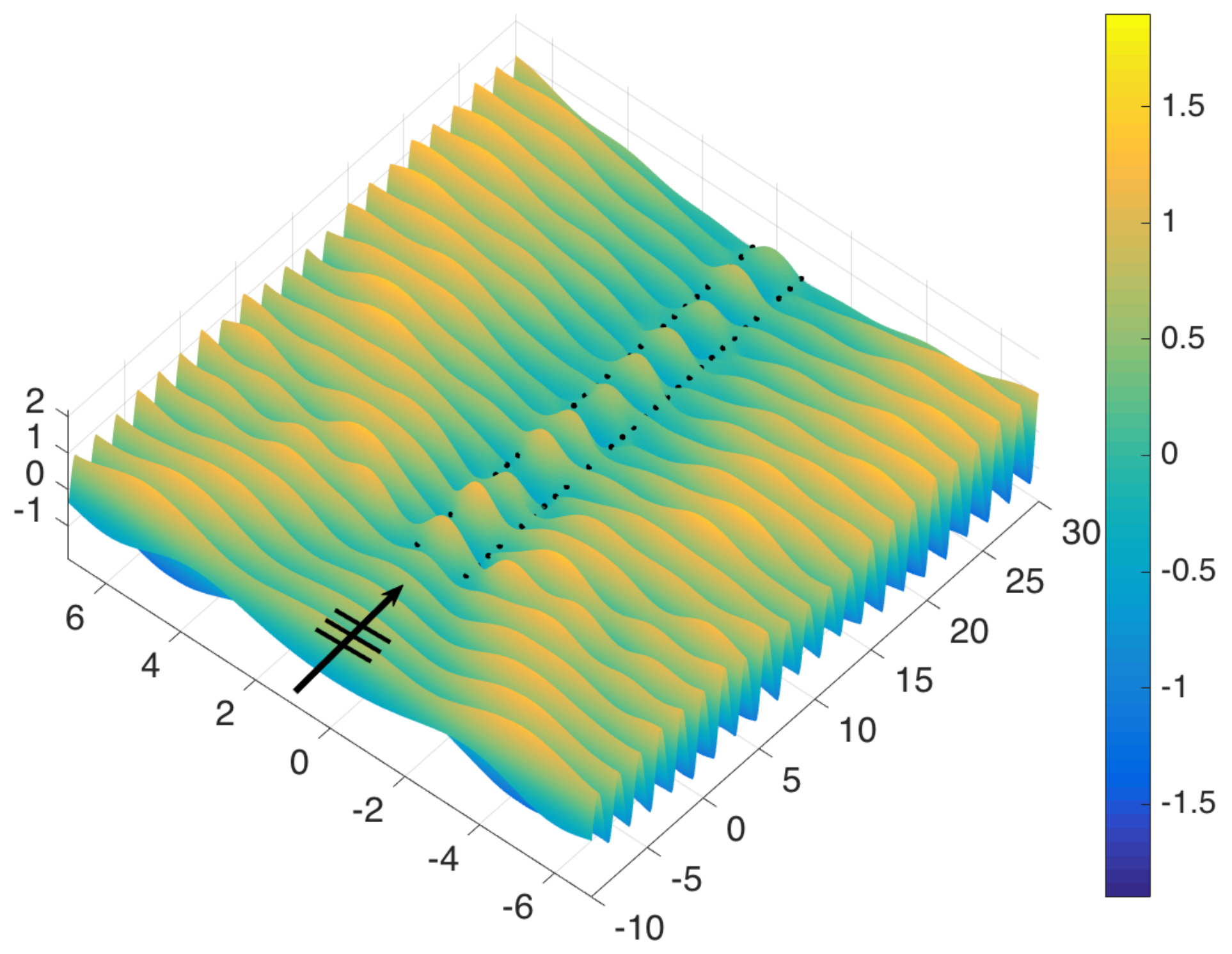}
 \put(-70,50) {(b)}
 
\includegraphics[width=.42\columnwidth]{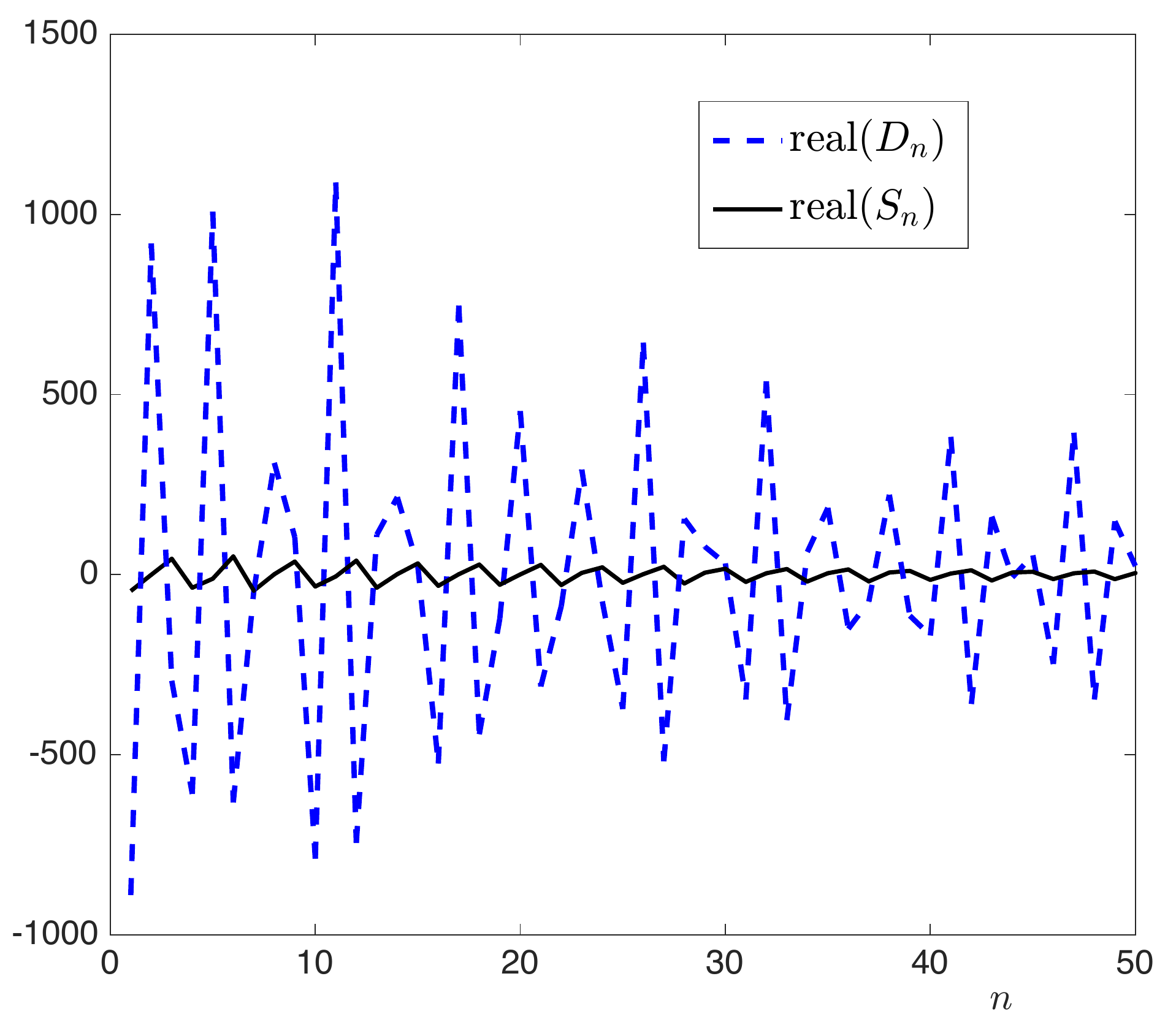}~~~
 \put(-70,58) {(c)}
\includegraphics[width=.42\columnwidth]{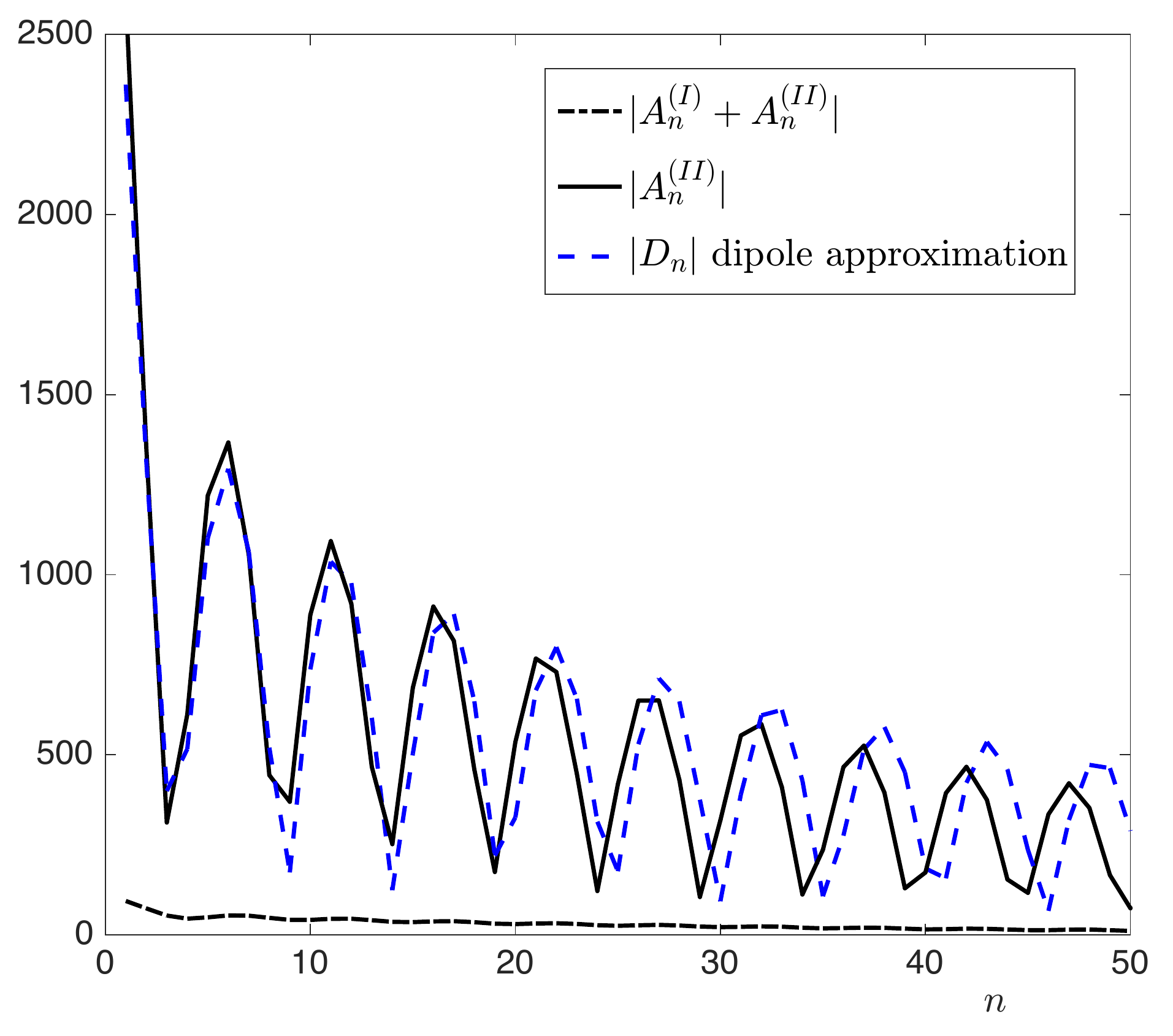}
 \put(-65,58) {(d)} 
\caption{Total displacement fields for $\psi = 0$, $\beta = 3.33$, $b = 1.3$ for (a) pair of unshifted gratings, (b) herringbone system with ${\bf s} = {\bf t} = (0.005,0)$, $\theta = 0$, $L = 100$. Real parts of the coefficients are illustrated in (c), with dipole ($D_n$, dashed blue) and source ($S_n$, solid black). (d) Moduli of scattering coefficients comparing the the dipole terms for the full herringbone and its approximation.
\label{hb_examp_ANGLES_1}
}
\end{center}
\end{figure}
In the herringbone system, formed of two arrays of sources and dipoles, the role of the dipole angle $\theta$ is greatly enhanced. For the same settings ($\beta = 3.33$ and $\psi = 0$), we construct a symmetric herringbone system with spacing $b = 1.3$ and initial shift vectors ${\bf s} = {\bf s}^- = (0.005,0)$. We impose the length $|{\bf s}| = 0.005$ to remain constant and vary the angle $\theta$ in the range $0 \le \theta \le \pi$. In this way, the dipole angle for the lower half of the herringbone, denoted by $\phi$ in Fig.~\ref{herr}, varies in the range $0 \ge \phi \ge -\pi$.

For the spacing $b = 1.3$, frequencies in the neighbourhood of $\beta = 3.33$ can be tuned to support waveguiding effects, as demonstrated by equation~(\ref{wg_model}) and in Fig.~\ref{hb_examp_3}. An example of blockage for a pair of gratings is obtained for $\beta = 3.33$ and $\psi = 0$ in Fig.~\ref{hb_examp_ANGLES_1}(a), where we observe a mode with significantly reduced resolution. The choice of normal incidence is a perfect regime to investigate the design possibilities of varying the dipole angle $\theta$, since each array of sources/dipoles is subject to ``head on" incidence.

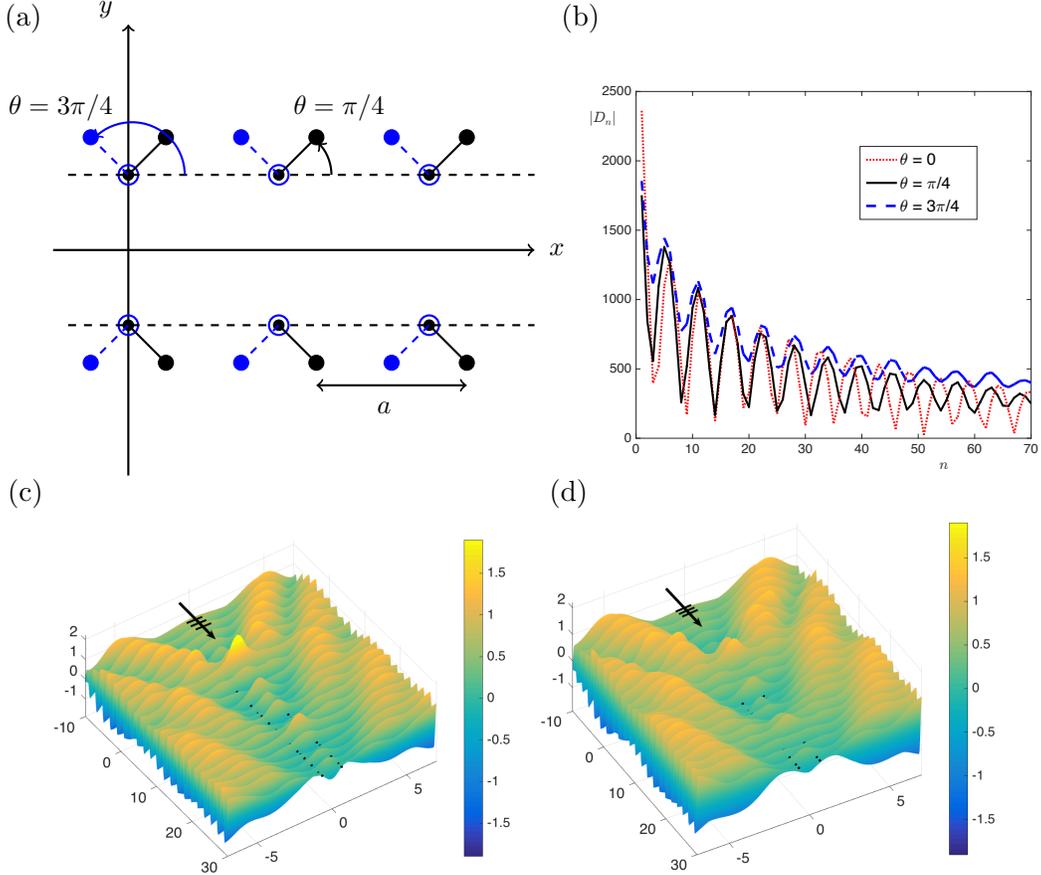
\begin{figure}[h]
\begin{center}
\begin{tikzpicture}
\draw  [->, thick] (-3.4,0.0) -- (3.0,0.0);
\draw  [->, thick] (-2.4,-3.0) -- (-2.4,3.0);
\node at (-2.7,3.2) {{\small $y$}};
\node at (3.3,0.0) {{\small$x$}};
\draw [fill] (-2.4,1.0) circle [radius=0.07] [black!];
\draw [blue!, thick] (-2.4,1.0) circle [radius=0.13];
\draw [fill] (-0.4,1.0) circle [radius=0.07];
\draw [blue!, thick] (-0.4,1.0) circle [radius=0.13];
\draw [fill] (1.6,1.0) circle [radius=0.07];
\draw [blue!, thick] (1.6,1.0) circle [radius=0.13];
\draw  [-, dashed,thick] (-3.2,1.0) -- (3.0,1.0);
\draw [blue!, thick] (-2.4,-1.0) circle [radius=0.13];
\draw [fill] (-2.4,-1.0) circle [radius=0.07];
\draw [blue!, thick] (-0.4,-1.0) circle [radius=0.13];
\draw [fill] (-0.4,-1.0) circle [radius=0.07];
\draw [fill] (1.6,-1.0) circle [radius=0.07];
\draw [blue!, thick] (1.6,-1.0) circle [radius=0.13];
\draw  [-, dashed,thick] (-3.2,-1.0) -- (3.0,-1.0);
\draw [fill] (-1.9,1.5) circle [radius=0.1];
\draw [fill] (0.1,1.5) circle [radius=0.1];
\draw [fill] (2.1,1.5) circle [radius=0.1];
\draw [fill] (-2.9,1.5) circle [radius=0.1] [blue!];
\draw [fill] (-0.9,1.5) circle [radius=0.1] [blue!];
\draw [fill] (1.1,1.5) circle [radius=0.1] [blue!];
\draw [fill] (-1.9,-1.5) circle [radius=0.1];
\draw [fill] (0.1,-1.5) circle [radius=0.1];
\draw [fill] (2.1,-1.5) circle [radius=0.1];
\draw [fill] (-2.9,-1.5) circle [radius=0.1] [blue!];
\draw [fill] (-0.9,-1.5) circle [radius=0.1] [blue!];
\draw [fill] (1.1,-1.5) circle [radius=0.1] [blue!];
\draw  [-, dashed, thick] (-2.4,-1.0) -- (-2.9,-1.5) [blue!];
\draw  [-, dashed, thick] (-0.4,-1.0) -- (-0.9,-1.5) [blue!];
\draw  [-, dashed, thick] (1.6,-1.0) -- (1.1,-1.5) [blue!];
\draw  [-, dashed, thick] (-2.4,1.0) -- (-2.9,1.5) [blue!];
\draw  [-, dashed, thick] (-0.4,1.0) -- (-0.9,1.5) [blue!];
\draw  [-, dashed, thick] (1.6,1.0) -- (1.1,1.5) [blue!];
\draw  [-, thick] (-2.4,1.0) -- (-1.9,1.5);
\draw  [-,  thick] (-0.4,1.0) -- (0.1,1.5);
\draw  [-, thick] (1.6,1.0) -- (2.1,1.5);
\draw  [-, thick] (-2.4,-1.0) -- (-1.9,-1.5);
\draw  [-, thick] (-0.4,-1.0) -- (0.1,-1.5);
\draw  [-, thick] (1.6,-1.0) -- (2.1,-1.5);
\draw  [<->, thick] (0.1,-1.8) -- (2.1,-1.8);
\node at (1.0,-2.1) {{\small $a$}};
\draw  [->, thick, blue!] (-1.65,1.0) arc (0:135:0.7);
\draw  [->, thick, black!] (0.3,1.0) arc (0:45:0.6);
\node at (-3.3,1.9) {{\small $\theta = 3\pi/4$}};
\node at (0.4,1.9) {{\small $\theta = \pi/4$}};
\end{tikzpicture}~~
 \put(-77,60) {(a)}
\includegraphics[width=.4\columnwidth]{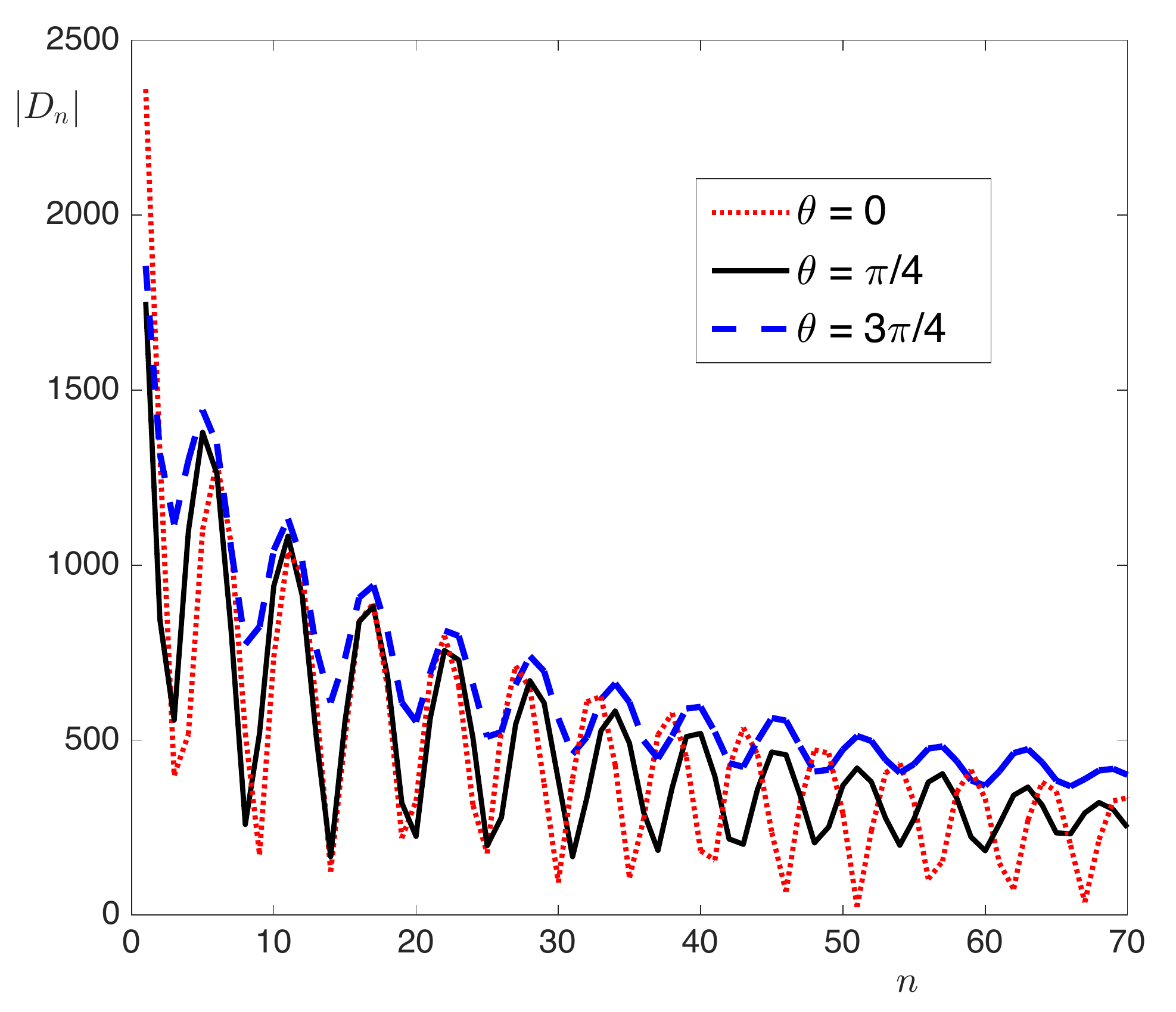}
 \put(-65,60) {(b)}
 
\includegraphics[width=.4\columnwidth]{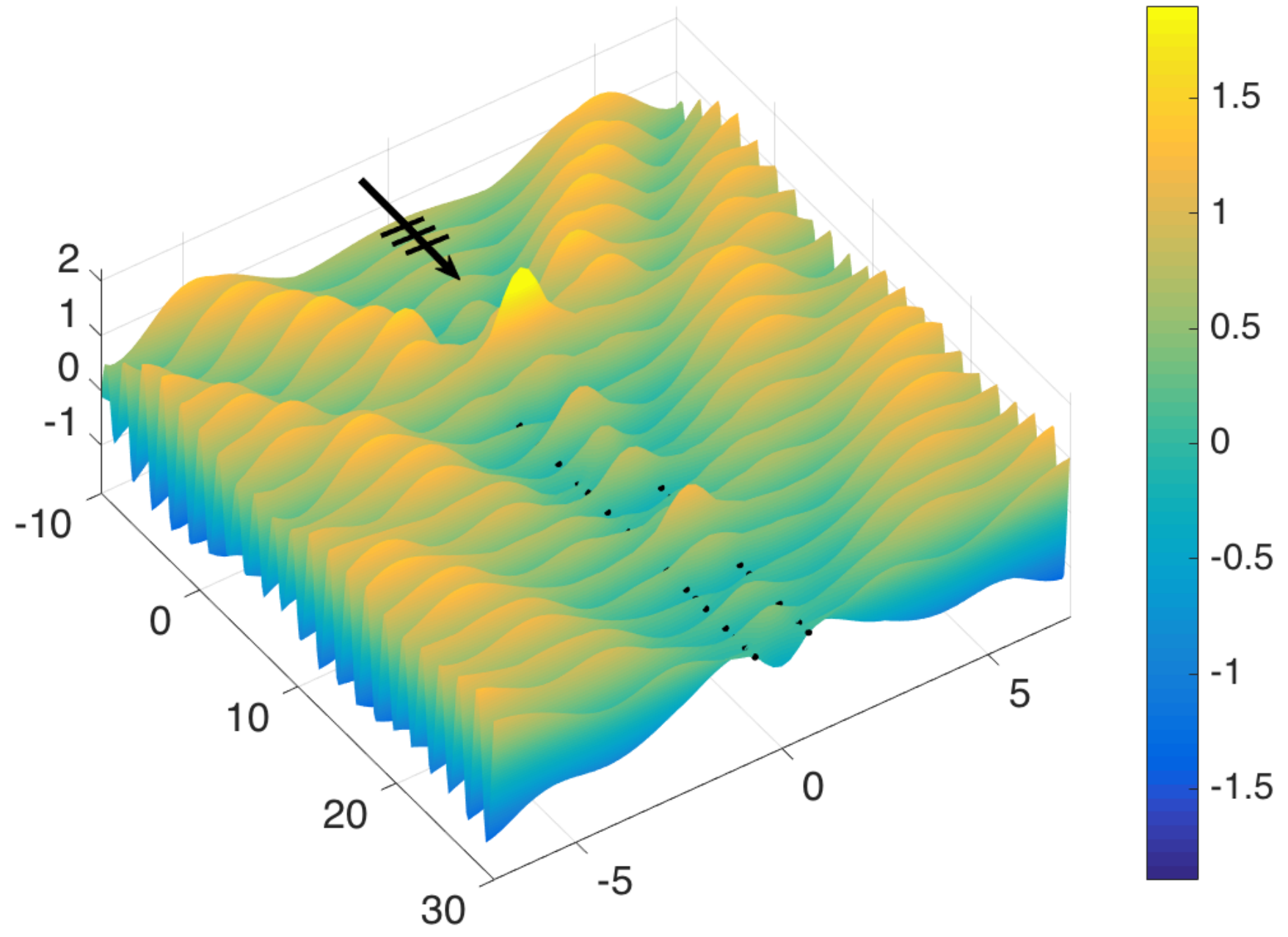}~~~
 \put(-70,50) {(c)}
 \includegraphics[width=.4\columnwidth]{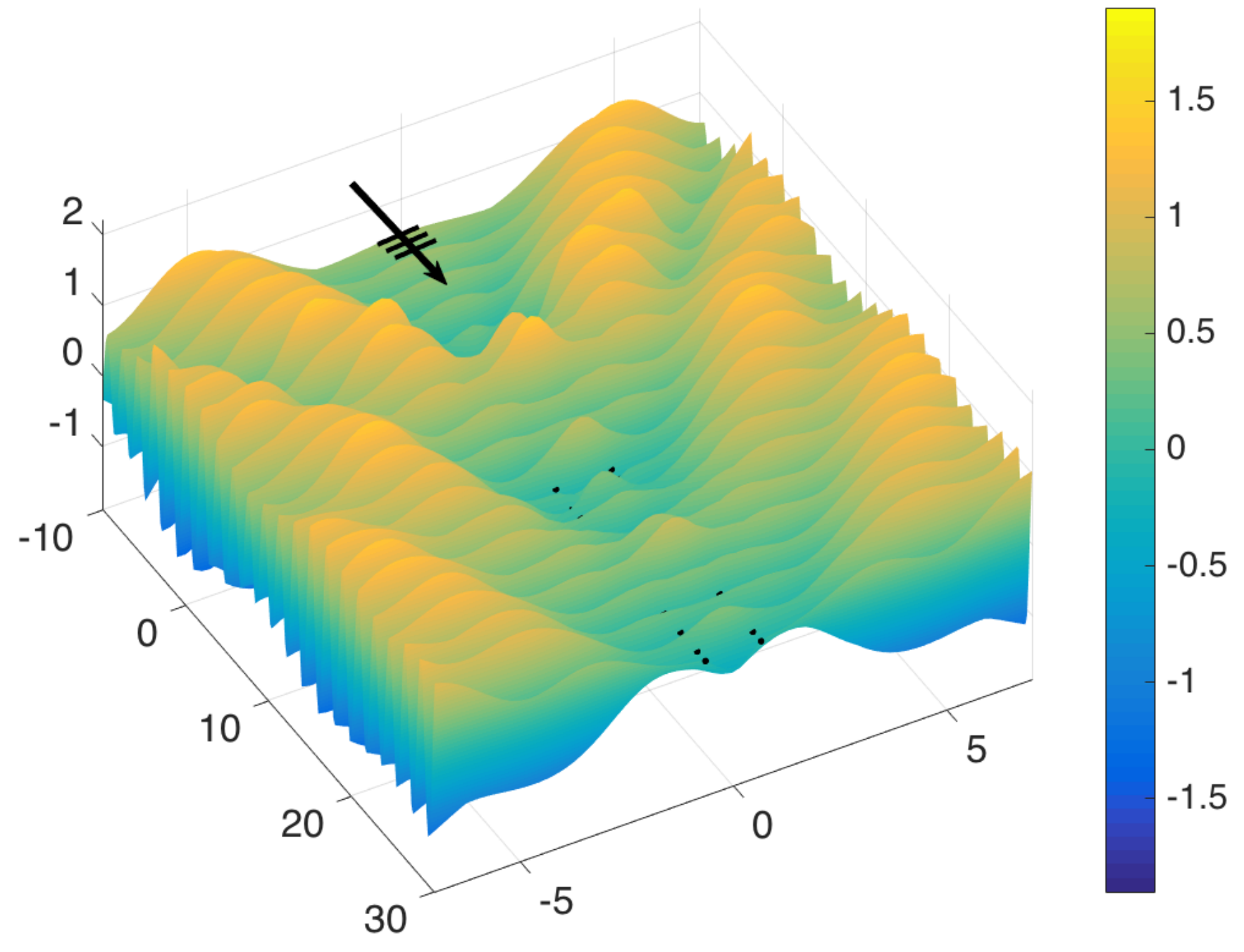}
 \put(-60,50) {(d)}
\caption{Symmetric herringbone system with $|{\bf s}| = 0.005, b = 1.3$, ${\bf t} = {\bf s}^-$ for normal incidence $\psi = 0$ and $\beta = 3.33$, $L = 100$. (a) Two configurations with $\theta = \pi/4$ and $\theta = 3\pi/4$. (b) Comparison of dipole coefficients $|D_n|$ for $\theta = 0, \pi/4, 3\pi/4$. (c,d) Total displacement fields for the herringbone systems with, respectively, $\theta = \pi/4$ and $\theta = 3\pi/4$.
\label{hb_examp_ANGLES}
}
\end{center}
\end{figure}

We consider the most basic addition of dipoles first, setting $s_2$ to be zero and $|{\bf s}| = 0.005$, aligning $\theta$ with angle of incidence $\psi = 0$. The effect is quite remarkable, with the herringbone system supporting waveguiding and localisation in the channel, as shown in Fig.~\ref{hb_examp_ANGLES_1}(b) for $L = 100$, with the first 30 pins shown. The waveguiding localisation is not as impressive as for the tuned design of $\psi = 0.805$ and ${\bf s} = (0.01,0.03)$ in Fig.~\ref{hb_examp_3}(a), but the ability to convert a blockage to a leaky waveguide by replacing sources with dipoles is an effect deserving of attention.

Figs.~\ref{hb_examp_ANGLES_1}(c,d) provide further verification of the dipole approximation method. In Fig.~\ref{hb_examp_ANGLES_1}(c), the real parts of the source and dipole coefficients are plotted for the first 50 points for the waveguide mode depicted in Fig.~\ref{hb_examp_ANGLES_1}(b).  The dominance of the dipole terms is clear, and the envelope function of the coefficients is consistent with the displacement field in Fig.~\ref{hb_examp_ANGLES_1}(b). The accuracy of the dipole approximations is also illustrated in Fig.~\ref{hb_examp_ANGLES_1}(d), where the moduli $|A_n^{\scriptsize \mbox{(II)}}|$ and $|D_n|$ are compared. Note once again the relative magnitudes for the source coefficients (dot-dashed curve).

We now consider varying the dipole orientation $\theta$. 
Fig.~\ref{hb_examp_ANGLES}(a) shows, schematically, two symmetric waveguides defined by $\theta = \pi/4,  3\pi/4$. We adopt a consistent classification for the entire figure with the solid (black) lines representing ${\bf s} = (0.0035,0.0035)$ (a convex entrance) and the dashed (blue) case being ${\bf s} = (-0.0035,0.0035)$ (the concave entrance). The convexity of the herringbone entrance concomitantly affects the localisation, one of the unique features of the model.

The moduli of the dipole coefficients $|D_n|$ are plotted in Fig.~\ref{hb_examp_ANGLES}(b), with the first seventy points shown ($L = 100$). We include those for $\theta = 0$, discussed in Fig.~\ref{hb_examp_ANGLES_1} and plotted here with a dot-dashed curve, and show that varying $\theta$ has clear effects on the scattering properties. The greater modulation of the amplitudes for $\theta = 0$ manifests in a clearly defined waveguide mode in Fig.~\ref{hb_examp_ANGLES_1}(b). The $\pi/2$ shift in $\theta$ from $\pi/4$ to $3\pi/4$ has an impact on the coefficients in Fig.~\ref{hb_examp_ANGLES}(b), and on the localisation in the vicinity of the herringbone entrance. The convex case favours strong localisation, as illustrated in Fig.~\ref{hb_examp_ANGLES}(c), where the displacement field for $\theta = \pi/4$ is plotted, compared with that for $\theta = 3\pi/4$ in Fig.~\ref{hb_examp_ANGLES}(d). 

Although the flexural wave fields are similar, there are visual differences in the localisation patterns observed within the gratings, with the peaks at the front of the system for $\theta = \pi/4$ significantly higher than for $\theta = 3\pi/4$. This effect is related to the convexity/concavity of the entrance to the herringbone system. For acute values of $\theta$, the symmetric herringbone supports greater localisation in the neighbourhood of the opening compared with the case of $\pi/2 < \theta < \pi$. It is also clear that the choice of a value of $\theta$ greater than $0$ leads to a reduction in the extent of waveguiding through the channel.

\section{Concluding remarks}
\label{conc}
In this article we have presented a new type of flexural waveguide designed in the form of a herringbone system. We have demonstrated that the herringbone system can significantly enhance the localisation effects, in terms of  amplitudes and focussing, compared with a simpler two-grating waveguide. We have also shown how the herringbone structure is capable of converting a grating pair's reflection mode into a waveguide mode for the same incident plane wave. 

This paper has introduced a novel type of approximation for waveguide modes in structured semi-infinite grating stacks.
It is based on the dipole approximation, which takes into account the relative positions and interactions within a structured waveguide such as the herringbone system.
This elegant asymptotic approximation is complemented by the derivation of the functional equation of Wiener-Hopf type and analysis of its kernel function. 

Several mathematical techniques were implemented, including a classical wave scattering method to derive a system of linear algebraic equations for the flexural displacement. The solution of this system was obtained in the form of scattering coefficients used to plot displacement fields that demonstrate waveguiding effects. A discrete Wiener-Hopf formulation yielded expressions for kernel matrices consisting of quasi-periodic Green's functions. The zeros of the determinant of such a kernel matrix correspond to Bloch modes for infinite grating systems, the knowledge of which was used to aid the solution of the corresponding semi-infinite scattering problems.

For the case when the shifted pairs consist of two closely spaced gratings, we implemented an alternative mathematical approach. The proximity of pairs of pins advocates the use of a dipole approximation. 
The validity of the approach for small $|{\bf s}|$ was illustrated with several examples. 
We showed that both the magnitude and the orientation of the dipoles are important for tuning the localisation effects. 


\appendix\newpage\markboth{Appendix}{Appendix}
\renewcommand{\thesection}{\Alph{section}}
\numberwithin{equation}{section}
\numberwithin{figure}{section}
\section{Dipole approximations for kernel matrix}
\label{sec:offdiag}
The structure of the kernel matrix in equation~(\ref{whopf0}) suggests that the off-diagonal entries $\hat G(\beta,k;{\boldsymbol \xi}^{(1)};{\boldsymbol \xi}^{(2)})$ can be approximated by expanding about the origin to obtain a dipole-like representation. 

For gratings I and II we define, respectively, 
${\boldsymbol \xi}^{(1)} = {\bf 0}$,  ${\boldsymbol \xi}^{(2)} = {\bf s}$. 
Then one can say, for ${\boldsymbol \xi}^{(2)} = {\boldsymbol \xi}^{(1)} + {\bf s}, |{\bf s}| \ll 1$, 
\begin{eqnarray}
\label{dip1}
\hat G(\beta,k; {\boldsymbol \xi}^{'(1)} , {\boldsymbol \xi}^{(2)}) 
& = & \hat G(\beta,k; {\boldsymbol \xi}^{'(1)};{\boldsymbol \xi}^{(1)} + {\bf s})  \simeq  \hat G(\beta,k;{\boldsymbol \xi}^{'(1)}; {\boldsymbol \xi}^{(1)}) + {\bf s}  \cdot \nabla \hat G(\beta,k;{\boldsymbol \xi}^{'(1)} ; {\boldsymbol \xi}^{(1)}), \nonumber  \\
&& \\
\hat G(\beta,k; {\boldsymbol \xi}^{'(2)}; {\boldsymbol \xi}^{(1)}) & = & \hat G(\beta,k;{\boldsymbol \xi}^{'(1)} + {\bf s}; {\boldsymbol \xi}^{(1)})  \simeq  \hat G(\beta,k;{\boldsymbol \xi}^{'(1)}; {\boldsymbol \xi}^{(1)})   + {\bf s} \cdot \nabla^{'} \hat G(\beta,k;{\boldsymbol \xi}^{'(1)}; {\boldsymbol \xi}^{(1)}),  \nonumber
\end{eqnarray}
where we have expanded around the lower grating's front pin position vector ${\boldsymbol \xi}^{(1)} = {\bf 0}$. The notation ${'}$ is used to distinguish between the arguments in both the expansions and the directional derivatives; $\nabla$ denotes that we are differentiating with respect to ${\boldsymbol \xi}^{(1)}$, whereas $\nabla^{'}$ signifies differentiation with respect to ${\boldsymbol \xi}^{'(1)}$.

Let us consider the directional derivative terms in~(\ref{dip1}), which require differentiation of the Green's function~(\ref{gf}). 
Define the function for the argument:
\begin{equation}
 \rho_I(j,{\boldsymbol \xi}^{(1)}) =  \left|ja{\bf e_1}  + {\boldsymbol \xi}^{'(1)} - {\boldsymbol \xi}^{(1)}\right|   =\sqrt{\left(ja + \xi^{'(1)}_1- \xi^{(1)}_1\right)^2 + \left(0 + \xi^{'(1)}_2 - \xi^{(1)}_2\right)^2}. 
 \label{rhodpm2}
\end{equation} 

Referring to~(\ref{dip1}), (\ref{rhodpm2}), for ${\boldsymbol \xi}^{(1)} = {\boldsymbol \xi}^{'(1)} = {\bf 0}$, $\rho_I = |j|a$, we obtain
\begin{equation}
\frac{\partial \hat G}{\partial \xi^{(1)}_1} 
\bigg|_{ {\boldsymbol \xi}^{'(1)} = {\boldsymbol \xi}^{(1)} = {\bf 0}} = \frac{-ja}{\rho_I} \frac{\partial \hat G}{\partial \rho_I} \bigg|_{ {\boldsymbol \xi}^{'(1)} = {\boldsymbol \xi}^{(1)} = {\bf 0}}
= - \frac{\partial \hat G}{\partial \rho_I} \bigg|_{ {\boldsymbol \xi}^{'(1)} = {\boldsymbol \xi}^{(1)} = {\bf 0}}; \quad\quad \frac{\partial \hat G}{\partial \xi_2^{(1)}}  \bigg|_{ {\boldsymbol \xi}^{'(1)} = {\boldsymbol \xi}^{(1)} = {\bf 0}} = 0. 
\label{derivs_part1}
\end{equation}
Using equations~(\ref{dip1}) and~(\ref{derivs_part1}), we deduce
\begin{eqnarray}
\label{dip_LG1}
\hat G_{12} & = & \hat G(\beta,k; {\boldsymbol \xi}^{'(1)}; {\boldsymbol \xi}^{(2)}) 
 \simeq  
\hat G(\beta,k;{\boldsymbol \xi}^{'(1)}; {\boldsymbol \xi}^{(1)}) + {\bf s}  \cdot \nabla \hat G(\beta,k;{\boldsymbol \xi}^{'(1)};{\boldsymbol \xi}^{(1)})   \\ 
& \simeq &  \frac{i}{8\beta^2}  \sum_{j=-\infty}^\infty \Bigg[ H_0^{(1)}(\beta |j|a) +\frac{2i}{\pi}K_0 (\beta |j|a) + \beta s_1 \left(H_1^{(1)}(\beta |j|a) +\frac{2 i}{\pi} K_1(\beta |j|a) \right) \Bigg] e^{i k ja}. \nonumber 
\end{eqnarray}
Similarly,
\begin{eqnarray}
\label{dip_LG2}
\hat G_{21} & = & \hat G(\beta,k; {\boldsymbol \xi}^{'(2)}; {\boldsymbol \xi}^{(1)}) 
 \simeq 
\hat G(\beta,k;{\boldsymbol \xi}^{'(1)}; {\boldsymbol \xi}^{(1)}) + {\bf s}  \cdot \nabla^{'} \hat G(\beta,k;{\boldsymbol \xi}^{'(1)}; {\boldsymbol \xi}^{(1)})   \\ 
& \simeq &  \frac{i}{8\beta^2}  \sum_{j=-\infty}^\infty \Bigg[ H_0^{(1)}(\beta |j|a) +\frac{2i}{\pi}K_0 (\beta |j|a) - \beta s_1 \left(H_1^{(1)}(\beta |j|a) +\frac{2 i}{\pi} K_1(\beta |j|a) \right) \Bigg] e^{i k ja}. \nonumber 
\end{eqnarray}
We note that the expressions for $\hat G_{12}$ and $\hat G_{21}$ differ only by a change in sign for the sum of the first order Bessel functions, and that the terms are independent of $s_2$. This is unsurprising since we are expanding around ${\boldsymbol \xi}^{(1)}$ on the $x$-axis, but the approximation is only valid for sufficiently small $|{\bf s}|$. One may also consider the dipole approximation associated with expanding around the vector ${\boldsymbol \xi}^d$ halfway between the two gratings, which yields the same expressions as for the lower grating expansions~(\ref{dip_LG1}), (\ref{dip_LG2}).

\begin{figure}[h]
\begin{center}
\includegraphics[width=.43\columnwidth]{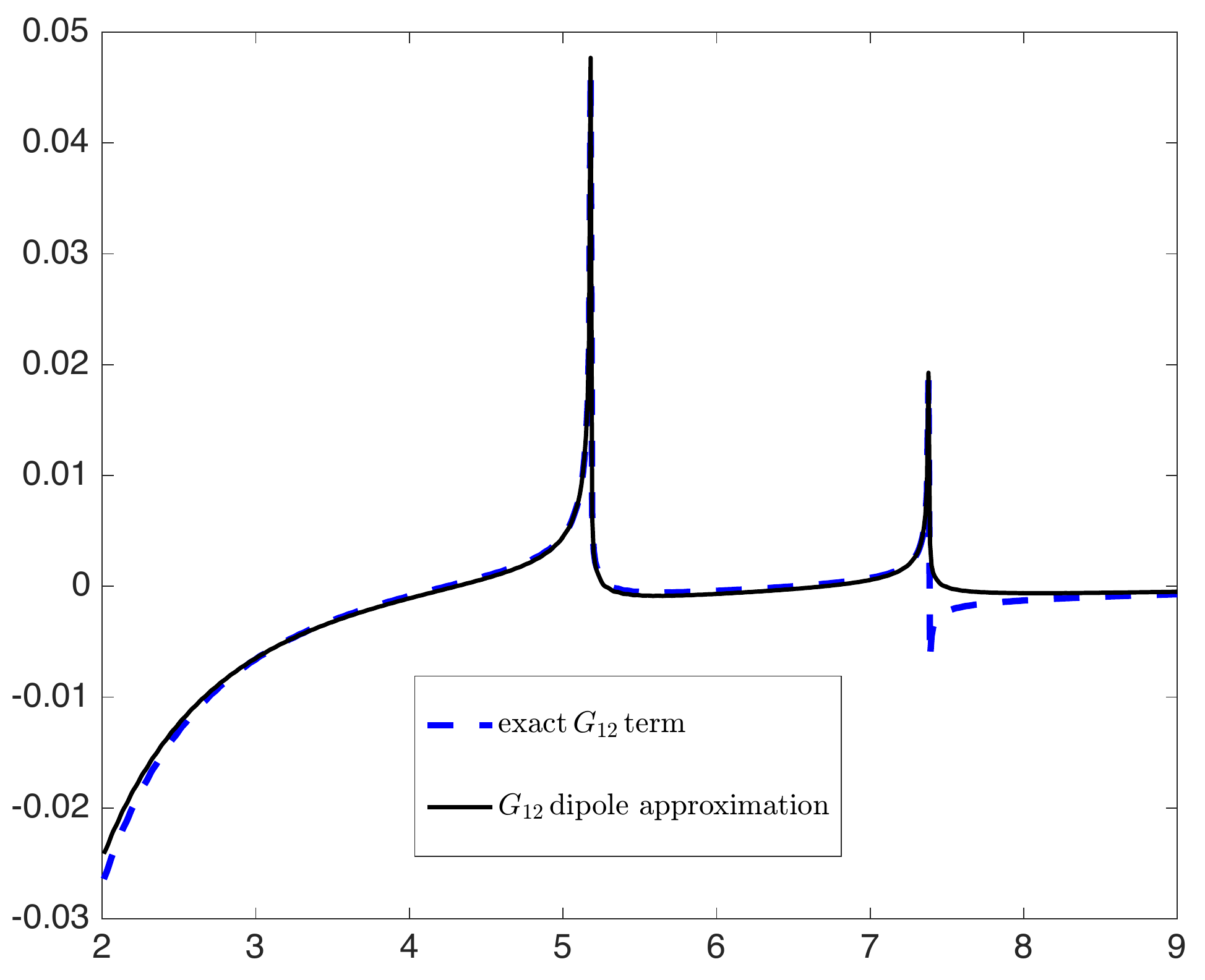}~
\includegraphics[width=.45\columnwidth]{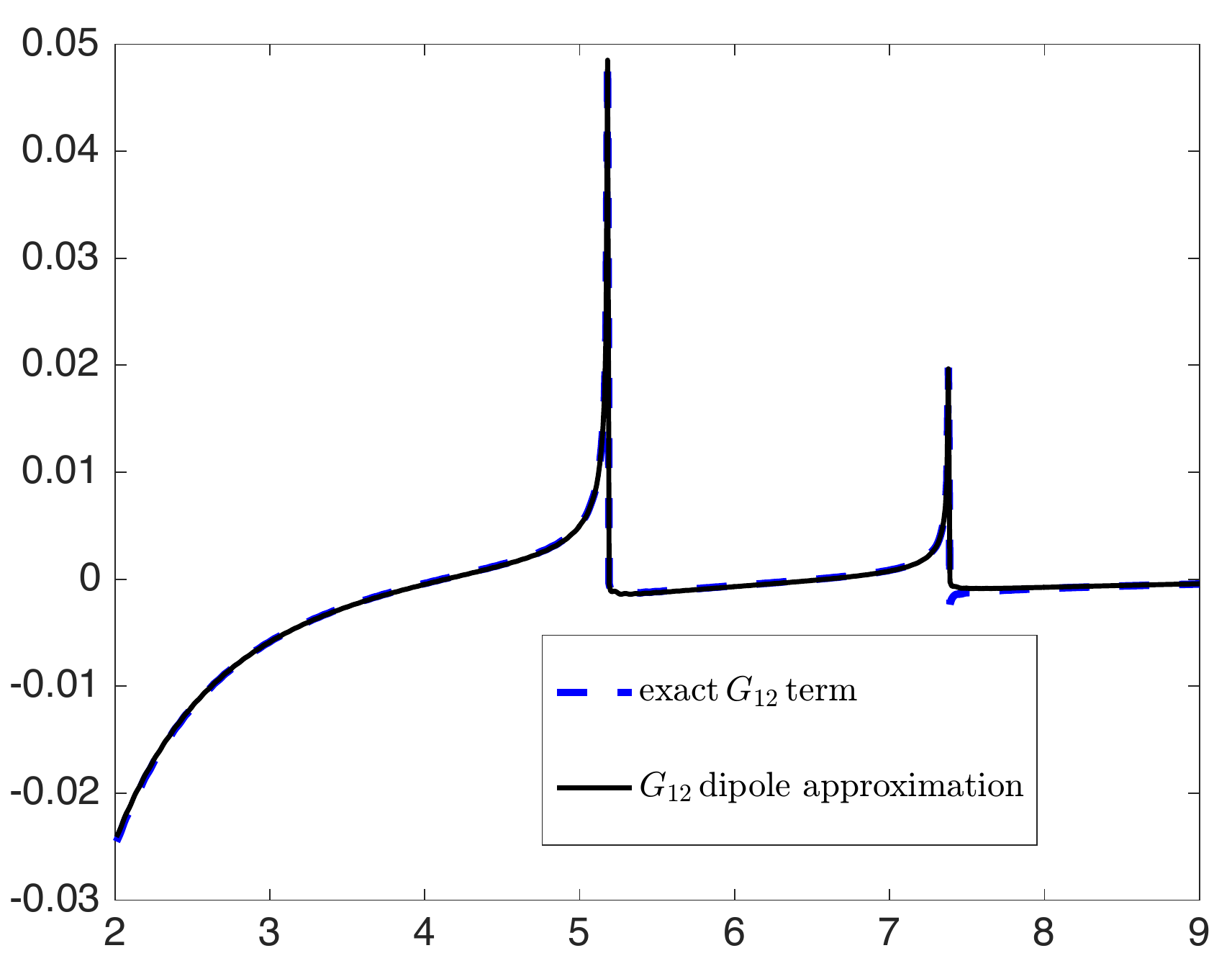}
\put(-140,57) {(a)}
\put(-70,57) {(b)}
\put(-100,-2) {\small $\beta$}
\put(-28,-2) {\small $\beta$}
\end{center}
\caption{ 
Comparison of approximating function (solid black curve) for real parts of the $G_{12}$ entries of the kernel matrix and their exact values (dashed blue curves) for $k_x = 1.1$ and $\beta$ in the range $2 \le \beta \le 9$ for two shift vectors ${\bf s}$: (a) ${\bf s} = (0.05, 0.025)$, (b) ${\bf s} = (0.01, 0.005)$.}
\label{g12comp}
\end{figure}

In Fig.~\ref{g12comp}, we illustrate the efficacy of this approach for a typical example with $k_x = 1.1$ for two choices of the shift vector ${\bf s} = (0.05, 0.025)$, ${\bf s} = (0.01, 0.005)$. The exact values for the real parts of the $G_{12}$ entry of the kernel matrix are plotted using a dashed (blue) curve, and those evaluated by the dipole-approximating function are plotted in solid (black), with ${\bf s} = (0.05, 0.025)$ in Fig.~\ref{g12comp}(a), and ${\bf s} = (0.01, 0.005)$ in Fig.~\ref{g12comp}(b). The improvement in the accuracy of the approximation for a smaller value of $|{\bf s}|$ is clearly evident in part (b).

\section{Evaluation of the coefficients for dipole approximation}
\label{sec:eval_dipole}
Recalling equations~(\ref{dip_als_1}),~(\ref{dip_als_2}) we give the explicit expressions required to evaluate the coefficients $S_n$, $D_n$ for a truncated semi-infinite grating (with truncation parameter $L$) using the wave scattering method. To determine ${\bf s}  \cdot \nabla_{\boldsymbol \xi} g(\beta; {\boldsymbol \xi};{\boldsymbol \xi^{(1)}})$, ${\bf s}  \cdot \nabla_{\boldsymbol \xi^{(1)}} g(\beta; {\boldsymbol \xi};{\boldsymbol \xi^{(1)}})$ we require the following functions and derivatives:
\begin{eqnarray}
 \quad \rho_{\xi} &  = & \left[\left(\xi_1 -  \xi_1^{(1)} \right)^2 + \left(\xi_2 -  \xi_2^{(1)} \right)^2 \right]^{\frac{1}{2}} , \quad  \frac{\partial g}{\partial \rho_{\xi}}  = - \frac{i}{8 \beta} \left[ H_1^{(1)} (\beta \rho_{\xi}) + \frac{2i}{\pi} K_1(\beta \rho_{\xi}) \right], 
\nonumber \\
  \frac{\partial \rho_{\xi}}{\partial \xi_{k}} &  = & \frac{\xi_k - \xi_k^{(1)}}{\rho_{\xi}} ; \quad \frac{\partial \rho_{\xi}}{\partial \xi_{k}^{(1)}} = \, - \frac{(\xi_k - \xi_k^{(1)})}{\rho_{\xi}}. 
\label{details} 
\end{eqnarray}
For the additional term ${\bf s}  \cdot \nabla_{\boldsymbol \xi} \left( {\bf s}  \cdot \nabla_{{\boldsymbol \xi^{(1)}}} g (\beta;{\boldsymbol \xi};{\boldsymbol \xi^{(1)}}) \right)$ in~(\ref{dip_als_2}), we also require the function $g_{\xi}^{(1)}$ and its derivative:
\begin{equation}
g_{\xi}^{(1)} = H_1^{(1)} (\beta \rho_{\xi}) + \frac{2i}{\pi} K_1 (\beta \rho_{\xi}); \frac{\partial g_{\xi}^{(1)}}{\partial \rho_{\xi}} = \beta H_0^{(1)} (\beta \rho_{\xi}) - \frac{H_1^{(1)}(\beta \rho_{\xi})}{\rho_{\xi}} - \frac{2i}{\pi} \left[\beta K_0(\beta \rho_{\xi}) + \frac{K_1(\beta \rho_{\xi})}{\rho_{\xi}} \right].
\label{greens_1_xi}
\end{equation}
Using the above formulae, equation~(\ref{dip_als_1}) becomes
\begin{eqnarray}
-e^{i\beta j a\cos\psi} & = & \frac{i}{8 \beta^2} \sum_{n=0}^{L-1} \bigg\{S_n \bigg[H_0^{(1)} \left(\beta |j-n|a \right)  +\frac{2i}{\pi} K_0 \left(\beta |j-n|a \right) \bigg] \label{dip_als_1_exp} \\
& + & D_n \frac{\beta s_1 (j - n)a}{|j - n|a} \bigg[  H_1^{(1)} \left(\beta |j-n|a \right)  +\frac{2i}{\pi} K_1 \left(\beta |j-n|a \right)\bigg] \bigg\} , \quad j = 0, 1, 2, \ldots
\nonumber 
\end{eqnarray}
Similarly, equation~(\ref{dip_als_2}) becomes
\begin{eqnarray}
&- &i \beta e^{i\beta j a\cos\psi} \left[s_1 \cos{\psi} + s_2 \sin{\psi} \right]  =  \frac{i}{8 \beta^2}  \bigg( \sum_{n=0}^{L-1}  -S_n \bigg\{ \frac{\beta s_1 (j - n)a}{|j - n|a} \bigg[  H_1^{(1)} \left(\beta |j-n|a \right) \bigg. \bigg. \nonumber \\ 
 &+& \frac{2i}{\pi} K_1 \left(\beta |j-n|a \right)\bigg] \bigg\}
 + \sum_{n=0}^{L-1} D_n \bigg\{ s_1^2 \bigg( \beta^2 \left[H_0^{(1)} \left(\beta |j-n|a \right)  -\frac{2i}{\pi} K_0 \left(\beta |j-n|a \right) \right]  \bigg. \nonumber \\ 
 & - & \frac{\beta}{|j - n|a} \left[H_1^{(1)} \left(\beta |j-n|a \right)  +\frac{2i}{\pi} K_1 \left(\beta |j-n|a \right)\right] \bigg) 
 + \frac{\beta s_2^2}{|j - n|a} \bigg[H_1^{(1)} \left(\beta |j-n|a \right) \bigg.  \nonumber \\ 
& + & \frac{2i}{\pi} K_1 \left(\beta |j-n|a \right) \bigg] \bigg\} \bigg),  \quad \quad \quad \quad j = 0, 1, 2, \ldots
\label{dip_als_2_exp}
\end{eqnarray}
We can express this system of equations in matrix form:
\begin{equation}
\left(
\begin{array}{c}
\textbf{F}^{(1)}\\
\textbf{F}^{(2)}\\
\end{array}
\right)= \frac{i}{8 \beta^2}
\left(
\begin{array}{cc}
\textbf{M}^{(1)}  & \textbf{M}^{(2)} \\
-\textbf{M}^{(2)} & \textbf{M}^{(3)}  \\
\end{array}
\right)
\left(
\begin{array}{c}
\textbf{S}\\
\textbf{D} \\
\end{array}
\right),
\label{dipole_me}
\end{equation}
where $\textbf{M}^{(1)}, \pm \textbf{M}^{(2)}, \textbf{M}^{(3)}$ are $L \times L$ block matrices and $\textbf{F}^{(1)}, \textbf{F}^{(2)}, \textbf{S}, \textbf{D}$ are $L \times 1$ column vectors with entries:
\begin{eqnarray}
M^{(1)}_{j n} & = & H_0^{(1)} \left(\beta |j-n|a \right)  +\frac{2i}{\pi} K_0 \left(\beta |j-n|a \right), \nonumber \\
M^{(2)}_{j n} & = & \frac{\beta s_1 (j - n)a}{|j - n|a} \bigg[  H_1^{(1)} \left(\beta |j-n|a \right)  +\frac{2i}{\pi} K_1 \left(\beta |j-n|a \right)\bigg], \nonumber \\
M^{(3)}_{j n} & = & s_1^2 \bigg( \beta^2 \left[H_0^{(1)} \left(\beta |j-n|a \right)  -\frac{2i}{\pi} K_0 \left(\beta |j-n|a \right) \right]  -   \frac{\beta}{|j - n|a} \bigg[H_1^{(1)} \left(\beta |j-n|a \right) \bigg. \bigg. \nonumber \\
& & + \frac{2i}{\pi} K_1 \left(\beta |j-n|a \right)\bigg] \bigg) 
  + \frac{\beta s_2^2}{|j - n|a} \left[H_1^{(1)} \left(\beta |j-n|a \right)  +\frac{2i}{\pi} K_1 \left(\beta |j-n|a \right)\right] ; \nonumber \\
F^{(1)}_j & = & -e^{i\beta j a\cos\psi}, \quad \quad F^{(2)}_j = -i \beta e^{i\beta j a\cos\psi} \left[s_1 \cos{\psi} + s_2 \sin{\psi} \right]. 
\label{dipole_me_entries}
\end{eqnarray}

To solve this system for a given incident field, the Bessel function derivatives require evaluation, in particular for the $\textbf{M}^{(3)}$ terms which introduce logarithmic singularities for the cases $j = n$ (the main diagonal entries in the block matrices).  The first derivative expressions in $\textbf{M}^{(2)}$ have leading order terms:
$$
\frac{i \beta \rho_{\xi} [-i \pi + 4 \ln{(\beta \rho_{\xi})} - 4 \ln{2} - 2 + 4\gamma]}{2 \pi} + \mathcal{O}\left((\beta \rho_{\xi})^3\right), \quad \rho_{\xi} = |j - n| a,
$$
where $\gamma$ represents the Euler constant, and which vanish for $j = n$. The second order derivatives may be expressed in the form:
\begin{equation}
\frac{i \beta^2}{\pi} \left[ 2 |{\bf s}|^2 \left(\ln{(\beta \rho_{\xi})} + \gamma - \ln{2} +\frac{\pi}{4i} \right) + (s_1^2 - s_2^2) \right] + \mathcal{O}\left((\beta \rho_{\xi})^2\right),
\label{regfn1st}
\end{equation}
which possesses a logarithmic term that grows slowly as $j \to n$. For sufficiently small $|{\bf s}|^2$, we  approximate these terms using 
\begin{equation}
-\frac{1}{8 \pi} \left[2 |{\bf s}|^2 \left(\ln{\left(\frac{\beta |{\bf s}|}{2}\right)} - 4\beta|{\bf s}| + \gamma - \ln{2} +\frac{\pi}{4i} \right) + (s_1^2 - s_2^2) \right] 
\label{regfn}
\end{equation}
for typical frequencies and shift vectors ${\bf s}$ that we consider here. The logarithmic term is replaced by the asymptotic term $$\ln{\left(\frac{\beta |{\bf s}|}{2}\right)} - 4\beta|{\bf s}|.$$ Note the change in the coefficient for~(\ref{regfn}) compared with~(\ref{regfn1st}), arising from the multiplication by the Green's function coefficient $i/(8\beta^2)$.

\section{Derivation of algebraic system for herringbone}
\label{app2}
Recalling from equation~(\ref{totfieldher}) that the total flexural displacement $u(x,y)$ is given by
\begin{multline}
u(x,y)=u_{\scriptsize{\mbox{inc}}}(x,y)+\sum_{n=0}^\infty A_n^{\scriptsize \mbox{(I)}} g(\beta; x,y; na_1,b/2) + \sum_{m=0}^\infty A_m^{\scriptsize \mbox{(II)}} g(\beta;x,y; s_1+ma_1,s_2+b/2)\\+\sum_{c=0}^\infty A_c^{\scriptsize \mbox{(III)}} g(\beta;x,y;ca_2,-b/2) + \sum_{d=0}^\infty A_d^{\scriptsize \mbox{(IV)}} g(\beta;x,y;t_1+da_2,t_2-b/2),
\label{totfieldherAPP}
\end{multline} 
the system of linear algebraic equations for the full pinned herringbone system is
\begin{multline}
-e^{i\beta[ja_1\cos\psi+(b/2)\sin\psi]}=\sum_{n=0}^\infty A_n^{\scriptsize \mbox{(I)}} g(\beta;ja_1,b/2;na_1,b/2) \\
+ \sum_{m=0}^\infty A_m^{\scriptsize \mbox{(II)}} g(\beta;ja_1,b/2;s_1+ma_1,s_2+b/2)
+\sum_{c=0}^\infty A_c^{\scriptsize \mbox{(III)}} g(\beta;ja_1,b/2;ca_2,-b/2) \\
+ \sum_{d=0}^\infty A_d^{\scriptsize \mbox{(IV)}} g(\beta;ja_1,b/2;t_1+da_2,t_2-b/2)  \\
j=0,1,2,\dots
\label{her1}
\end{multline}

\begin{multline}
-e^{i\beta[(s_1+la_1)\cos\psi+(s_2+b/2)\sin\psi]}=\sum_{n=0}^\infty A_n^{\scriptsize \mbox{(I)}} g(\beta;s_1+la_1,s_2+b/2;na_1,b/2) +\\  \sum_{m=0}^\infty A_m^{\scriptsize \mbox{(II)}} g(\beta;s_1+la_1,s_2+b/2;s_1+ma_1,s_2+b/2)+\sum_{c=0}^\infty A_c^{\scriptsize \mbox{(III)}} g(\beta;s_1+la_1,s_2+b/2;ca_2,-b/2)\\  + \sum_{d=0}^\infty A_d^{\scriptsize \mbox{(IV)}} g(\beta;s_1+la_1,s_2+b/2;t_1+da_2,t_2-b/2) \\ l=0,1,2, \ldots
\label{her2}
\end{multline}

\begin{multline}
-e^{i\beta[pa_2\cos\psi-(b/2)\sin\psi]}=\sum_{n=0}^\infty A_n^{\scriptsize \mbox{(I)}} g(\beta;pa_2,-b/2;na_1,b/2) \\
+ \sum_{m=0}^\infty A_m^{\scriptsize \mbox{(II)}} g(\beta;pa_2,-b/2;s_1+ma_1,s_2+b/2) +\sum_{c=0}^\infty A_c^{\scriptsize \mbox{(III)}} g(\beta;pa_2,-b/2;ca_2,-b/2) \\
+ \sum_{d=0}^\infty A_d^{\scriptsize \mbox{(IV)}} g(\beta;pa_2,-b/2;t_1+da_2,t_2-b/2) \\ 
p=0,1,2,\ldots
\label{her3}
\end{multline}

\begin{multline}
-e^{i\beta[(t_1+qa_2)\cos\psi+(t_2-b/2)\sin\psi]}=\sum_{n=0}^\infty A_n^{\scriptsize \mbox{(I)}} g(\beta;t_1+qa_2,t_2-b/2;na_1,b/2) +\\  \sum_{m=0}^\infty A_m^{\scriptsize \mbox{(II)}} g(\beta;t_1+qa_2,t_2-b/2;s_1+ma_1,s_2+b/2)+\sum_{c=0}^\infty A_c^{\scriptsize \mbox{(III)}} g(\beta;t_1+qa_2,t_2-b/2;ca_2,-b/2) \\ + \sum_{d=0}^\infty A_d^{\scriptsize \mbox{(IV)}} g(\beta;t_1+qa_2,t_2-b/2;t_1+da_2,t_2-b/2) \\ q=0,1,2,\ldots
\label{her4}
\end{multline}

The discrete Wiener-Hopf approach implemented in Section~\ref{kerher} is repeated here. We introduce similar notation for $N, M, C, D \in \mathbb{Z}$:

\begin{equation}
u(Na_1, b/2)=\Bigg \{ 
\begin{array}{c l}  
0, &  \quad\quad N \ge 0 \\
\\
B_N^{\scriptsize \mbox{(I)}}, &  \quad\quad N  <  0
\label{bb1n}
\end{array} 
\end{equation}

\begin{equation}
\hspace{-1.3cm}
u(s_1+Ma_1, s_2+b/2)=\Bigg \{ 
\begin{array}{c l}  
0, &  \quad\quad M \ge 0 \\
\\
B_M^{\scriptsize \mbox{(II)}}, &  \quad\quad M  <  0
\end{array}  
\label{bb2m}
\end{equation}

\begin{equation}
u(Ca_2, -b/2)=\Bigg \{ 
\begin{array}{c l}  
0, &  \quad\quad C \ge 0 \\
\\
B_C^{\scriptsize \mbox{(III)}}, &  \quad\quad C  <  0
\end{array}  
\label{bb3r}
\end{equation}

\begin{equation}
\hspace{-1.2cm}
u(t_1+Da_2,t_2 -b/2)=\Bigg \{ 
\begin{array}{c l}  
0, & \quad\quad D \ge 0 \\
\\
B_D^{\scriptsize \mbox{(IV)}}, & \quad\quad D  <  0
\end{array}  
\label{bb4s}
\end{equation}

\begin{equation}
u_{\scriptsize{\mbox{inc}}}(Na_1, b/2)=F_N^{\scriptsize \mbox{(I)}},~~~~~~\quad
u_{\scriptsize{\mbox{inc}}}(s_1+Ma_1, s_2+b/2)=F_M^{\scriptsize \mbox{(II)}};
\label{f12h}
\end{equation} 
\begin{equation}
u_{\scriptsize{\mbox{inc}}}(Ca_2, -b/2)=F_C^{\scriptsize \mbox{(I)}},~~~~~~\quad
u_{\scriptsize{\mbox{inc}}}(t_1+Da_2, t_2-b/2)=F_D^{\scriptsize \mbox{(II)}}.
\label{f12hh}
\end{equation}

Here $B_N^{\scriptsize \mbox{(I)}}$ to $B_D^{\scriptsize \mbox{(IV)}}$  represent the unknown amplitudes of the total flexural displacement at the points $(Na_1, b/2)$,  $(s_1+Ma_1, s_2+b/2)$, $(Ca_2, -b/2)$ and $(t_1+Da_2, t_2-b/2)$  for, respectively, $N,M,C,D < 0$ i.e. in the ``reflection'' region to the left of each grating pair. The field incident at the pins is denoted by $F_N^{\scriptsize \mbox{(I)}}, F_M^{\scriptsize \mbox{(II)}}, F_C^{\scriptsize \mbox{(III)}}, F_D^{\scriptsize \mbox{(IV)}}$.

Similar to the preceding case of Section~\ref{kerher}, we consider the displacement (\ref{totfieldherAPP}) at four field points  ${\bf r} = (Na_1,b/2)$, ${\bf r} = (s_1+Ma_1,s_2+b/2)$, ${\bf r} = (Ca_2,-b/2)$ and ${\bf r} = (t_1+Da_2,t_2-b/2)$. As the definitions~(\ref{bb1n})-(\ref{f12hh}) suggest, the scattering coefficients are extended for $m,n,c,d < 0$, where they are evaluated to be zero since there are no pins present in this region. Assuming the symmetric herringbone case ($a_1 = a_2$), after applying the discrete Fourier Transform to equations~(\ref{totfieldherAPP})-(\ref{f12hh}) using Fourier variable $k$ and indices of summation $N$, $M$, $C$ and $D$, we obtain:

\begin{multline}
\sum_{N=-\infty}^\infty B_N^{\scriptsize \mbox{(I)}}e^{ikNa}=\sum_{N=-\infty}^\infty F_N^{\scriptsize \mbox{(I)}}e^{ikNa}+\sum_{n=-\infty}^\infty A_n^{\scriptsize \mbox{(I)}}e^{ikna}\sum_{j=-\infty}^\infty g(\beta;ja,b/2;0,b/2)e^{ikja}
\\  
+\sum_{m=-\infty}^\infty A_m^{\scriptsize \mbox{(II)}}e^{ikma}\sum_{j=-\infty}^\infty g(\beta;ja,b/2;s_1,s_2+b/2)e^{ikja}
\\
+\sum_{c=-\infty}^\infty A_c^{\scriptsize \mbox{(III)}}e^{ikca}\sum_{j=-\infty}^\infty g(\beta;ja,b/2;0,-b/2) e^{ikja}
\\
+\sum_{d=-\infty}^\infty A_d^{\scriptsize \mbox{(IV)}} e^{ikda} \sum_{j=-\infty}^\infty g(\beta;ja,b/2;t_1,t_2-b/2)e^{ikja},
\label{totfieldher111}
\end{multline}

\begin{multline}
\sum_{M=-\infty}^\infty B_M^{\scriptsize \mbox{(II)}}e^{ikMa}=\sum_{M=-\infty}^\infty F_M^{\scriptsize \mbox{(II)}}e^{ikMa}+\sum_{n=-\infty}^\infty A_n^{\scriptsize \mbox{(I)}}e^{ikna}\sum_{j=-\infty}^\infty g(\beta;s_1+ja,s_2+b/2;0,b/2)e^{ikja}
\\
+ \sum_{m=-\infty}^\infty A_m^{\scriptsize \mbox{(II)}}e^{ikma} \sum_{j=-\infty}^\infty g(\beta;s_1+ja,s_2+b/2;s1,s_2+b/2)e^{ikja}\\
+\sum_{c=-\infty}^\infty A_c^{\scriptsize \mbox{(III)}}e^{ikca}\sum_{j=-\infty}^\infty g(\beta;s_1+ja,s_2+b/2;0,-b/2) e^{ikja}
\\
+ \sum_{d=-\infty}^\infty A_d^{\scriptsize \mbox{(IV)}}e^{ikda}\sum_{j=-\infty}^\infty g(\beta;s_1+ja,s_2+b/2;t_1,t_2-b/2)e^{ikja},
\label{totfieldher222}
\end{multline}

\begin{multline}
\sum_{C=-\infty}^\infty B_C^{\scriptsize \mbox{(III)}}e^{ikCa}=\sum_{C=-\infty}^\infty F_C^{\scriptsize \mbox{(III)}}e^{ikCa}+\sum_{n=-\infty}^\infty A_n^{\scriptsize \mbox{(I)}}e^{ikna}\sum_{j=-\infty}^\infty g(\beta;ja,-b/2;0,b/2)e^{ikja}
\\
+  \sum_{m=-\infty}^\infty A_m^{\scriptsize \mbox{(II)}}e^{ikma}\sum_{j=-\infty}^\infty g(\beta;ja,-b/2;s_1,s_2+b/2)e^{ikja}
\\
+\sum_{c=-\infty}^\infty A_c^{\scriptsize \mbox{(III)}}e^{ikca}\sum_{j=-\infty}^\infty g(\beta;ja,-b/2;0,-b/2) e^{ikja}
\\
+ \sum_{d=-\infty}^\infty A_d^{\scriptsize \mbox{(IV)}}e^{ikda}\sum_{j=-\infty}^\infty g(\beta;ja,-b/2;t_1,t_2-b/2)e^{ikja},
\label{totfieldher333}
\end{multline}

\begin{multline}
\sum_{D=-\infty}^\infty B_D^{\scriptsize \mbox{(IV)}}e^{ikDa}=\sum_{D=-\infty}^\infty F_D^{\scriptsize \mbox{(IV)}}e^{ikDa}+\sum_{n=-\infty}^\infty A_n^{\scriptsize \mbox{(I)}}e^{ikna}\sum_{j=-\infty}^\infty g(\beta;t_1+ja,t_2-b/2;0,b/2)e^{ikja}
\\
+  \sum_{m=-\infty}^\infty A_m^{\scriptsize \mbox{(II)}}e^{ikma}\sum_{j=-\infty}^\infty g(\beta;t_1+ja,t_2-b/2;s_1,s_2+b/2)e^{ikja}\\
+ \sum_{c=-\infty}^\infty A_c^{\scriptsize \mbox{(III)}}e^{ikca}\sum_{j=-\infty}^\infty g(\beta;t_1+ja,t_2-b/2;0,-b/2) e^{ikja}
\\
+ \sum_{d=-\infty}^\infty A_d^{(IV)}e^{ikda}\sum_{j=-\infty}^\infty g(\beta;t_1+ja,t_2-b/2;t_1,t_2-b/2)e^{ikja}.
\label{totfieldher444}
\end{multline}
The resulting system is an extended version of that for the shifted pair presented in Section~\ref{kerher}.

\section{Waveguide modes}
\label{wg_table}
In Table~\ref{tab:etao} we show computed values for the optimal grating separation $b^*$ for a waveguide consisting of an unshifted pair of infinite gratings. The parameter $\eta^*$ represents an idealised grating separation for first order waveguide modes, determined by the approximate waveguide model for the Helmholtz operator (neglecting the evanescent modes for the biharmonic case) and by~(\ref{wg_model}).  
For gratings with unit periodicity ($a = 1$) and an incident plane wave characterised by $\psi$, the optimised spectral parameter value $\beta^*$ is obtained for maximum single grating reflectance~\cite{has2014}. Using equation~(\ref{kx}), a corresponding $k_x^*$ value is obtained. The various pairs of $(\beta^*, k_x^*)$ values are used to determine resonant trapped modes, with excellent agreement with the approximate waveguide model, as shown by columns 4 and 5 of Table~\ref{tab:etao}.

\begin{table}[h]
\begin{center}
\caption{Resonant frequencies $\beta^*$ and wavenumbers $k_x^*$, and the corresponding optimised grating separation $b^*$ for pairs of unshifted infinite gratings for various angles of incidence $\psi$.}
~~
\label{tab:etao}
\begin{tabular}{ccccc}
\hline\noalign{\smallskip}
$\psi$ & $\beta^*$ & $k_x^*$ & $\eta^*$ & $b^*$  \\
\noalign{\smallskip}\hline\noalign{\smallskip}
$0$ & $4.456001$ & 0  & $0.705025$ & 0.705251 \\ 
$\pi/60$ &$4.438147$ & 0.232275 &0.708833 & 0.709101  \\
$\pi/30$ & 4.387466  &  0.458615    &$0.719982$ & 0.720367 \\ 
$\pi/20$ &$4.311191$ & 0.674419 &0.73779 &  0.738331 \\
$\pi/15$ & $4.217801$  & 0.87693    &$0.761482$ & 0.762182  \\  
$\pi/12$ & 4.11476 & 1.06498 & $0.790427$  & 0.79126 \\ 
$\pi/10$ & $4.007707$ & 1.23845 & $0.824228$  & 0.825150\\  
$7\pi/60$ & $3.900536$  & 1.39783     &$0.862728$ & 0.863689 \\ 
$2\pi/15$ & 3.79580 & 1.54389    &$0.905975$ & 0.906927\\ 
$3\pi/20$ & 3.6950925  & 1.67754  &$0.954209$ & 0.955108 \\  
$\pi/6$ & $3.599363$  & 1.79968  &$1.00784$ & 1.00866 \\ 
$11\pi/60$ & $3.509134$ & 1.91121  &$1.06748$ &1.06818\\
$\pi/5$ & 3.424645   & 2.01296  &$1.133905$ & 1.134490\\ 
$\pi/4$ &3.205694  & 2.26677  &$1.38593$ &   1.386185  \\ 
$\pi/3$ &2.94716   & 2.55232   &$2.131946$ &2.131958      \\ 
  \noalign{\smallskip}\hline
  \end{tabular}
    \end{center}
  \end{table}


\begin{thebibliography}{3}

\bibitem{arlt} Arlt G, Sasko P. 1980. Domain configuration and equilibrium size of domains in BaTiO$_3$ ceramics. {\it J. Appl. Phys.} {\bf 51}, 4956-4960.

\bibitem{tsou} Tsou NT, Potnis PR, Huber JE. 2011. Classification of laminate domain patterns in ferroelectrics. {\it Phys. Rev. B} {\bf 83}, 184120.

\bibitem{glasgow} Farooq MU, Villaurrutia R, MacLaren I, Burnett TL, Comyn TP, Bell AJ, Kungl H, Hoffman MJ. 2008. Electron backscatter diffraction mapping of herringbone domain structures in tetragonal piezoelectrics. {\it J. Appl. Phys.} {\bf 104}, 024111.

\bibitem{vyas} Vyas VS, Gutzler R, Nuss J, Kern K, Lotsch BV. 2014. Optical gap in herringbone and $\pi$- stacked crystals of $[1]$benzothieno[3,2-$b$]benzothiopene and its brominated derivative. {\it Cryst. Eng. Comm.} {\bf 16}, 7389-7392.

\bibitem{wang} Wang C, Dong H, Li H, Zhao H, Meng Q, Hu W. 2010. Dibenzothiopene derivatives: from herringbone to lamellar packing motif. {\it Cryst. Growth Des.} {\bf 10}, 4155-4160.

\bibitem{zhang} Zhang Y {\it et al.} 2016. Probing carrier transport and structure-property relationship of highly ordered organic semiconductors at the two-dimensional mobilities. {\it Phys. Rev. Lett.} {\bf 116}, 016602.

\bibitem{RCM_ABM_NVM}
McPhedran RC, Movchan AB, Movchan NV. 2009. 
Platonic crystals: Bloch bands, neutrality and defects.
{\it Mech. Mater.} {\bf 41}, 356-363.

\bibitem{mead}
Mead DJ. 1996. Wave propagation in continuous periodic structures:Research contributions from Southampton, 1964-1995. {\it J. Sound Vib.} {\bf 190}:(3), 495-524.


\bibitem{Evans} 
Evans DV, Porter R. 2007. Penetration of flexural
waves through a periodically constrained thin elastic plate in {\it vacuo} and floating
on water. {\it J. Eng. Maths.} {\bf 58}, 317-337.

\bibitem{MJAS_RP_TDW}
Smith MJA, Porter R, Williams TD. 2012. The effect on bending waves by defects in pinned elastic plates. {\it J. Sound Vib.} {\bf 331}, 5087-5106.

\bibitem{anton1} Antonakakis T, Craster RV. 2012.  
High-frequency asymptotics for microstructured thin elastic plates and platonics. {\it Proc. R. Soc. A} {\bf 468}, 1408-1427.

\bibitem{anton2} Antonakakis T, Craster RV, Guenneau S. 2014. Moulding and shielding flexural waves in elastic plates. {\it EPL} {\bf 105}, doi:10.1209/0295-5075/105/54004. 

\bibitem{ABM_NVM_RCM}
Movchan AB, Movchan NV, McPhedran RC. 2007.
Bloch-Floquet bending waves in perforated thin plates. 
{\it Proc. R. Soc. A} {\bf 463}, 2505-2518.

\bibitem{has2014} Haslinger SG, Movchan AB, Movchan NV, McPhedran RC. 2014. Symmetry and resonant modes in platonic grating stacks. {\it Waves in Random and Complex
Media}. {\bf 24}:(2), 126-148.

\bibitem{Dirac}
McPhedran RC, Movchan AB, Movchan NV, Brun M, Smith MJA. 2015. `Parabolic' trapped modes and steered Dirac cones in platonic crystals.
{\it Proc. R. Soc. A} {\bf 471}, 20140746.

\bibitem{Evans_Porter}
Evans DV, Porter R. 2008. Flexural waves on a pinned semi-infinite thin elastic plate. {\it Wave Motion.} {\bf 45}, 745-757.

\bibitem{SGH_RVC_ABM_NVM_ISJ}
Haslinger SG, Craster RV, Movchan AB, Movchan NV, Jones IS.  2015. 
Dynamic interfacial trapping of flexural waves in structured plates. 
{\it Proc. R. Soc. A} {\bf 472}, 20150658. 

\bibitem{SGH_NVM_ABM_ISJ_RVC}
Haslinger SG, Movchan NV, Movchan AB, Jones IS, Craster RV.  2017. 
Controlling flexural waves in semi-infinite platonic crystals with resonator-type scatterers. 
{\it Q. J. Mech. Appl. Math.} {\bf 70}:(3), 216-247.


\bibitem{ISJ_NVM_ABM} Jones IS, Movchan NV, Movchan AB. 2016. Blockage and guiding of flexural waves in a semi-infinite double grating. {\it Math. Meth. Appl. Sci.} {\bf 40}:(9), 3265-3282.

\bibitem{hills1}  Hills NL, Karp SN. 1965. Semi-infinite diffraction gratings I. {\it Comm. Pure Appl. Math.} {\bf 18}, 203-233. 

\bibitem{foldy} Foldy LL. 1945. The multiple scattering of waves I, General theory of isotropic scattering by randomly distributed scatterers. {\it Phys. Rev.} {\bf 67}, 107-119. 

\bibitem{linton} Linton CM, Martin PA. 2004. Semi-infinite arrays of isotropic point scatterers - a unified approach. {\it SIAM J. Appl. Math.} {\bf 64}, 1035-1056.

\bibitem{mahan} Mahan GD, Obermair G. 1969. Polaritons at surfaces. {\it Phys. Rev.} {\bf 183}:(3), 834-841.

\bibitem{mead_ca} Mead CA. Exactly soluble model for crystal with spatial dispersion. {\it Phys. Rev. B} {\bf 15}:(2), 519-532.

\bibitem{belov}
Belov PA, Simovski CR. 2006. Boundary conditions for interfaces of electromagnetic crystals and the generalized Ewald-Oseen extinction principle. {\it Phys. Rev. B} {\bf 73}, 045102.

\bibitem{albani}
Albani M, Capolino F. 2011. Wave dynamics by a plane wave on a half-space metamaterial made
of plasmonic nanospheres: a discrete Wiener-Hopf formulation. {\it J. Opt. Soc. Amer. B} {\bf 28}, 2174-2185.

\bibitem{tymis}
Tymis N, Thompson I.  2011. Low frequency scattering by a semi-infinite lattice of cylinders. 
{\it Q. J. Mech. Appl. Math.} {\bf 64}, 171-195.

\bibitem{tymis2}
Tymis N, Thompson I.  2014. Scattering by a semi-infinite lattice and the excitation of Bloch waves. 
{\it Q. J. Mech. Appl. Math.} {\bf 67}, 469-503.


\bibitem{MJAS_MHM_RCM}
Smith MJA, Meylan MH, McPhedran RC. 2014. Density of states for platonic crystals and clusters. {\it SIAM J. Appl. Math.} {\bf 74}, 1551-1570.



\bibitem{Abram}
Abramowitz M, Stegun IA.  1965. {\em Handbook of
Mathematical Functions with Formulas, Graphs, and Mathematical
Tables}. Dover Reprint.



\end{thebibliography}
\end{document}